# Seeking for ultrashort "non-bonded" hydrogen-hydrogen contacts in some rigid hydrocarbons and their chlorinated derivatives

Rohoullah Firouzi[1] and Shant Shahbazian[2]

[1]Department of Chemistry, Chemistry and Chemical Engineering Research Center of Iran, P.O. Box 14968-13151, Tehran, Iran.

E-mails: RFirouzi@ccerci.ac.ir and firouzi.chemist@yahoo.com

[2]Faculty of Chemistry, Shahid Beheshti University, G. C. , Evin, Tehran, Iran, 19839, P.O. Box 19395-4716. Tel/Fax: 98-21-22431661

E-mail: chemist_shant@yahoo.com

**Abstract**

In this communication a systematic computational survey on some rigid hydrocarbon skeletons, e.g. half-cage pentacyclododecanes and tetracyclododecanes, and their chlorinated derivatives in order to seek for the so-called ultrashort "non-bonded" hydrogen-hydrogen contacts is done. It is demonstrated that upon a proper choice and modifications of the main hydrocarbon backbones, and addition of some chlorine atoms instead of the original hydrogen atoms in parts of the employed hydrocarbons, the resulting strain triggers structural changes yielding ultrashort hydrogen-hydrogen contacts with inter-nuclear distances as small as 1.38 Å. Such ultrashort contacts are clearly less than the world record of an ultrashort non-bonded hydrogen-hydrogen contact, 1.56 Å, very recently realized experimentally by Pascal and coworkers in *in,in*-bis(hydrosilane) [J. Am. Chem. Soc. 135, 13235 (2013)]. The resulting computed structures as well as the developed methodology for structure design open the door for constructing a proper set of molecules for future studies on the nature of the so-called non-bonded hydrogen-hydrogen interactions that is now an active and controversial area of research.

**Introduction**

Recently, Pascal and coworkers have reported the synthesis and characterization of novel macrobicyclic structure, *in,in*-bis(hydrosilane), claiming a new "world record" for the so-called non-bonded H…H distance, ~1.56 Å [1]. This discovery from both experimental and computational viewpoints is important. From experimental perspective, it is a real synthetic breakthrough since it bypassed the previous world record of the non-bonded H…H contacts, 1.61 Å, reported by Ermer and coworkers almost 28 years ago in a derivative of half-cage pentacyclododecane [2], by a new "molecular design" (for more on the unique synthetic and design approach of Pascal's school see [3-6] while for a perspective on the history of ultrashort non-bonded H…H contacts see [2,7-13]). On the other hand, from theoretical standpoint, such ultrashort H…H contacts are also of particular importance since the nature of the so-called non-bonded H…H interactions has been a matter of disputes and exchanges since 2003 [14-23]; from the "classical" viewpoint the interactions are exclusively destabilizing as one expects from the steric effects [24-27] (the orthodox van der Waals radius of hydrogen atom dictates ~2.4 Å separation as the optimal distance for non-bonded H…H contacts) while claims have been made by the quantum theory of atoms in molecules (QTAIM) analysis that these interactions are bonded and stabilizing [14,18,20]. Currently, the controversy is ongoing, questioning/confirming the classical view on the nature of such H…H interactions using various quantum chemical analyses though the whole problem may be seen as part of the intricate problem of rigorous elucidation of the nature of stabilizing/destabilizing interactions of atoms (or groups of atoms) of a molecule [28-61].

In contrast to this rich background, the aim of this communication is modest, just trying to introduce a new set of molecular structures, albeit not exotic structures, with ultrashort non-bonded H…H contacts leaving the questions on the nature of H…H interactions as well as their QTAIM analysis into a future study (accordingly, the word "non-bonded" is just used in this communication to discriminate these H…H contacts from the dihydrogen bonds that the nature of their H…H interactions has been clarified previously and completely different from the contacts considered in this communication (*vide infra*)). In order to reach this goal, the main and modified hydrocarbon half-cage pentacyclododecane of Ermer [2] as well as tetracyclododecane [9,27] are used as backbones, and in a series of structural modifications, new motifs are proposed and then geometrically optimized to see whether it is possible to introduce structures with non-bonded H…H contacts below 1.56 Å. Accordingly, the half-cage pentacyclododecane backbone is a particularly promising target to start with since its "rigidity" prevents conformational variations that hamper the ultrashort H…H contacts (see the footnote 10 in [1] for an analysis of this point in *in,in*-bis(hydrosilane)). Our guide in this search and the basis of "design strategy" [62] has been the classical paradigm of the steric effects; steric destabilization through using large atoms in parts of a molecule as well as employing small rings may energetically favor ultrashort H…H contacts in other part of a molecule yielding local minima. Whether these local minima are synthesizable quantitatively in usual laboratory conditions is not our concern in this study [63,64], but the aim is constructing a set of molecules that reveal a pattern of "semi-continuous" variations of non-bonded H…H contacts in the range of ultrashort distances making them an ideal set for future comparative QTAIM and bonding analysis.

**Computational guideline**

In order to start the survey, at first step the structures 1 and 2 (see Figure 1 for numbering of species) were considered at various computational levels; since concomitant X-ray and neutron diffraction analyses have been done on these two molecules [9], the derived experimental H…H distances are reliable and may be used as references for the selection of proper density functionals. Inspection of Table 1, containing the computed and experimentally derived H…H distances, reveals that from a large number of considered density functionals/basis set combinations, M06L/6-311++G(d,p) [65-67] works relatively better though none of density functional/basis set combinations seems completely capable of reproducing both distances. Accordingly, to make the computational results more reliable, not only M06L but also the well-known HF, B3LYP [68] and MP2 [69] methods were also used in conjunction with 6-311++G(d,p) basis set for geometry optimization and force constant (vibrational frequencies) calculations employing a modified version of GAMESS suite of programs [70] (using more extended basis sets for larger members of this set of molecules was computationally intractable thus 6-311++G(d,p) was preferred for the homogeneity of calculations, though for MP2 method the force constants, because of computational cost, were just derived with 6-31G(d) basis set). Figure 1 depicts the final optimized structures while Tables 2, 3, S1-S4 gather the computed H…H distances as well as the C-H internuclear distances, the X-C-H (X = H, Cl) angles, and the symmetric and asymmetric C-H stretching vibrational frequencies of the $CH_2$ and CHCl units that their hydrogen atoms are involved in ultrashort contacts (Supplementary information contains the full Cartesian coordinates of all derived optimized geometries at all considered computational levels and Tables S1, S2, S3 and S4). The main focus of this communication is on the patterns of variations of the derived H…H distances in this set of molecules and their relationship with other geometrical parameters, offered in Table 3 at M06L/6-311++G(d,p) computational level (for other computational levels see Tables S1-S4), that are depicted in Figure 2. Since five distinct computational levels have been employed in this study, except otherwise stated, in all discussions an arithmetic mean value, derived from computed H…H distances at various computational levels, is reported in the text as a trustable value.

**Results and discussion**

To perform detailed comparisons, the structures depicted in Figure 1 are categorized into subsets at first step and next, general correlations emerging from considering the whole set are considered briefly. Accordingly, first the structures 1-3 are compared at first and then the subset of 4-6 is considered to comprehend the role of attaching "peripheral" small carbon rings to the structure 1. Subsequently, the structures 6-9 are compared to unravel the role of chlorination and concomitant induced steric effect; the subset containing structures 10-15 are used for the same propose. Finally, the structures 16-18 are detailed to check the efficiency of half-cage structures and their chlorinated derivatives to squeeze H…H contacts. The geometrical changes which are induced in the $CH_2$ and CClH units involved in the ultrashort H…H contacts are the main focus of this communication; in the rest of this communication, C-H bonds that their hydrogen is involved in ultrashort contacts are termed "inner" C-H, while the C-H bonds that are directed toward outside, far from contact region, are called "outer" C-H. Other geometric parameters which affected by steric effects in these molecules are mentioned in discussions focused on each subset. Furthermore, from various vibrational normal modes those mainly originating from C-H vibrations involved in the ultrashort H…H contacts have been selected and considered; this engages the symmetric and anti-symmetric stretching vibrations of the two target C-H bonds. These high frequency C-H stretching modes were previously observed experimentally and their origin has been attributed to H…H steric effects [71,72].

The structures 1-3 and their derivatives have been characterized experimentally so comparison with the computed geometrical parameters and vibrational frequencies are feasible. However, in the case of 3 that is a derivative of 2 since solely the X-ray diffraction data have been used for structure elucidation [13], uncertainty remains on the reliability of the estimated H…H distance. The estimated H…H distance, albeit determined indirectly, is just ~1.62 Å which is much shorter than that observed in 2, 1.71 Å [9]. This bold difference, ~0.09 Å, which must seemingly originates from the steric effect associated with the bulk moiety attached to 3, is not in line with values derived from present calculations that collectively point to ~1.70 Å near to the value derived for 2 both experimentally and computationally. Since neutron diffraction analysis has not been done on 3, one may conclude that the estimated H…H distance in 3 needs to be viewed skeptically. Inspection of Tables 3 and S1-S4 demonstrates that in this subset, in response to the contraction of H…H distance, the inner C-H inter-nuclear distances as well as the H-C-H angles all contract while both the symmetric and anti-symmetric vibrations shift to higher wavenumbers. These geometrical and vibrational patterns are typical for molecules containing ultrashort H…H contacts, observed both experimentally and computationally, and are interpreted in steric paradigm as routes toward reorganization of geometry in order to avoid, as far as possible, the unfavorable/destabilizing steric effects originating from H…H contacts (for a detailed but transparent discussion see particularly pages 76-84 in [27]).

The case of 3 casts doubt that simple design strategy based on attaching a single bulky moiety to main skeleton being efficient enough to reach ultrashort H…H contacts. Thus, the design strategy was switched to modify what was used successfully by Ermer, the half-cage pentacyclododecanes [2], to produce ultrashort H…H distances; the structures 4-6 were proposed and optimized seeking the role of peripheral ring size contraction on the H…H contacts. Inspection of Table 2 indeed confirms that this is an efficient strategy and the H…H distance reduces from ~1.73 Å in 4 to ~1.60 Å in 5 and then to ~1.52 Å in 6 with an overall ~0.2 Å shortening. Even in this initial try, 6 bypasses the world record of ~1.56 Å, though there is no question that it is largely under strain even if no negative force constants were observed; as a witness for large strain, the variations of C-C distances at MP2/6-311++G(d,p) level range from ~1.66 Å in the corners to ~1.46 Å in the center of the molecule. The geometrical data in Tables 3 and S1-S4 also conform to this picture since from 4 to 6 the inner C-H inter-nuclear distance shortens considerably, whereas the outer C-H inter-nuclear distance is almost the same observed in 1, and the H-C-H angle also contracts appreciably. Both vibrational modes also shift to larger wavenumbers in this subset reaching to ~3220 and ~3301 $cm^{-1}$ for anti-symmetric and symmetric modes, respectively, at M06L/6-311++G(d,p) level that are the largest in the considered chlorine-free hydrocarbons.

Invoking structures with smaller H…H distances, bulky chlorine atoms, close to each other, were replaced with smaller hydrogen atoms consecutively in the corners of 6 trying to add more steric interactions yielding structures 7 and 8. Obviously, the strategy works well and the H…H distances further shorten to ~1.38 Å in 7 and then to ~1.17 Å in 8. However, structure 7 is quite deformed and C-C distances at the MP2/6-311++G(d,p) level range from ~1.69 Å in the corners to ~1.46 Å in the center of the molecule while 8 reveals serious signs of instability since at all considered computational levels no local minimum was found. Interestingly, animation of the corresponding mode of vibrations in 8 reveals that the imaginary mode involves the two target hydrogen atoms that are trying to avoid the "head to head" arrangement. Re-optimization of 8 relaxing the head to head arrangement yields 9 that is now a stable structure at HF, B3LYP and MP2/6-31G(d) levels but yet a saddle point at the M06L/6-311++G(d,p) level. The H…H distance is now reproduced in a range of ~1.26-1.28 Å and even with a conservative estimate, based on MP2/6-31G(d) optimized geometry, it seems H…H distance is not above ~1.30 Å (because of the number of chlorine atoms it was not feasible to optimize this structure at the MP2/6-311++G(d,p) level). However, 9 is highly deformed and C-C distances at the B3LYP/6-311++G(d,p) level range from ~1.90 Å in the corners to ~1.48 Å in the center of the molecule; with such huge deformations it is hard to trust the results derived from "single reference" computational methods. Nevertheless, the strategy of using some chlorine atoms, close to each other, and the imposed steric effects seems to be an efficient method to produce ultrashort non-bonded H…H contacts. Inspection of Tables 3 and S1-S4 also reveals the same pattern observed for previous subset namely, shortening of inner C-H distances, contraction of Cl-C-H angle and elevation of vibrational frequencies; interestingly, upon relaxing the head to head arrangement going from 8 to 9 all geometrical and vibrational data clearly reveal that 9 experiences less strain than 8.

Since the strategy of replacing chlorine with hydrogen atoms works well, the same design strategy was used to introduce 10, which has been inspired from the experimentally discovered 1 [9,27], then constructing congested chlorine substituted derivatives, the structures 11-15. The H…H computed distances in 10, ~1.72 Å, is very similar to 2 while for 11 with four chlorine atoms in the corners this distance diminishes to ~1.57 Å that is almost equal to the one estimated in *in,in*-bis(hydrosilane) [1]. Although C-C inter-nuclear distance variations in 11 are much more narrowed than 7 and 9, just ~1.55-1.60 Å, but the angles around carbon atoms at the corners are largely distorted from the standard tetrashedral angle revealing a strained species [73]. Adding two more chlorine atoms to the central carbons, yielding 12, diminishes the H…H computed distances further, ~1.51 Å. Further chlorination of 12 gives 13 with H…H distance ~1.43 Å while adding another chlorine atom yields 14 with a clear sign of instability; a negative force constant at all the considered computational levels. Indeed, similar to the case of 8, animation of this mode demonstrates that the problematic vibrational mode results from head to head arrangement, so a relaxed structure was optimized yielding the stable local minimum 15. In contrast to the previous structures the computed H…H distances from density functionals, ~1.38 Å, is boldly shorter than MP2, ~1.43 Å; however, even assuming a mean value, ~1.40 Å, this is shorter than the contacts of the previously considered structures of this subset. One may conclude that the main rational behind the design works well though the resulting congested molecules are clearly strained; once again this is best illustrated inspecting Tables 3, S1-S4 that reveals the same patterns observed for previous chlorinated subset.

As a final attempt, the structures 16 and 17, similar to the one characterized by Ermer [5], were considered and as it is evident from Table 2, the former is unstable revealing imaginary normal modes while an octa chlorinated derivative of 17 yields 18 revealing another ultrashort H…H distance, ~1.40 Å. Nevertheless, clear signs of strain are observed in 18 considering the C-C distances that are scattered in the range from ~1.52 Å to ~1.65 Å at the MP2/6-311++G(d,p) level.

In order to have a more general picture Figure 2 depicts the inner C-H distances, the X-C-H angles as well as the C-H vibrational frequencies in four panels and the emerging patterns point to evident correlations. Upon the decrease of H…H contact, the geometrical relaxations of the $CH_2$ and CClH units are markedly interconnected and molecules even with different skeletons respond similarly; namely, contractions of the inner C-H distances and the X-C-H angles as well as concomitant elevation of the C-H stretching force constants. However, in contrast to this similarity of "local" response of the involved units, as also emphasized by Allinger [27], because of "propagation of stress" the effect of ultrashort H…H contact may be quite "non-local"; in other words, units distant from the center of stress may experience significant geometrical distortions. Furthermore, the non-local fingerprint of the stress depends on the nature of hydrocarbon skeleton and reveals itself differently in different skeletons. Such "long range" structural effects of ultrashort H…H contacts deserved to be considered in detail in carefully designed molecular structures since they are directly relevant to the problematic nature of H…H interactions and related controversy.

**Conclusion and Prospects**

This study suggests that by a proper combination of rigid hydrocarbon skeletons and some bulky substitutions, close to each other to make congested systems, designing ultrashort non-bonded H…H contacts is feasible at least computationally (of course the main challenge is experimental realization of these or similar structures that is far from trivial). Evidently, even with a conservative estimate, there is a large gap between the world's new record of H…H distance, ~1.56 Å, with what is possible to achieve in this series of species, ~1.38 Å; however, there is no reason to believe that considered structural motifs are the most proper candidates for reaching ultrashort H…H contacts and searching for smaller contacts yet remains an ongoing challenge. Indeed, it has been demonstrated that in the case of dihydrogen bonds, H…H contacts with ultrashort distances, ~1.0-1.3 Å, are conceivable [74-77] though the electrostatic stabilizing interactions between hydrogen atoms are a major driving force in molecules with dihydrogen bonds that are absent in the case of hydrocarbons considered in this study (the typical experimentally characterized dihydrogen based ultrashort H…H distances are usually larger than 1.6 Å [78-80]).

As mentioned previously, elucidating the nature of interactions for the present so-called non-bonded H…H contacts is of great importance. The design strategy used in this study as well as the observed correlations in Figure 2 are all consistently interpreted within the steric paradigm however, a detailed study is needed to firmly establish the nature of H…H interactions. A primary QTAIM analysis on all the considered structures demonstrated that a bond critical point always appears between such H…H contacts thus the two hydrogen atomic basins share an inter-atomic surface [14,23]. Therefore, this series of species as well as other structures generated with the same design strategy are an ideal set to unravel the pattern of variations of topological indexes and basin (atomic) properties upon the "semi-continuous" shorting of H…H contacts as will be disclosed in a future study.

**Acknowledgments**

The authors are grateful to Masumeh Gharabaghi and Shahin Sowlati for their detailed reading of a previous draft of this paper and helpful suggestions.

**Figure Captions**

**Figure 1.** The schematic representation of all the optimized structures. For clarity of the displaying non-bonded H…H contacts, the two involved hydrogen atoms are displayed by gray balls.

**Figure 2.** The "inner" C-H inter-nuclear distances, the H-C-H and Cl-C-H angles, and the C-H symmetric and anti-symmetric stretching frequencies versus the H…H distances, all computed at the M06L/6-311++G(d,p) level.

**Table 1.** The H…H inter-nuclear distance (Å) of structures 1 and 2 computed using Hartree-Fock (HF) and various density functional methods with 6-311++G(d,p) basis set (The number of imaginary frequencies has been indicated in parentheses).

| Struct. | HF | B3LYP | BP86 | M06L | M062X | B97D | BMK | CAM-B3LYP | HSEh1PBE | tHCTH | TPSSH | VSXC | wB97D | Neutron[*] |
|---|---|---|---|---|---|---|---|---|---|---|---|---|---|---|
| *1* | 1.830 | 1.797 | 1.785 | **1.779** | 1.801 | 1.786 | 1.797 | 1.789 | 1.777 | 1.785 | 1.786 | 1.844 (1) | 1.787 | **1.754** |
| *2* | 1.748 | 1.713 | 1.704 | **1.715** | 1.745 | 1.701 | 1.747 | 1.716 | 1.708 | 1.693 | 1.715 | 1.803 (1) | 1.716 | **1.713** |

[*] Extracted from the neutron diffraction analysis [9]

**Table 2.** The H…H inter-nuclear distances (Å) for the set of considered molecules.[a]

| Structure | HF/6-311++G(d,p) | B3LYP/6-311++G(d,p) | M06L/6-311++G(d,p) | MP2/6-31G(d) | MP2/6-311++G(d,p)[b] |
|---|---|---|---|---|---|
| *1*  | 1.830     | 1.797     | 1.779     | 1.797     | 1.780 |
| *2*  | 1.748     | 1.713     | 1.715     | 1.735     | 1.717 |
| *3*  | 1.731     | 1.704     | 1.709     | 1.725[b]  | 1.707 |
| *4*  | 1.781     | 1.750     | 1.742     | 1.749     | 1.726 |
| *5*  | 1.628     | 1.604     | 1.604     | 1.608     | 1.587 |
| *6*  | 1.536     | 1.522     | 1.525     | 1.531     | 1.512 |
| *7*  | 1.385     | 1.385     | 1.381     | 1.379[b]  | 1.360 |
| *8*  | ---       | 1.178 (-1)| 1.168 (-2)| ---       | --- |
| *9*  | 1.279     | 1.258     | 1.283 (-2)| 1.315[b]  | --- |
| *10* | 1.758     | 1.728     | 1.721     | 1.734     | 1.717 |
| *11* | 1.606     | 1.585     | 1.567     | 1.582     | 1.558 |
| *12* | 1.532     | 1.519     | 1.504     | 1.519     | 1.500 |
| *13* | 1.454     | 1.442     | 1.430     | 1.442     | 1.422 |
| *14* | 1.370 (-1)| 1.362 (-1)| 1.348 (-1)| 1.359 (-1)| 1.340 |
| *15* | 1.426     | 1.388     | 1.381     | 1.437     | 1.432 |
| *16* | 1.497     | 1.492 (-2)| 1.494 (-1)| 1.499 (-2)| 1.483 |
| *17* | 1.797     | 1.768     | 1.764     | 1.777     | 1.759 |
| *18* | 1.430     | 1.420     | 1.403     | 1.411[b]  | 1.383 |

[a] The number of imaginary frequencies has been indicated between parentheses.

[b] Force constants have not been determined for these optimized structures.

**Table 3.** The C-H inter-nuclear distances (in Å), the X-C-H angles (in degrees) and C-H stretching frequencies (in cm$^{-1}$) of the CH$_2$/CClH unit, all computed at M06L/6-311++G(d,p) computational level.

| Structure | C-H Inter-nuclear distances [a] | X-C-H angle [b] | Anti-symmetrical C-H stretching vibrations | Symmetrical C-H stretching vibrations |
|---|---|---|---|---|
| *1* | 1.087 (1.095) | 107.5 | 3147 | 3184 |
| *2* | 1.084 (1.095) | 107.2 | 3168 | 3209 |
| *3* | 1.083 (1.094) | 107.1 | 3176 | 3218 |
| *4* | 1.087 (1.098) | 106.9 | 3140 | 3183 |
| *5* | 1.081 (1.097) | 105.5 | 3180 | 3247 |
| *6* | 1.075 (1.096) | 103.4 | 3220 | 3301 |
| *7* | 1.063 | 95.3 | 3328 | 3424 |
| *8* | 1.041 | 85.4 | 3537 | 3719 |
| *9* | 1.047 | 87.6 | 3473 | 3558 |
| *10* | 1.085 (1.093) | 108.3 | 3152 | 3196 |
| *11* | 1.078 | 102.3 | 3208 | 3249 |
| *12* | 1.075 | 101.5 | 3236 | 3293 |
| *13* | 1.070 | 99.7 | 3220 | 3406 |
| *14* | 1.064 | 98.2 | 3333 | 3420 |
| *15* | 1.066 | 98.6 | 3328 | 3397 |
| *16* | 1.076 (1.097) | 103.3 | 3189 | 3290 |
| *17* | 1.087 (1.096) | 106.1 | 3133 | 3166 |
| *18* | 1.070 | 96.2 | 3264 | 3329 |

[a] The C-H distances of "outer" CH bonds are given in the parenthesis.

[b] For structures 1-6, 10, 16, and 17 X is hydrogen whereas it is chlorine in the rest of structures.

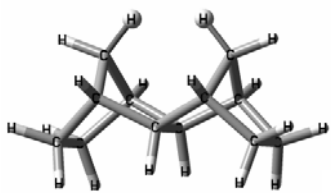
1

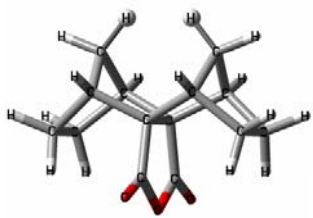
2

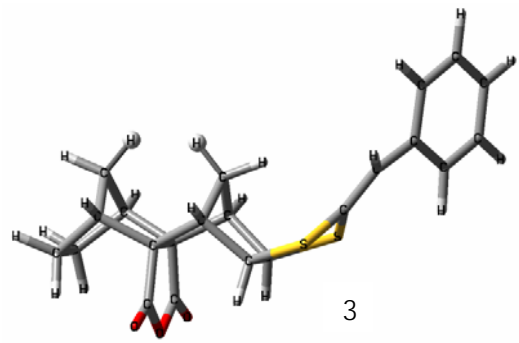
3

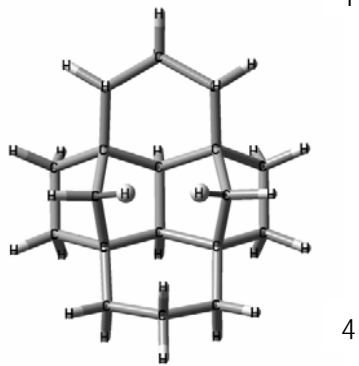
4

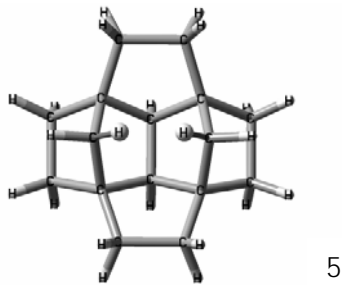
5

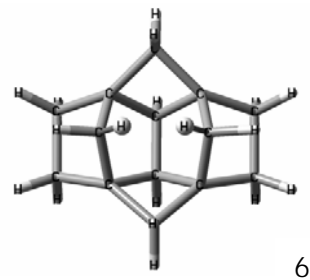
6

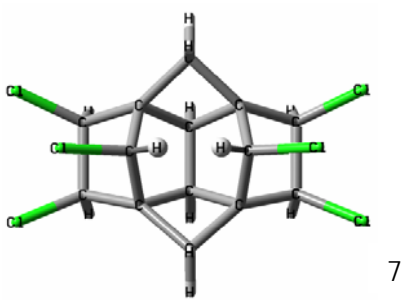
7

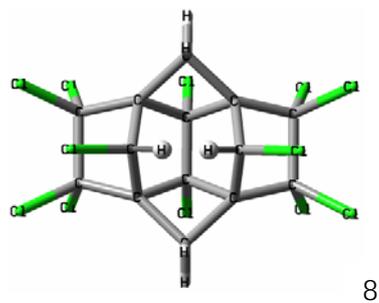
8

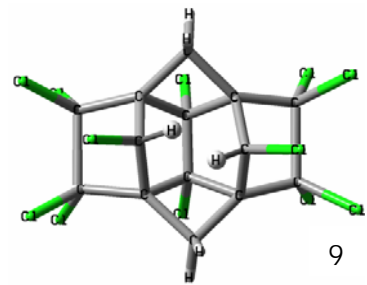
9

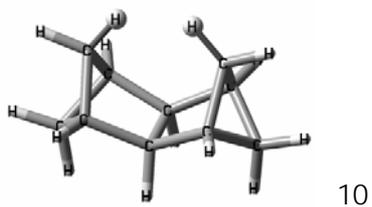
10

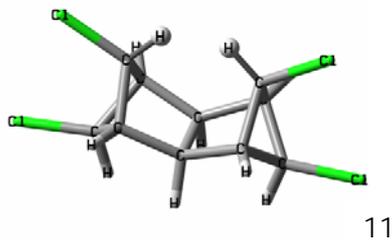
11

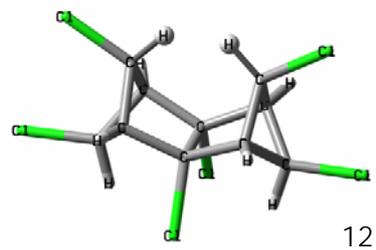
12

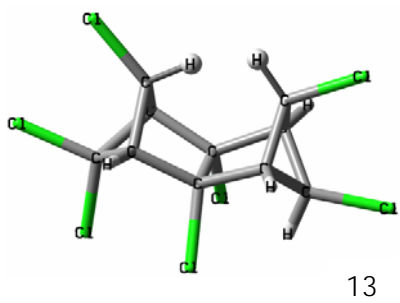
13

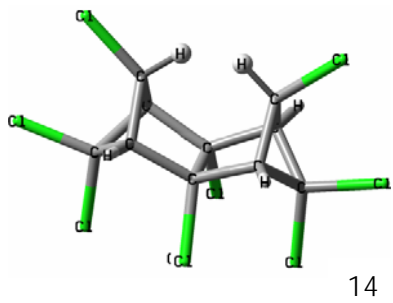
14

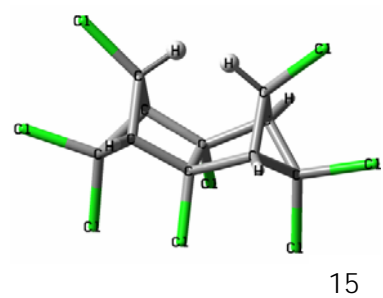
15

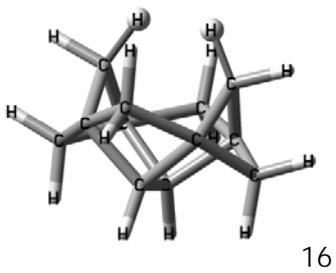
16

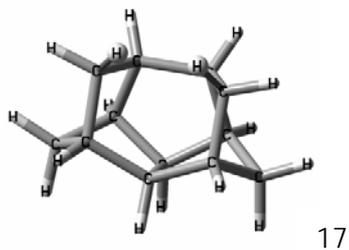
17

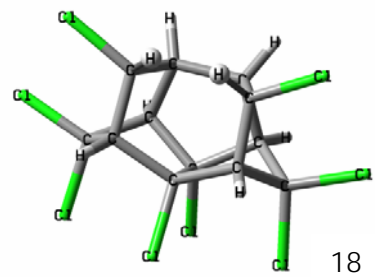
18

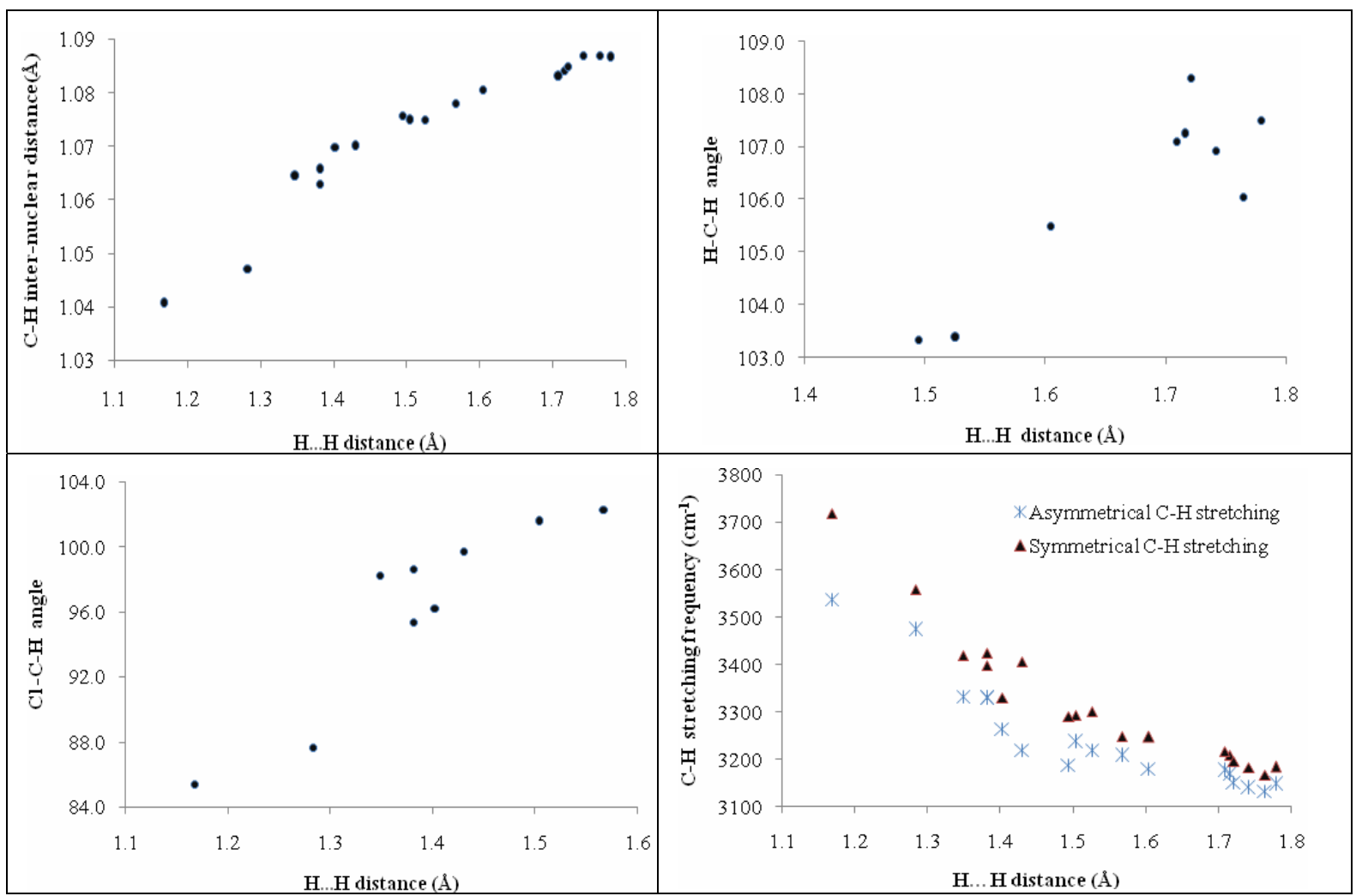

**Figure-2**

# Supporting Information

Seeking for ultrashort "non-bonded" hydrogen-hydrogen contacts in some rigid hydrocarbons and their chlorinated derivatives

*Rohoullah Firouzi[1] and Shant Shahbazian[2]*

[1]*Department of Chemistry, Chemistry and Chemical Engineering Research Center of Iran, P.O. Box 14968-13151, Tehran, Iran.*

E-mails: RFirouzi@ccerci.ac.ir and firouzi.chemist@yahoo.com

[2]*Faculty of Chemistry, Shahid Beheshti University, G. C. , Evin, Tehran, Iran, 19839, P.O. Box 19395-4716. Tel/Fax: 98-21-22431661*

E-mail: chemist_shant@yahoo.com

# Cartesian coordinates for all the species

The Cartesian coordinate of optimized 1 compound at HF/6-311++g(d,p).

------------------------------------------------------------

| Atomic Number | Coordinates (Angstroms) | | |
|---|---|---|---|
| | X | Y | Z |

------------------------------------------------------------

| | | | |
|---|---|---|---|
| 6 | -1.343710 | -1.129260 | 0.294964 |
| 6 | 0.000002 | -0.784026 | -0.384655 |
| 6 | 1.343706 | -1.129247 | 0.294989 |
| 6 | 2.449148 | -0.780923 | -0.727184 |
| 6 | -1.572285 | -0.000025 | 1.314500 |
| 6 | -1.343705 | 1.129244 | 0.294997 |
| 6 | -0.000001 | 0.784029 | -0.384649 |
| 6 | 1.343710 | 1.129257 | 0.294973 |
| 6 | 2.449153 | 0.780907 | -0.727193 |
| 6 | 1.572284 | 0.000014 | 1.314501 |
| 1 | -1.403438 | -2.143431 | 0.675127 |
| 1 | 0.000016 | -1.204064 | -1.385235 |
| 1 | 1.403426 | -2.143409 | 0.675174 |
| 1 | 3.408713 | -1.168543 | -0.399135 |
| 1 | -2.591992 | -0.000028 | 1.691388 |
| 1 | -0.915130 | -0.000046 | 2.167184 |
| 1 | -1.403425 | 2.143404 | 0.675191 |
| 1 | -0.000015 | 1.204074 | -1.385226 |
| 1 | 1.403437 | 2.143425 | 0.675143 |
| 1 | 3.408723 | 1.168528 | -0.399158 |
| 1 | 2.251608 | 1.198475 | -1.708992 |
| 1 | 0.915127 | 0.000028 | 2.167184 |
| 1 | 2.591991 | 0.000015 | 1.691390 |
| 1 | 2.251609 | -1.198507 | -1.708978 |
| 6 | -2.449145 | 0.780930 | -0.727180 |
| 6 | -2.449156 | -0.780901 | -0.727197 |
| 1 | -2.251603 | 1.198519 | -1.708972 |
| 1 | -3.408709 | 1.168551 | -0.399131 |
| 1 | -3.408726 | -1.168520 | -0.399163 |
| 1 | -2.251614 | -1.198464 | -1.708998 |

------------------------------------------------------------

The Cartesian coordinate of optimized 1 compound at MP2/6-31G(d).
-----------------------------------------------------------

| Atomic Number | Coordinates (Angstroms) | | |
|---|---|---|---|
| | X | Y | Z |

-----------------------------------------------------------

| | | | |
|---|---|---|---|
| 6 | 1.338021 | 1.129409 | 0.296037 |
| 6 | -0.000001 | 0.785095 | -0.387149 |
| 6 | -1.338021 | 1.129406 | 0.296044 |
| 6 | -2.440747 | 0.780446 | -0.726176 |
| 6 | 1.564378 | 0.000005 | 1.313494 |
| 6 | 1.338021 | -1.129406 | 0.296044 |
| 6 | 0.000001 | -0.785095 | -0.387149 |
| 6 | -1.338021 | -1.129409 | 0.296037 |
| 6 | -2.440745 | -0.780442 | -0.726184 |
| 6 | -1.564378 | -0.000005 | 1.313494 |
| 1 | 1.396654 | 2.153819 | 0.681278 |
| 1 | -0.000004 | 1.210354 | -1.399739 |
| 1 | -1.396652 | 2.153813 | 0.681292 |
| 1 | -3.409160 | 1.173577 | -0.397073 |
| 1 | 2.594581 | 0.000005 | 1.692185 |
| 1 | 0.898304 | 0.000009 | 2.173246 |
| 1 | 1.396652 | -2.153813 | 0.681292 |
| 1 | 0.000004 | -1.210354 | -1.399739 |
| 1 | -1.396654 | -2.153819 | 0.681278 |
| 1 | -3.409158 | -1.173581 | -0.397091 |
| 1 | -2.235941 | -1.203783 | -1.715256 |
| 1 | -0.898304 | -0.000010 | 2.173246 |
| 1 | -2.594581 | -0.000006 | 1.692185 |
| 1 | -2.235950 | 1.203799 | -1.715244 |
| 6 | 2.440747 | -0.780446 | -0.726176 |
| 6 | 2.440745 | 0.780442 | -0.726184 |
| 1 | 2.235950 | -1.203798 | -1.715244 |
| 1 | 3.409160 | -1.173577 | -0.397073 |
| 1 | 3.409158 | 1.173581 | -0.397090 |
| 1 | 2.235941 | 1.203784 | -1.715256 |

-----------------------------------------------------------

The Cartesian coordinate of optimized 1 compound at MP2/6-311++g(d,p).

---

| Atomic Number | Coordinates (Angstroms) | | |
|---|---|---|---|
| | X | Y | Z |

---

| | | | |
|---|---|---|---|
| 6 | 0.043987 | 1.752814 | 0.296940 |
| 6 | -0.588122 | 0.523387 | -0.388760 |
| 6 | -1.735815 | -0.247074 | 0.297134 |
| 6 | -2.208616 | -1.304237 | -0.727574 |
| 6 | 1.039264 | 1.167899 | 1.316908 |
| 6 | 1.735815 | 0.247074 | 0.297134 |
| 6 | 0.588122 | -0.523387 | -0.388760 |
| 6 | -0.043987 | -1.752814 | 0.296940 |
| 6 | -1.039264 | -2.344867 | -0.727786 |
| 6 | -1.039264 | -1.167899 | 1.316908 |
| 1 | -0.683447 | 2.474698 | 0.682142 |
| 1 | -0.902398 | 0.802982 | -1.402760 |
| 1 | -2.537132 | 0.391697 | 0.682521 |
| 1 | -3.144260 | -1.768689 | -0.397486 |
| 1 | 1.724552 | 1.937900 | 1.694003 |
| 1 | 0.591701 | 0.665173 | 2.171059 |
| 1 | 2.537132 | -0.391697 | 0.682521 |
| 1 | 0.902398 | -0.802982 | -1.402760 |
| 1 | 0.683447 | -2.474698 | 0.682142 |
| 1 | -1.391907 | -3.328192 | -0.397912 |
| 1 | -0.584223 | -2.464517 | -1.716521 |
| 1 | -0.591701 | -0.665173 | 2.171059 |
| 1 | -1.724552 | -1.937900 | 1.694003 |
| 1 | -2.380560 | -0.866062 | -1.716217 |
| 6 | 2.208616 | 1.304237 | -0.727574 |
| 6 | 1.039264 | 2.344867 | -0.727786 |
| 1 | 2.380560 | 0.866062 | -1.716217 |
| 1 | 3.144260 | 1.768689 | -0.397486 |
| 1 | 1.391907 | 3.328192 | -0.397912 |
| 1 | 0.584223 | 2.464517 | -1.716521 |

---

The Cartesian coordinate of optimized 1 compound at B3LYP/6-311++g(d,p).

```
------------------------------------------------------------
Atomic          Coordinates (Angstroms)
Number        X         Y         Z
------------------------------------------------------------
   6      -1.134974   1.349067   0.296288
   6      -0.790443   0.000000  -0.386090
   6      -1.134974  -1.349067   0.296288
   6      -0.784388  -2.458756  -0.730692
   6       0.000000   1.573017   1.320915
   6       1.134974   1.349067   0.296288
   6       0.790443   0.000000  -0.386090
   6       1.134974  -1.349067   0.296288
   6       0.784388  -2.458756  -0.730692
   6       0.000000  -1.573017   1.320915
   1      -2.155285   1.409634   0.681048
   1      -1.210994   0.000000  -1.395934
   1      -2.155285  -1.409634   0.681048
   1      -1.176967  -3.425201  -0.403360
   1       0.000000   2.596187   1.710275
   1       0.000000   0.898694   2.171745
   1       2.155285   1.409634   0.681048
   1       1.210994   0.000000  -1.395934
   1       2.155285  -1.409634   0.681048
   1       1.176967  -3.425201  -0.403360
   1       1.204380  -2.254384  -1.718781
   1       0.000000  -0.898694   2.171745
   1       0.000000  -2.596187   1.710275
   1      -1.204380  -2.254384  -1.718781
   6       0.784388   2.458756  -0.730692
   6      -0.784388   2.458756  -0.730692
   1       1.204380   2.254384  -1.718781
   1       1.176967   3.425201  -0.403360
   1      -1.176967   3.425201  -0.403360
   1      -1.204380   2.254384  -1.718781
------------------------------------------------------------
```

The Cartesian coordinate of optimized 1 compound at M06L/6-311++g(d,p).
------------------------------------------------------------

| Atomic Number | Coordinates (Angstroms) | | |
|---|---|---|---|
| | X | Y | Z |

------------------------------------------------------------

| | | | |
|---|---|---|---|
| 6 | -1.125922 | 1.337399 | 0.293850 |
| 6 | -0.783229 | 0.000000 | -0.386422 |
| 6 | -1.125922 | -1.337399 | 0.293850 |
| 6 | -0.777435 | -2.438659 | -0.723614 |
| 6 | 0.000000 | 1.559896 | 1.312334 |
| 6 | 1.125922 | 1.337399 | 0.293850 |
| 6 | 0.783229 | 0.000000 | -0.386422 |
| 6 | 1.125922 | -1.337399 | 0.293850 |
| 6 | 0.777435 | -2.438659 | -0.723614 |
| 6 | 0.000000 | -1.559896 | 1.312334 |
| 1 | -2.147989 | 1.395378 | 0.678293 |
| 1 | -1.204690 | 0.000000 | -1.398674 |
| 1 | -2.147989 | -1.395378 | 0.678293 |
| 1 | -1.171913 | -3.406002 | -0.398834 |
| 1 | 0.000000 | 2.585362 | 1.697641 |
| 1 | 0.000000 | 0.889515 | 2.167460 |
| 1 | 2.147989 | 1.395378 | 0.678293 |
| 1 | 1.204690 | 0.000000 | -1.398674 |
| 1 | 2.147989 | -1.395378 | 0.678293 |
| 1 | 1.171913 | -3.406002 | -0.398834 |
| 1 | 1.200350 | -2.235434 | -1.711830 |
| 1 | 0.000000 | -0.889515 | 2.167460 |
| 1 | 0.000000 | -2.585362 | 1.697641 |
| 1 | -1.200350 | -2.235434 | -1.711830 |
| 6 | 0.777435 | 2.438659 | -0.723614 |
| 6 | -0.777435 | 2.438659 | -0.723614 |
| 1 | 1.200350 | 2.235434 | -1.711830 |
| 1 | 1.171913 | 3.406002 | -0.398834 |
| 1 | -1.171913 | 3.406002 | -0.398834 |
| 1 | -1.200350 | 2.235434 | -1.711830 |

------------------------------------------------------------

The Cartesian coordinate of optimized 2 compound at HF/6-311++g(d,p).

```
------------------------------------------------------------
Atomic         Coordinates (Angstroms)
Number         X          Y          Z
------------------------------------------------------------
   6      -1.348635   0.907368   1.135005
   6      -0.000006   0.216652   0.772486
   6       1.348584   0.907437   1.135002
   6       2.517562  -0.045403   0.781104
   6      -1.539976   1.933366   0.000011
   6      -1.348632   0.907386  -1.134999
   6      -0.000011   0.216651  -0.772486
   6       1.348587   0.907418  -1.135008
   6       2.517561  -0.045421  -0.781088
   6       1.539889   1.933422  -0.000012
   1      -1.382034   1.283473   2.149086
   6       0.000032  -1.253111   1.136948
   1       1.381965   1.283560   2.149077
   1       3.442249   0.367975   1.167195
   1      -2.551132   2.328312   0.000012
   1      -0.874357   2.775131   0.000019
   1      -1.382024   1.283507  -2.149074
   6      -0.000012  -1.253111  -1.136947
   1       1.381975   1.283524  -2.149089
   1       3.442250   0.367945  -1.167188
   1       2.423019  -1.036336  -1.206288
   1       0.874240   2.775164  -0.000020
   1       2.551030   2.328404  -0.000013
   1       2.423023  -1.036309   1.206327
   6      -2.517582  -0.045489  -0.781104
   6      -2.517581  -0.045506   0.781088
   1      -2.423017  -1.036393  -1.206326
   1      -3.442281   0.367864  -1.167195
   1      -3.442282   0.367834   1.167188
   1      -2.423013  -1.036419   1.206288
   8       0.000025  -2.001644   0.000000
   8       0.000079  -1.756502   2.193906
   8       0.000110  -1.756503  -2.193905
------------------------------------------------------------
```

The Cartesian coordinate of optimized 2 compound at MP2/6-31G(d).

```
------------------------------------------------------------
Atomic        Coordinates (Angstroms)
Number         X         Y         Z
------------------------------------------------------------
   6       -1.344016   0.910586   1.134747
   6        0.000029   0.225092   0.772698
   6        1.344163   0.910429   1.134751
   6        2.495672  -0.060117   0.780269
   6       -1.540756   1.934638   0.000001
   6       -1.344014   0.910590  -1.134748
   6        0.000029   0.225092  -0.772698
   6        1.344165   0.910424  -1.134750
   6        2.495673  -0.060122  -0.780261
   6        1.541012   1.934454  -0.000002
   1       -1.377811   1.287739   2.162027
   6       -0.000077  -1.234899   1.156764
   1        1.377994   1.287587   2.162028
   1        3.436721   0.340124   1.172260
   1       -2.564813   2.328714   0.000001
   1       -0.867461   2.786130   0.000004
   1       -1.377805   1.287747  -2.162026
   6       -0.000088  -1.234899  -1.156764
   1        1.378000   1.287577  -2.162029
   1        3.436723   0.340117  -1.172253
   1        2.377431  -1.057693  -1.212430
   1        0.867810   2.786020  -0.000005
   1        2.565113   2.328419  -0.000001
   1        2.377430  -1.057686   1.212444
   6       -2.495603  -0.059859  -0.780269
   6       -2.495604  -0.059863   0.780261
   1       -2.377425  -1.057440  -1.212439
   1       -3.436617   0.340455  -1.172270
   1       -3.436619   0.340448   1.172262
   1       -2.377426  -1.057446   1.212426
   8       -0.000185  -2.030217   0.000000
   8       -0.000214  -1.736146   2.255014
   8       -0.000196  -1.736147  -2.255014
------------------------------------------------------------
```

The Cartesian coordinate of optimized 2 compound at MP2/6-311++g(d,p).

```
------------------------------------------------------------
Atomic         Coordinates (Angstroms)
Number         X          Y          Z
------------------------------------------------------------
  6        -1.344015   0.911493   1.138030
  6        -0.000049   0.223743   0.773817
  6         1.344212   0.910952   1.138780
  6         2.499208  -0.058896   0.782604
  6        -1.538479   1.936830  -0.000701
  6        -1.343592   0.911023  -1.138823
  6        -0.000493   0.223279  -0.773866
  6         1.343523   0.910970  -1.137865
  6         2.499109  -0.058591  -0.782549
  6         1.539308   1.936591   0.000511
  1        -1.376123   1.290584   2.163423
  6        -0.000826  -1.238107   1.155222
  1         1.375776   1.289683   2.164333
  1         3.439075   0.345240   1.172278
  1        -2.562726   2.330057  -0.001039
  1        -0.857656   2.781936  -0.000723
  1        -1.374762   1.289889  -2.164340
  6         0.000538  -1.238639  -1.154708
  1         1.375428   1.290187  -2.163233
  1         3.438739   0.346187  -1.172114
  1         2.379312  -1.058138  -1.209193
  1         0.858969   2.782101   0.000799
  1         2.563783   2.329204   0.000256
  1         2.379082  -1.058527   1.208946
  6        -2.499137  -0.058305  -0.783266
  6        -2.499136  -0.058412   0.781967
  1        -2.379403  -1.057855  -1.209893
  1        -3.438850   0.346276  -1.172852
  1        -3.439086   0.345697   1.171457
  1        -2.379073  -1.058098   1.208240
  8        -0.000249  -2.029873   0.000413
  8        -0.001608  -1.738241   2.245298
  8         0.001417  -1.739137  -2.244619
------------------------------------------------------------
```

The Cartesian coordinate of optimized 2 compound at B3LYP/6-311++g(d,p).
------------------------------------------------------------

| Atomic Number | Coordinates (Angstroms) | | |
|---|---|---|---|
| | X | Y | Z |

------------------------------------------------------------

| | | | |
|---|---|---|---|
| 6 | -1.354416 | 0.913606 | 1.140293 |
| 6 | -0.000277 | 0.215727 | 0.779087 |
| 6 | 1.353290 | 0.914725 | 1.140270 |
| 6 | 2.525505 | -0.043188 | 0.784390 |
| 6 | -1.540153 | 1.944498 | 0.000010 |
| 6 | -1.354422 | 0.913613 | -1.140280 |
| 6 | -0.000284 | 0.215726 | -0.779087 |
| 6 | 1.353284 | 0.914716 | -1.140283 |
| 6 | 2.525502 | -0.043192 | -0.784401 |
| 6 | 1.538226 | 1.945758 | -0.000011 |
| 1 | -1.386794 | 1.293611 | 2.161411 |
| 6 | 0.000345 | -1.254066 | 1.158200 |
| 1 | 1.385375 | 1.294759 | 2.161387 |
| 1 | 3.458241 | 0.368270 | 1.175959 |
| 1 | -2.555390 | 2.351355 | 0.000014 |
| 1 | -0.857565 | 2.784792 | 0.000010 |
| 1 | -1.386804 | 1.293623 | -2.161396 |
| 6 | 0.000318 | -1.254067 | -1.158199 |
| 1 | 1.385364 | 1.294743 | -2.161403 |
| 1 | 3.458237 | 0.368264 | -1.175975 |
| 1 | 2.421726 | -1.041233 | -1.211765 |
| 1 | 0.854982 | 2.785519 | -0.000012 |
| 1 | 2.553146 | 2.353408 | -0.000016 |
| 1 | 2.421727 | -1.041226 | 1.211760 |
| 6 | -2.525907 | -0.045195 | -0.784390 |
| 6 | -2.525904 | -0.045198 | 0.784401 |
| 1 | -2.421384 | -1.043150 | -1.211762 |
| 1 | -3.458953 | 0.365563 | -1.175954 |
| 1 | -3.458950 | 0.365561 | 1.175968 |
| 1 | -2.421383 | -1.043155 | 1.211769 |
| 8 | 0.001397 | -2.036169 | 0.000000 |
| 8 | 0.001681 | -1.763882 | 2.235781 |
| 8 | 0.001646 | -1.763883 | -2.235780 |

------------------------------------------------------------

The Cartesian coordinate of optimized 2 compound at M06L/6-311++g(d,p).

```
------------------------------------------------------------
Atomic          Coordinates (Angstroms)
Number          X          Y          Z
------------------------------------------------------------
  6        -1.343473   0.909939   1.131118
  6         0.000041   0.219998   0.772056
  6         1.343636   0.909774   1.131129
  6         2.496932  -0.051814   0.777431
  6        -1.534265   1.933046  -0.000004
  6        -1.343464   0.909943  -1.131128
  6         0.000045   0.219998  -0.772056
  6         1.343645   0.909769  -1.131120
  6         2.496935  -0.051822  -0.777409
  6         1.534558   1.932853   0.000003
  1        -1.376845   1.288172   2.154869
  6        -0.000049  -1.239797   1.154310
  1         1.377046   1.288007   2.154878
  1         3.435930   0.345460   1.170659
  1        -2.553583   2.332156  -0.000008
  1        -0.857571   2.779978   0.000000
  1        -1.376828   1.288179  -2.154878
  6        -0.000052  -1.239798  -1.154309
  1         1.377063   1.287997  -2.154871
  1         3.435935   0.345445  -1.170638
  1         2.383415  -1.049361  -1.207920
  1         0.857966   2.779866  -0.000001
  1         2.553924   2.331841   0.000007
  1         2.383415  -1.049350   1.207950
  6        -2.496872  -0.051510  -0.777431
  6        -2.496875  -0.051517   0.777409
  1        -2.383468  -1.049060  -1.207949
  1        -3.435824   0.345871  -1.170661
  1        -3.435829   0.345858   1.170639
  1        -2.383468  -1.049069   1.207920
  8        -0.000212  -2.025190   0.000000
  8        -0.000280  -1.744677   2.234825
  8        -0.000225  -1.744678  -2.234825
------------------------------------------------------------
```

The Cartesian coordinate of optimized 3 compound at HF/6-311++g(d,p).
------------------------------------------------------------
| Atomic | Coordinates (Angstroms) | | |
| Number | X | Y | Z |
------------------------------------------------------------

| Atomic Number | X | Y | Z |
|---|---|---|---|
| 6  | -0.740410 | -0.999365 |  0.159371 |
| 6  | -2.294422 | -0.893284 |  0.246033 |
| 6  | -3.178563 | -2.024886 | -0.362471 |
| 6  | -4.664608 | -1.734112 | -0.033946 |
| 6  | -0.431726 | -0.518186 | -1.269734 |
| 6  | -1.215642 |  0.799071 | -1.133111 |
| 6  | -2.618795 |  0.330816 | -0.636712 |
| 6  | -3.654521 | -0.224967 | -1.662020 |
| 6  | -4.992372 | -0.495875 | -0.928722 |
| 6  | -3.210993 | -1.687586 | -1.866691 |
| 1  | -0.350856 | -1.962786 |  0.454276 |
| 6  | -2.773904 | -0.487370 |  1.624477 |
| 1  | -2.854988 | -3.019780 | -0.087050 |
| 1  | -5.264554 | -2.589072 | -0.322906 |
| 1  |  0.625074 | -0.343558 | -1.414040 |
| 1  | -0.745883 | -1.169481 | -2.060321 |
| 1  | -1.251452 |  1.448755 | -1.995451 |
| 6  | -3.253463 |  1.315860 |  0.323372 |
| 1  | -3.756097 |  0.388070 | -2.547617 |
| 1  | -5.753808 | -0.738964 | -1.660515 |
| 1  | -5.367376 |  0.348745 | -0.365296 |
| 1  | -2.301485 | -1.842683 | -2.413989 |
| 1  | -3.983719 | -2.257285 | -2.372747 |
| 1  | -4.860476 | -1.563385 |  1.016773 |
| 6  | -0.453608 |  1.424267 |  0.069542 |
| 6  | -0.120589 |  0.179938 |  0.961105 |
| 1  | -1.044096 |  2.146467 |  0.611356 |
| 1  | -0.537227 |  0.282112 |  1.950950 |
| 8  | -3.305807 |  0.764497 |  1.567109 |
| 8  | -2.729970 | -1.082899 |  2.630629 |
| 8  | -3.656109 |  2.395488 |  0.120450 |
| 16 |  1.047073 |  2.342315 | -0.400181 |
| 16 |  1.670983 | -0.012841 |  1.250165 |
| 6  |  2.245451 |  1.098011 | -0.002382 |
| 6  |  3.438244 |  1.090948 | -0.583580 |
| 6  |  4.584811 |  0.194186 | -0.320006 |
| 1  |  3.597158 |  1.820541 | -1.360656 |
| 6  |  4.937436 | -0.239114 |  0.953938 |
| 6  |  5.371797 | -0.209040 | -1.398952 |
| 6  |  6.455193 | -1.047344 | -1.215229 |
| 6 | 6.784110 | -1.489294 | 0.057663 |
| 6 | 6.025215 | -1.077987 | 1.139111 |
| 1 | 4.382203 | 0.094258 | 1.810499 |
| 1 | 5.126116 | 0.132918 | -2.389427 |
| 1 | 7.043900 | -1.352936 | -2.061845 |
| 1 | 7.628887 | -2.138339 | 0.204091 |
| 1 | 6.282998 | -1.399980 | 2.132315 |

---

The Cartesian coordinate of optimized 3 compound at MP2/6-31G(d).

---

| Atomic Number | Coordinates (Angstroms) | | |
|---|---|---|---|
|  | X | Y | Z |

---

|   |   |   |   |
|---|---|---|---|
| 6 | -0.647505 | -0.925922 | 0.294211 |
| 6 | -2.199107 | -0.875797 | 0.332926 |
| 6 | -3.021446 | -2.096026 | -0.163132 |
| 6 | -4.518199 | -1.814434 | 0.111013 |
| 6 | -0.309220 | -0.575638 | -1.165314 |
| 6 | -1.158180 | 0.706585 | -1.188667 |
| 6 | -2.547751 | 0.231992 | -0.683352 |
| 6 | -3.528413 | -0.468972 | -1.662787 |
| 6 | -4.867636 | -0.698219 | -0.922196 |
| 6 | -3.026785 | -1.926069 | -1.694597 |
| 1 | -0.222880 | -1.846549 | 0.705719 |
| 6 | -2.718803 | -0.363520 | 1.655164 |
| 1 | -2.665156 | -3.051276 | 0.235603 |
| 1 | -5.095261 | -2.724368 | -0.084468 |
| 1 | 0.756196 | -0.360004 | -1.293081 |
| 1 | -0.571626 | -1.326462 | -1.902953 |
| 1 | -1.196298 | 1.273811 | -2.123548 |
| 6 | -3.246693 | 1.295448 | 0.129882 |
| 1 | -3.630076 | 0.048968 | -2.621887 |
| 1 | -5.617620 | -1.048554 | -1.639143 |
| 1 | -5.276425 | 0.205383 | -0.461503 |
| 1 | -2.083389 | -2.101412 | -2.201690 |
| 1 | -3.771509 | -2.585005 | -2.158226 |
| 1 | -4.730261 | -1.528303 | 1.145011 |
| 6 | -0.479566 | 1.486782 | -0.031823 |
| 6 | -0.116754 | 0.353612 | 0.992542 |
| 1 | -1.133347 | 2.238177 | 0.419820 |
| 1  | -0.574600 | 0.532616 | 1.969403 |
| 8  | -3.324162 | 0.890297 | 1.472232 |
| 8  | -2.668625 | -0.876842 | 2.746437 |
| 8  | -3.697637 | 2.356753 | -0.226663 |
| 16 | 0.997957 | 2.415245 | -0.556507 |
| 16 | 1.677755 | 0.266074 | 1.329784 |
| 6  | 2.179323 | 1.192015 | -0.086080 |
| 6  | 3.350886 | 1.057006 | -0.755729 |
| 6  | 4.452676 | 0.165961 | -0.398300 |
| 1  | 3.476063 | 1.661913 | -1.654432 |
| 6  | 4.809388 | -0.102311 | 0.934851 |
| 6  | 5.196511 | -0.439292 | -1.428380 |
| 6  | 6.236183 | -1.318141 | -1.135248 |
| 6  | 6.562318 | -1.600692 | 0.194871 |
| 6  | 5.850775 | -0.984547 | 1.225767 |
| 1  | 4.309140 | 0.422307 | 1.745184 |
| 1  | 4.941058 | -0.225543 | -2.465032 |
| 1  | 6.794788 | -1.783567 | -1.944026 |
| 1  | 7.375711 | -2.284664 | 0.424606 |
| 1  | 6.115489 | -1.179160 | 2.262432 |

-----------------------------------------------------------

The Cartesian coordinate of optimized 3 compound at MP2/6-311++g(d,p).
-----------------------------------------------------------
| Atomic | Coordinates (Angstroms) | | |
|---|---|---|---|
| Number | X | Y | Z |

-----------------------------------------------------------
| | | | |
|---|---|---|---|
| 6 | -0.625996 | -0.910475 | 0.301326 |
| 6 | -2.178241 | -0.876174 | 0.340536 |
| 6 | -2.986331 | -2.109026 | -0.150755 |
| 6 | -4.488499 | -1.845499 | 0.126000 |
| 6 | -0.292097 | -0.566957 | -1.164149 |
| 6 | -1.156035 | 0.709742 | -1.197654 |
| 6 | -2.539882 | 0.226107 | -0.680786 |
| 6 | -3.517471 | -0.486599 | -1.656436 |
| 6 | -4.854249 | -0.730787 | -0.910554 |
| 6 | -2.996496 | -1.940082 | -1.685562 |
| 1 | -0.188176 | -1.820618 | 0.719258 |
| 6 | -2.702323 | -0.361167 | 1.661279 |
| 1  | -2.615287 | -3.058053 |  0.246574 |
| 1  | -5.054320 | -2.761707 | -0.070810 |
| 1  |  0.771502 | -0.340770 | -1.292522 |
| 1  | -0.554850 | -1.327137 | -1.892055 |
| 1  | -1.200632 |  1.267479 | -2.136612 |
| 6  | -3.245030 |  1.286350 |  0.133495 |
| 1  | -3.624968 |  0.026798 | -2.616139 |
| 1  | -5.600699 | -1.092498 | -1.624938 |
| 1  | -5.267236 |  0.167389 | -0.443348 |
| 1  | -2.048349 | -2.097606 | -2.189354 |
| 1  | -3.735411 | -2.608274 | -2.144948 |
| 1  | -4.700395 | -1.554026 |  1.158340 |
| 6  | -0.479525 |  1.505809 | -0.048558 |
| 6  | -0.108277 |  0.383625 |  0.986809 |
| 1  | -1.136908 |  2.256228 |  0.399041 |
| 1  | -0.568758 |  0.571053 |  1.960726 |
| 8  | -3.317268 |  0.883632 |  1.473470 |
| 8  | -2.646889 | -0.864702 |  2.748202 |
| 8  | -3.702713 |  2.336488 | -0.221237 |
| 16 |  0.990874 |  2.436581 | -0.586364 |
| 16 |  1.685219 |  0.313453 |  1.323663 |
| 6  |  2.171012 |  1.217567 | -0.109415 |
| 6  |  3.344405 |  1.072716 | -0.781702 |
| 6  |  4.431561 |  0.166735 | -0.409341 |
| 1  |  3.466718 |  1.655811 | -1.694562 |
| 6  |  4.789150 | -0.073283 |  0.932169 |
| 6  |  5.159862 | -0.483472 | -1.427636 |
| 6  |  6.177334 | -1.386825 | -1.113004 |
| 6  |  6.500945 | -1.643492 |  0.226919 |
| 6  |  5.808962 | -0.978374 |  1.245653 |
| 1  |  4.302120 |  0.482636 |  1.728954 |
| 1  |  4.906341 | -0.288578 | -2.467932 |
| 1  |  6.720944 | -1.887483 | -1.910081 |
| 1  |  7.295055 | -2.343400 |  0.472579 |
| 1  |  6.074106 | -1.149778 |  2.285761 |

----------------------------------------------------------

The Cartesian coordinate of optimized 3 compound at B3LYP/6-311++g(d,p).

------------------------------------------------------------

| Atomic Number | Coordinates (Angstroms) | | |
|---|---|---|---|
| | X | Y | Z |

------------------------------------------------------------

| | | | |
|---|---|---|---|
| 6 | 0.776464 | -1.031973 | -0.001864 |
| 6 | 2.337574 | -0.920178 | -0.096716 |
| 6 | 3.230240 | -1.932264 | 0.700084 |
| 6 | 4.721374 | -1.674933 | 0.340271 |
| 6 | 0.445882 | -0.331911 | 1.334949 |
| 6 | 1.217465 | 0.962721 | 0.997258 |
| 6 | 2.639975 | 0.443586 | 0.588871 |
| 6 | 3.672530 | 0.066404 | 1.705959 |
| 6 | 5.025809 | -0.300373 | 1.032564 |
| 6 | 3.234172 | -1.354661 | 2.136519 |
| 1 | 0.398836 | -2.042720 | -0.145236 |
| 6 | 2.822230 | -0.744327 | -1.525096 |
| 1 | 2.917393 | -2.969597 | 0.583118 |
| 1 | 5.335325 | -2.471781 | 0.764861 |
| 1 | -0.623826 | -0.148417 | 1.439700 |
| 1 | 0.766128 | -0.850567 | 2.227923 |
| 1 | 1.235588 | 1.747488 | 1.751290 |
| 6 | 3.274814 | 1.286230 | -0.503433 |
| 1 | 3.755704 | 0.818914 | 2.489819 |
| 1 | 5.791396 | -0.411054 | 1.803168 |
| 1 | 5.391402 | 0.456730 | 0.337957 |
| 1 | 2.303771 | -1.425013 | 2.684972 |
| 1 | 4.005037 | -1.832645 | 2.747265 |
| 1 | 4.920560 | -1.667149 | -0.731958 |
| 6 | 0.461493 | 1.381062 | -0.303349 |
| 6 | 0.152501 | 0.001336 | -0.991796 |
| 1 | 1.055451 | 2.024557 | -0.950122 |
| 1 | 0.579062 | -0.054486 | -1.991764 |
| 8 | 3.355382 | 0.538493 | -1.682348 |
| 8 | 2.796677 | -1.509502 | -2.438398 |
| 8 | 3.669994 | 2.409897 | -0.466975 |
| 16 | -1.070765 | 2.348458 | 0.003737 |
| 16 | -1.652977 | -0.263760 | -1.258267 |
| 6 | -2.247147 | 1.027792 | -0.197332 |
| 6 | -3.470117 | 1.119291 | 0.366628 |
| 6 | -4.628681 | 0.232884 | 0.256493 |
| 1 | -3.616712 | 1.973592 | 1.022349 |
| 6 | -4.843820 | -0.668754 | -0.800864 |
| 6 | -5.603464 | 0.295634 | 1.271781 |
| 6 | -6.722076 | -0.527853 | 1.251643 |
| 6 | -6.906473 | -1.433657 | 0.206419 |
| 6 | -5.965088 | -1.493938 | -0.819197 |
| 1 | -4.153753 | -0.709795 | -1.633287 |
| 1 | -5.469348 | 0.997805 | 2.088194 |
| 1 | -7.453068 | -0.461775 | 2.049842 |
| 1 | -7.780095 | -2.075048 | 0.185526 |
| 1 | -6.109322 | -2.178011 | -1.648098 |

---

The Cartesian coordinate of optimized 3 compound at M06L/6-311++g(d,p).

---

| Atomic Number | Coordinates (Angstroms) | | |
|---|---|---|---|
| | X | Y | Z |

---

| | | | |
|---|---|---|---|
| 6 | 0.760945 | -1.017998 | 0.015095 |
| 6 | 2.308258 | -0.916886 | -0.077229 |
| 6 | 3.188623 | -1.913879 | 0.725715 |
| 6 | 4.665163 | -1.665588 | 0.354070 |
| 6 | 0.426241 | -0.300448 | 1.328586 |
| 6 | 1.208061 | 0.971521 | 0.978484 |
| 6 | 2.614770 | 0.442962 | 0.582635 |
| 6 | 3.638684 | 0.080976 | 1.693862 |
| 6 | 4.974616 | -0.294573 | 1.019265 |
| 6 | 3.200394 | -1.321511 | 2.143918 |
| 1 | 0.375710 | -2.029884 | -0.117618 |
| 6 | 2.789247 | -0.770492 | -1.500437 |
| 1 | 2.870447 | -2.952947 | 0.621745 |
| 1 | 5.284725 | -2.458298 | 0.780026 |
| 1 | -0.645126 | -0.102482 | 1.415370 |
| 1 | 0.729762 | -0.808472 | 2.235186 |
| 1 | 1.228940 | 1.770353 | 1.721055 |
| 6 | 3.250045 | 1.266080 | -0.511854 |
| 1 | 3.728316 | 0.847172 | 2.466307 |
| 1 | 5.750727 | -0.393509 | 1.781889 |
| 1 | 5.335946 | 0.455610 | 0.312242 |
| 1 | 2.271372 | -1.381329 | 2.698891 |
| 1 | 3.973890 | -1.793825 | 2.757370 |
| 1 | 4.855007 | -1.675067 | -0.721634 |

```
 6      0.467578    1.374374   -0.323465
 6      0.154926   -0.002942   -0.988081
 1      1.073788    2.005313   -0.977074
 1      0.591008   -0.078134   -1.986471
 8      3.329011    0.504955   -1.680719
 8      2.756575   -1.552692   -2.399707
 8      3.651401    2.388456   -0.485952
16     -1.047778    2.342162   -0.024937
16     -1.636448   -0.266205   -1.245786
 6     -2.213903    1.032311   -0.207739
 6     -3.439728    1.131665    0.357652
 6     -4.584636    0.246106    0.251958
 1     -3.582661    1.991494    1.008980
 6     -4.780566   -0.679997   -0.786223
 6     -5.571901    0.323098    1.251737
 6     -6.678576   -0.510650    1.239069
 6     -6.840236   -1.443227    0.217176
 6     -5.889313   -1.516324   -0.795447
 1     -4.085083   -0.724555   -1.616093
 1     -5.450464    1.048313    2.051828
 1     -7.420621   -0.432690    2.027323
 1     -7.707109   -2.095422    0.202227
 1     -6.019011   -2.220191   -1.611430
-----------------------------------------------------------
```

The Cartesian coordinate of optimized 4 compound at HF/6-311++g(d,p).
--------------------------------------------------------

| Atomic Number | Coordinates (Angstroms) | | |
|---|---|---|---|
| | X | Y | Z |

--------------------------------------------------------

| Atomic Number | X | Y | Z |
|---|---|---|---|
| 6 | 1.167751 | 1.317962 | 0.099336 |
| 6 | 0.748664 | 0.000000 | 0.783376 |
| 6 | 1.167751 | -1.317961 | 0.099337 |
| 6 | 0.737151 | -2.444895 | 1.069674 |
| 6 | 0.096551 | 1.538030 | -0.987846 |
| 6 | -1.115856 | 1.316481 | -0.059144 |
| 6 | -0.786868 | 0.000000 | 0.678300 |
| 6 | -1.115856 | -1.316481 | -0.059144 |
| 6 | -0.823256 | -2.443997 | 0.962285 |
| 6 | 0.096552 | -1.538030 | -0.987846 |
| 6 | 2.635080 | 1.284210 | -0.313404 |
| 1 | 1.118787 | 0.000000 | 1.805062 |
| 6 | 2.635081 | -1.284209 | -0.313403 |
| 1 | 1.145512 | -3.399236 | 0.748046 |
| 1 | 0.121717 | 2.559679 | -1.362978 |
| 1 | 0.144915 | 0.890603 | -1.845519 |
| 6 | -2.515987 | 1.282762 | -0.704551 |
| 1 | -1.283553 | 0.000000 | 1.643228 |
| 6 | -2.515986 | -1.282762 | -0.704552 |
| 1 | -1.183850 | -3.396772 | 0.584971 |
| 1 | -1.308282 | -2.272310 | 1.918717 |
| 1 | 0.144916 | -0.890603 | -1.845519 |
| 1 | 0.121719 | -2.559679 | -1.362978 |
| 1 | 1.087629 | -2.270272 | 2.082404 |
| 6 | -0.823257 | 2.443997 | 0.962286 |
| 6 | 0.737151 | 2.444896 | 1.069673 |
| 1 | -1.308282 | 2.272310 | 1.918718 |
| 1 | -1.183851 | 3.396772 | 0.584971 |
| 1 | 1.145511 | 3.399237 | 0.748045 |
| 1 | 1.087629 | 2.270273 | 2.082403 |
| 6 | -3.293823 | -0.000001 | -0.365845 |
| 6 | 2.994593 | 0.000000 | -1.080549 |
| 1 | 3.253081 | 1.342425 | 0.583447 |
| 1 | 2.884352 | 2.155999 | -0.915841 |
| 1 | 2.884353 | -2.155999 | -0.915840 |
| 1 | 3.253081 | -1.342424 | 0.583448 |
| 1 | -2.418780 | 1.353211 | -1.786727 |
| 1 | -3.100975 | 2.146413 | -0.397471 |
| 1 | -3.100974 | -2.146414 | -0.397474 |
| 1 | -2.418778 | -1.353210 | -1.786728 |
| 1 | -3.547248 | -0.000001 | 0.692664 |
| 1 | -4.239129 | -0.000001 | -0.903216 |
| 1 | 2.488320 | 0.000000 | -2.043495 |
| 1 | 4.059558 | 0.000000 | -1.300244 |

---

The Cartesian coordinate of optimized 4 compound at MP2/6-31G(d).

---

| Atomic Number | Coordinates (Angstroms) | | |
|---|---:|---:|---:|
| | X | Y | Z |

---

| | | | |
|---|---:|---:|---:|
| 6 | 1.168035 | 1.313013 | 0.101620 |
| 6 | 0.749158 | 0.000001 | 0.787843 |
| 6 | 1.168038 | -1.313011 | 0.101623 |
| 6 | 0.733965 | -2.436601 | 1.070619 |
| 6 | 0.101851 | 1.530473 | -0.986553 |
| 6 | -1.113934 | 1.310743 | -0.064177 |
| 6 | -0.788081 | -0.000001 | 0.679227 |
| 6 | -1.113930 | -1.310745 | -0.064179 |
| 6 | -0.825396 | -2.434179 | 0.959519 |
| 6 | 0.101857 | -1.530474 | -0.986552 |
| 6 | 2.630889 | 1.277171 | -0.308354 |
| 1 | 1.126557 | 0.000003 | 1.821808 |
| 6 | 2.630893 | -1.277167 | -0.308350 |
| 1 | 1.150508 | -3.398550 | 0.745388 |
| 1 | 0.128439 | 2.563575 | -1.362956 |
| 1 | 0.152162 | 0.874318 | -1.852638 |
| 6 | -2.507362 | 1.273813 | -0.713878 |
| 1 | -1.293264 | -0.000002 | 1.655485 |
| 6 | -2.507356 | -1.273816 | -0.713884 |
| 1 | -1.194193 | -3.394614 | 0.578074 |
| 1 | -1.321575 | -2.253285 | 1.920624 |
| 1 | 0.152169 | -0.874320 | -1.852637 |
| 1 | 0.128447 | -2.563576 | -1.362954 |
| 1 | 1.091703 | -2.255207 | 2.091117 |
| 6 | -0.825399 | 2.434177 | 0.959521 |
| 6 | 0.733962 | 2.436604 | 1.070615 |
| 1 | -1.321575 | 2.253282 | 1.920626 |
| 1 | -1.194199 | 3.394611 | 0.578077 |
| 1 | 1.150502 | 3.398553 | 0.745382 |
| 1 | 1.091704 | 2.255212 | 2.091113 |
| 6 | -3.282433 | -0.000004 | -0.346521 |
| 6 | 2.971081 | 0.000001 | -1.091728 |
| 1 | 3.253743 | 1.313892 | 0.598451 |
| 1 | 2.886209 | 2.167970 | -0.899275 |
| 1 | 2.886215 | -2.167967 | -0.899267 |
| 1 | 3.253746 | -1.313883 | 0.598455 |
| 1 | -2.404674 | 1.313834 | -1.807451 |
| 1 | -3.091941 | 2.158885 | -0.429446 |
| 1 | -3.091932 | -2.158893 | -0.429463 |
| 1 | -2.404663 | -1.313827 | -1.807458 |
| 1 | -3.498037 | -0.000008 | 0.729249 |
| 1 | -4.254335 | -0.000005 | -0.856130 |
| 1 | 2.425298 | -0.000001 | -2.043683 |
| 1 | 4.037960 | 0.000003 | -1.348952 |

-----------------------------------------------------------

The Cartesian coordinate of optimized 4 compound at MP2/6-311++g(d,p).
-----------------------------------------------------------
Atomic         Coordinates (Angstroms)
Number          X          Y          Z
-----------------------------------------------------------
| | | | |
|---|---|---|---|
| 6 | -1.171130 | -1.313332 | 0.103127 |
| 6 | -0.750997 | 0.000217 | 0.791471 |
| 6 | -1.170874 | 1.313477 | 0.102278 |
| 6 | -0.735401 | 2.437444 | 1.073435 |
| 6 | -0.104283 | -1.528829 | -0.988796 |
| 6 | 1.115098 | -1.310980 | -0.066991 |
| 6 | 0.789779 | -0.000068 | 0.680660 |
| 6 | 1.115350 | 1.310985 | -0.066117 |
| 6 | 0.828269 | 2.436799 | 0.957879 |
| 6 | -0.103197 | 1.528552 | -0.988946 |
| 6 | -2.635777 | -1.278269 | -0.306088 |
| 1 | -1.125940 | 0.000475 | 1.825758 |
| 6 | -2.635315 | 1.278259 | -0.307460 |
| 1 | -1.150990 | 3.399475 | 0.747615 |
| 1 | -0.131142 | -2.562684 | -1.363506 |
| 1 | -0.154840 | -0.863548 | -1.847988 |
| 6 | 2.509094 | -1.272735 | -0.718212 |
| 1 | 1.297631 | -0.000493 | 1.654836 |
| 6 |  2.510026 |  1.273144 | -0.716053 |
| 1 |  1.191107 |  3.396855 |  0.570071 |
| 1 |  1.324248 |  2.251586 |  1.918142 |
| 1 | -0.152948 |  0.862885 | -1.847864 |
| 1 | -0.129732 |  2.562268 | -1.364142 |
| 1 | -1.086198 |  2.246841 |  2.094461 |
| 6 |  0.829013 | -2.437088 |  0.957179 |
| 6 | -0.734635 | -2.436736 |  1.074448 |
| 1 |  1.326193 | -2.252470 |  1.916921 |
| 1 |  1.190765 | -3.397240 |  0.568580 |
| 1 | -1.150999 | -3.398816 |  0.749751 |
| 1 | -1.084152 | -2.245461 |  2.095789 |
| 6 |  3.287268 | -0.000427 | -0.339066 |
| 6 | -2.968977 | -0.000360 | -1.096949 |
| 1 | -3.258698 | -1.306101 |  0.601752 |
| 1 | -2.890000 | -2.170021 | -0.896196 |
| 1 | -2.889120 |  2.169510 | -0.898514 |
| 1 | -3.258419 |  1.307296 |  0.600220 |
| 1 |  2.402580 | -1.298235 | -1.812284 |
| 1 |  3.088381 | -2.162174 | -0.437226 |
| 1 |  3.089616 |  2.161562 | -0.432493 |
| 1 |  2.404600 |  1.301213 | -1.810163 |
| 1 |  3.489555 | -0.001460 |  0.740120 |
| 1 |  4.261765 | -0.000348 | -0.842983 |
| 1 | -2.409815 | -0.000958 | -2.041946 |
| 1 | -4.033306 | -0.000270 | -1.363505 |

---

The Cartesian coordinate of optimized 4 compound at B3LYP/6-311++g(d,p).

---

| Atomic | Coordinates (Angstroms) | | |
|--------|-----------|-----------|-----------|
| Number | X | Y | Z |

---

|   |           |           |           |
|---|-----------|-----------|-----------|
| 6 |  1.174852 |  1.323984 |  0.099189 |
| 6 |  0.752474 |  0.000002 |  0.787831 |
| 6 |  1.174859 | -1.323980 |  0.099195 |
| 6 |  0.740369 | -2.455609 |  1.074217 |
| 6 |  0.097389 |  1.539995 | -0.992636 |
| 6 | -1.122550 |  1.322499 | -0.059951 |
| 6 | -0.791385 | -0.000001 | 0.681997 |
| 6 | -1.122542 | -1.322503 | -0.059955 |
| 6 | -0.826237 | -2.453783 | 0.967318 |
| 6 | 0.097402 | -1.539996 | -0.992634 |
| 6 | 2.644493 | 1.288064 | -0.315236 |
| 1 | 1.125646 | 0.000005 | 1.817628 |
| 6 | 2.644500 | -1.288054 | -0.315228 |
| 1 | 1.153931 | -3.415970 | 0.750719 |
| 1 | 0.123356 | 2.565486 | -1.379095 |
| 1 | 0.144819 | 0.875167 | -1.848847 |
| 6 | -2.525356 | 1.286889 | -0.706714 |
| 1 | -1.292091 | -0.000005 | 1.654793 |
| 6 | -2.525343 | -1.286896 | -0.706729 |
| 1 | -1.193846 | -3.412580 | 0.589155 |
| 1 | -1.314109 | -2.274813 | 1.929847 |
| 1 | 0.144834 | -0.875170 | -1.848846 |
| 1 | 0.123375 | -2.565488 | -1.379091 |
| 1 | 1.095073 | -2.275310 | 2.092957 |
| 6 | -0.826245 | 2.453781 | 0.967321 |
| 6 | 0.740361 | 2.455615 | 1.074210 |
| 1 | -1.314111 | 2.274807 | 1.929853 |
| 1 | -1.193862 | 3.412575 | 0.589161 |
| 1 | 1.153917 | 3.415976 | 0.750705 |
| 1 | 1.095073 | 2.275321 | 2.092948 |
| 6 | -3.306653 | -0.000010 | -0.368923 |
| 6 | 3.005140 | 0.000004 | -1.085139 |
| 1 | 3.267467 | 1.346308 | 0.588021 |
| 1 | 2.896047 | 2.167846 | -0.919476 |
| 1 | 2.896062 | -2.167838 | -0.919461 |
| 1 | 3.267474 | -1.346287 | 0.588031 |
| 1 | -2.426377 | 1.358915 | -1.796731 |
| 1 | -3.113424 | 2.158274 | -0.398634 |
| 1 | -3.113404 | -2.158293 | -0.398672 |
| 1 | -2.426353 | -1.358901 | -1.796746 |
| 1 | -3.562900 | -0.000017 | 0.696999 |
| 1 | -4.258930 | -0.000011 | -0.909503 |
| 1 | 2.492465 | -0.000002 | -2.053914 |
| 1 | 4.076699 | 0.000006 | -1.310588 |

-----------------------------------------------------------

The Cartesian coordinate of optimized 4 compound at M06L/6-311++g(d,p).
------------------------------------------------------------
| Atomic | Coordinates (Angstroms) | | |
| Number | X | Y | Z |
------------------------------------------------------------

| Atomic Number | X | Y | Z |
|---|---|---|---|
| 6 | 1.166295 | 1.314325 | 0.102879 |
| 6 | 0.745233 | 0.000003 | 0.786996 |
| 6 | 1.166306 | -1.314319 | 0.102888 |
| 6 | 0.730337 | -2.436585 | 1.067971 |
| 6 | 0.103346 | 1.529474 | -0.987405 |
| 6 | -1.110554 | 1.311916 | -0.065610 |
| 6 | -0.784800 | -0.000002 | 0.675599 |
| 6 | -1.110542 | -1.311922 | -0.065617 |
| 6 | -0.821458 | -2.434237 | 0.954769 |
| 6 | 0.103366 | -1.529476 | -0.987402 |
| 6 | 2.627307 | 1.275254 | -0.306611 |
| 1 | 1.121272 | 0.000009 | 1.818985 |
| 6 | 2.627319 | -1.275240 | -0.306598 |
| 1 | 1.148628 | -3.397220 | 0.747111 |
| 1 | 0.131060 | 2.558330 | -1.368502 |
| 1 | 0.156537 | 0.870883 | -1.850480 |
| 6 | -2.504860 | 1.272077 | -0.708166 |
| 1 | -1.295616 | -0.000008 | 1.646393 |
| 6 | -2.504842 | -1.272085 | -0.708187 |
| 1 | -1.191517 | -3.393255 | 0.575944 |
| 1 | -1.319678 | -2.255715 | 1.913790 |
| 1 | 0.156563 | -0.870888 | -1.850478 |
| 1 | 0.131087 | -2.558333 | -1.368496 |
| 1 | 1.085825 | -2.257891 | 2.088185 |
| 6 | -0.821467 | 2.434233 | 0.954774 |
| 6 | 0.730328 | 2.436593 | 1.067960 |
| 1 | -1.319676 | 2.255705 | 1.913800 |
| 1 | -1.191539 | 3.393248 | 0.575955 |
| 1 | 1.148609 | 3.397229 | 0.747090 |
| 1 | 1.085829 | 2.257908 | 2.088171 |
| 6 | -3.276758 | -0.000013 | -0.346898 |
| 6 | 2.964439 | 0.000004 | -1.085266 |
| 1 | 3.250784 | 1.316754 | 0.598982 |
| 1 | 2.888017 | 2.163867 | -0.895202 |
| 1 | 2.888037 | -2.163858 | -0.895179 |
| 1 | 3.250794 | -1.316725 | 0.598997 |
| 1 | -2.406249 | 1.321586 | -1.800969 |
| 1 | -3.089052 | 2.154605 | -0.422803 |
| 1 | -3.089025 | -2.154629 | -0.422855 |
| 1 | -2.406217 | -1.321565 | -1.800991 |

```
1       -3.497283  -0.000024   0.728990
1       -4.249093  -0.000016  -0.850242
1        2.412944  -0.000003  -2.035328
1        4.024989   0.000008  -1.357330
--------------------------------------------------------
```

The Cartesian coordinate of optimized 5 compound at HF/6-311++g(d,p).
---------------------------------------------------------

| Atomic Number | Coordinates (Angstroms) | | |
|---|---|---|---|
| | X | Y | Z |

---------------------------------------------------------

| | | | |
|---|---|---|---|
| 6 | 1.224626 | 1.144449 | 0.061157 |
| 6 | -0.000001 | 0.747262 | -0.751341 |
| 6 | -1.224629 | 1.144446 | 0.061156 |
| 6 | -2.452553 | 0.797852 | -0.824454 |
| 6 | 1.425034 | 0.000002 | 1.084998 |
| 6 | 1.224629 | -1.144446 | 0.061156 |
| 6 | 0.000001 | -0.747262 | -0.751341 |
| 6 | -1.224626 | -1.144449 | 0.061157 |
| 6 | -2.452551 | -0.797859 | -0.824454 |
| 6 | -1.425034 | -0.000002 | 1.084998 |
| 6 | 0.796324 | 2.535839 | 0.579751 |
| 1 | -0.000001 | 1.237496 | -1.721489 |
| 6 | -0.796331 | 2.535838 | 0.579750 |
| 1 | -3.377081 | 1.174510 | -0.398102 |
| 1 | 2.450571 | 0.000003 | 1.447916 |
| 1 | 0.813910 | 0.000000 | 1.961920 |
| 6 | 0.796331 | -2.535838 | 0.579750 |
| 1 | 0.000001 | -1.237496 | -1.721489 |
| 6 | -0.796324 | -2.535839 | 0.579751 |
| 1 | -3.377078 | -1.174520 | -0.398102 |
| 1 | -2.368857 | -1.199474 | -1.829870 |
| 1 | -0.813910 | 0.000000 | 1.961920 |
| 1 | -2.450571 | -0.000003 | 1.447916 |
| 1 | -2.368861 | 1.199466 | -1.829871 |
| 6 | 2.452553 | -0.797852 | -0.824454 |
| 6 | 2.452551 | 0.797859 | -0.824454 |
| 1 | 2.368861 | -1.199466 | -1.829871 |
| 1 | 3.377081 | -1.174510 | -0.398102 |
| 1 | 3.377078 | 1.174520 | -0.398102 |
| 1 | 2.368857 | 1.199474 | -1.829870 |
| 1 | 1.154284 | 3.319738 | -0.082180 |
| 1 | 1.194354 | 2.753466 | 1.566289 |
| 1 | -1.194363 | 2.753464 | 1.566287 |
| 1 | -1.154292 | 3.319735 | -0.082182 |
| 1 | 1.194363 | -2.753465 | 1.566287 |
| 1 | 1.154291 | -3.319735 | -0.082183 |
| 1 | -1.154284 | -3.319738 | -0.082179 |
| 1 | -1.194354 | -2.753466 | 1.566289 |

---------------------------------------------------------

The Cartesian coordinate of optimized 5 compound at MP2/6-31G(d).
-------------------------------------------------------------
Atomic          Coordinates (Angstroms)
Number           X          Y          Z
-------------------------------------------------------------
    6      -1.223086  -1.144713   0.064413
    6       0.000006  -0.747973  -0.753903
    6       1.223106  -1.144694   0.064409
    6       2.443918  -0.796957  -0.824897
    6      -1.421020  -0.000009   1.084801
    6      -1.223106   1.144694   0.064409
    6      -0.000006   0.747973  -0.753903
    6       1.223086   1.144713   0.064412
    6       2.443905   0.796999  -0.824894
    6       1.421020   0.000009   1.084801
    6      -0.795221  -2.532095   0.580103
    1       0.000009  -1.248632  -1.733207
    6       0.795264  -2.532085   0.580096
    1       3.378812  -1.178260  -0.397625
    1      -2.458990  -0.000018   1.447618
    1      -0.803888  -0.000002   1.972901
    6      -0.795264   2.532085   0.580095
    1      -0.000009   1.248632  -1.733207
    6       0.795221   2.532095   0.580103
    1       3.378792   1.178316  -0.397619
    1       2.351303   1.207013  -1.837663
    1       0.803889   0.000002   1.972901
    1       2.458990   0.000018   1.447618
    1       2.351322  -1.206968  -1.837668
    6      -2.443918   0.796957  -0.824897
    6      -2.443905  -0.796999  -0.824894
    1      -2.351322   1.206968  -1.837667
    1      -3.378812   1.178260  -0.397624
    1      -3.378792  -1.178317  -0.397619
    1      -2.351303  -1.207013  -1.837663
    1      -1.157417  -3.319735  -0.092147
    1      -1.202460  -2.754052   1.573409
    1       1.202516  -2.754043   1.573397
    1       1.157465  -3.319714  -0.092163
    1      -1.202517   2.754045   1.573396
    1      -1.157464   3.319714  -0.092165
    1       1.157418   3.319735  -0.092145
    1       1.202459   2.754050   1.573411
-------------------------------------------------------------

The Cartesian coordinate of optimized 5 compound at MP2/6-311++g(d,p).
------------------------------------------------------------

| Atomic Number | Coordinates (Angstroms) | | |
|---|---|---|---|
| | X | Y | Z |

------------------------------------------------------------

| | | | |
|---|---|---|---|
| 6 | 1.224657 | 1.147166 | 0.064059 |
| 6 | -0.000022 | 0.750249 | -0.758476 |
| 6 | -1.224730 | 1.147094 | 0.064045 |
| 6 | -2.447621 | 0.799213 | -0.826274 |
| 6 | 1.419622 | 0.000027 | 1.086337 |
| 6 | 1.224723 | -1.147101 | 0.064043 |
| 6 | 0.000039 | -0.750239 | -0.758491 |
| 6 | -1.224645 | -1.147152 | 0.064048 |
| 6 | -2.447556 | -0.799401 | -0.826278 |
| 6 | -1.419638 | -0.000030 | 1.086337 |
| 6 | 0.797372 | 2.535890 | 0.582643 |
| 1 | -0.000033 | 1.254936 | -1.734828 |
| 6 | -0.797541 | 2.535852 | 0.582623 |
| 1 | -3.381497 | 1.177408 | -0.394487 |
| 1 | 2.458571 | 0.000049 | 1.447324 |
| 1 | 0.793315 | 0.000003 | 1.968123 |
| 6 | 0.797528 | -2.535854 | 0.582609 |
| 1 | 0.000051 | -1.254917 | -1.734848 |
| 6 | -0.797377 | -2.535864 | 0.582651 |
| 1 | -3.381390 | -1.177675 | -0.394474 |
| 1 | -2.348491 | -1.205252 | -1.840098 |
| 1 | -0.793337 | 0.000000 | 1.968126 |
| 1 | -2.458593 | -0.000064 | 1.447309 |
| 1 | -2.348579 | 1.205073 | -1.840091 |
| 6 | 2.447623 | -0.799227 | -0.826280 |
| 6 | 2.447575 | 0.799381 | -0.826272 |
| 1 | 2.348578 | -1.205074 | -1.840104 |
| 1 | 3.381488 | -1.177445 | -0.394493 |
| 1 | 3.381416 | 1.177653 | -0.394483 |
| 1 | 2.348500 | 1.205232 | -1.840092 |
| 1 | 1.157661 | 3.323613 | -0.090626 |
| 1 | 1.201123 | 2.750079 | 1.578953 |
| 1 | -1.201326 | 2.750029 | 1.578922 |
| 1 | -1.157858 | 3.323544 | -0.090666 |
| 1 | 1.201332 | -2.750048 | 1.578896 |
| 1 | 1.157825 | -3.323540 | -0.090698 |
| 1 | -1.157693 | -3.323597 | -0.090591 |
| 1 | -1.201119 | -2.750020 | 1.578972 |

------------------------------------------------------------

The Cartesian coordinate of optimized 5 compound at B3LYP/6-311++g(d,p).
------------------------------------------------------------

| Atomic Number | Coordinates (Angstroms) | | |
|---|---|---|---|
| | X | Y | Z |

------------------------------------------------------------

| | | | |
|---|---|---|---|
| 6 | 1.232671 | 1.152225 | 0.062314 |
| 6 | -0.000009 | 0.750274 | -0.755170 |
| 6 | -1.232701 | 1.152197 | 0.062308 |
| 6 | -2.462306 | 0.801889 | -0.830440 |
| 6 | 1.430882 | 0.000014 | 1.089500 |
| 6 | 1.232701 | -1.152197 | 0.062309 |
| 6 | 0.000009 | -0.750275 | -0.755170 |
| 6 | -1.232671 | -1.152225 | 0.062313 |
| 6 | -2.462286 | -0.801954 | -0.830435 |
| 6 | -1.430882 | -0.000014 | 1.089500 |
| 6 | 0.800030 | 2.544830 | 0.584270 |
| 1 | -0.000012 | 1.243149 | -1.733176 |
| 6 | -0.800097 | 2.544815 | 0.584258 |
| 1 | -3.394441 | 1.180963 | -0.403210 |
| 1 | 2.461390 | 0.000027 | 1.462472 |
| 1 | 0.802067 | 0.000002 | 1.966231 |
| 6 | 0.800097 | -2.544815 | 0.584256 |
| 1 | 0.000013 | -1.243149 | -1.733176 |
| 6 | -0.800030 | -2.544830 | 0.584272 |
| 1 | -3.394411 | -1.181048 | -0.403201 |
| 1 | -2.372510 | -1.206489 | -1.842573 |
| 1 | -0.802067 | -0.000002 | 1.966230 |
| 1 | -2.461390 | -0.000027 | 1.462471 |
| 1 | -2.372538 | 1.206420 | -1.842580 |
| 6 | 2.462306 | -0.801889 | -0.830440 |
| 6 | 2.462287 | 0.801953 | -0.830434 |
| 1 | 2.372537 | -1.206420 | -1.842580 |
| 1 | 3.394441 | -1.180964 | -0.403210 |
| 1 | 3.394412 | 1.181047 | -0.403199 |
| 1 | 2.372511 | 1.206489 | -1.842572 |
| 1 | 1.160788 | 3.336135 | -0.079921 |
| 1 | 1.200947 | 2.761830 | 1.578149 |
| 1 | -1.201033 | 2.761820 | 1.578128 |
| 1 | -1.160858 | 3.336104 | -0.079950 |
| 1 | 1.201036 | -2.761824 | 1.578125 |
| 1 | 1.160856 | -3.336102 | -0.079956 |
| 1 | -1.160790 | -3.336137 | -0.079915 |
| 1 | -1.200944 | -2.761825 | 1.578153 |

------------------------------------------------------------

The Cartesian coordinate of optimized 5 compound at M06L/6-311++g(d,p).

```
------------------------------------------------------------
Atomic          Coordinates (Angstroms)
Number          X          Y          Z
------------------------------------------------------------
   6        -1.222721  -1.142969   0.063241
   6         0.000013  -0.745283  -0.753005
   6         1.222763  -1.142930   0.063234
   6         2.443026  -0.793474  -0.820014
   6        -1.419598  -0.000019   1.084070
   6        -1.222763   1.142930   0.063234
   6        -0.000013   0.745283  -0.753005
   6         1.222721   1.142969   0.063241
   6         2.442998   0.793564  -0.820006
   6         1.419599   0.000020   1.084070
   6        -0.792407  -2.527385   0.575673
   1         0.000018  -1.247251  -1.728791
   6         0.792500  -2.527364   0.575658
   1         3.375975  -1.175170  -0.393725
   1        -2.454699  -0.000038   1.447668
   1        -0.802051  -0.000003   1.970781
   6        -0.792500   2.527364   0.575656
   1        -0.000018   1.247251  -1.728791
   6         0.792407   2.527385   0.575675
   1         3.375933   1.175289  -0.393711
   1         2.356306   1.202301  -1.832340
   1         0.802052   0.000003   1.970781
   1         2.454700   0.000038   1.447667
   1         2.356344  -1.202204  -1.832351
   6        -2.443026   0.793474  -0.820013
   6        -2.442998  -0.793564  -0.820006
   1        -2.356345   1.202204  -1.832350
   1        -3.375975   1.175170  -0.393724
   1        -3.375932  -1.175289  -0.393711
   1        -2.356306  -1.202301  -1.832340
   1        -1.156235  -3.316013  -0.091853
   1        -1.198047  -2.754380   1.566564
   1         1.198168  -2.754364   1.566536
   1         1.156336  -3.315970  -0.091891
   1        -1.198170   2.754368   1.566532
   1        -1.156334   3.315968  -0.091896
   1         1.156237   3.316015  -0.091848
   1         1.198045   2.754376   1.566567
------------------------------------------------------------
```

The Cartesian coordinate of optimized 6 compound at HF/6-311++g(d,p).

---

| Atomic Number | Coordinates (Angstroms) | | |
|---|---|---|---|
| | X | Y | Z |

---

| | | | |
|---|---|---|---|
| 6 | -1.098355 | -1.152327 | 0.149819 |
| 6 | 0.000006 | -0.726267 | -0.782994 |
| 6 | 1.098365 | -1.152321 | 0.149826 |
| 6 | 2.421779 | -0.829743 | -0.628935 |
| 6 | -1.306900 | -0.000004 | 1.172407 |
| 6 | -1.098364 | 1.152321 | 0.149823 |
| 6 | -0.000003 | 0.726267 | -0.782995 |
| 6 | 1.098356 | 1.152327 | 0.149820 |
| 6 | 2.421761 | 0.829756 | -0.628960 |
| 6 | 1.306896 | 0.000005 | 1.172411 |
| 1 | 0.000010 | -1.285556 | -1.711912 |
| 6 | 0.000007 | -2.271871 | 0.591342 |
| 1 | 3.323902 | -1.179227 | -0.140385 |
| 1 | -2.343840 | -0.000006 | 1.494200 |
| 1 | -0.767839 | -0.000002 | 2.087774 |
| 1 | -0.000006 | 1.285555 | -1.711912 |
| 6 | -0.000007 | 2.271871 | 0.591340 |
| 1 | 3.323883 | 1.179274 | -0.140435 |
| 1 | 2.421136 | 1.206119 | -1.646678 |
| 1 | 0.767827 | 0.000003 | 2.087774 |
| 1 | 2.343832 | 0.000007 | 1.494213 |
| 1 | 2.421178 | -1.206136 | -1.646642 |
| 6 | -2.421779 | 0.829744 | -0.628934 |
| 6 | -2.421759 | -0.829757 | -0.628961 |
| 1 | -2.421181 | 1.206139 | -1.646640 |
| 1 | -3.323902 | 1.179224 | -0.140381 |
| 1 | -3.323881 | -1.179276 | -0.140436 |
| 1 | -2.421134 | -1.206120 | -1.646680 |
| 1 | 0.000005 | -2.593056 | 1.626634 |
| 1 | 0.000012 | -3.139190 | -0.057588 |
| 1 | -0.000010 | 3.139190 | -0.057590 |
| 1 | -0.000006 | 2.593057 | 1.626632 |

---

The Cartesian coordinate of optimized 6 compound at MP2/6-31G(d).

```
------------------------------------------------------------
Atomic          Coordinates (Angstroms)
Number        X           Y           Z
------------------------------------------------------------
  6         -1.101960   -1.154054    0.156522
  6          0.000013   -0.727149   -0.784898
  6          1.101982   -1.154035    0.156538
  6          2.412505   -0.829078   -0.630958
  6         -1.306287   -0.000002    1.175574
  6         -1.101974    1.154048    0.156521
  6          0.000001    0.727152   -0.784901
  6          1.101966    1.154052    0.156532
  6          2.412473    0.829086   -0.630978
  6          1.306274    0.000012    1.175596
  1          0.000025   -1.300309   -1.719852
  6          0.000018   -2.267412    0.587019
  1          3.328995   -1.180414   -0.144839
  1         -2.356903   -0.000007    1.495370
  1         -0.765363    0.000003    2.105924
  1          0.000005    1.300309   -1.719856
  6         -0.000008    2.267423    0.587007
  1          3.328970    1.180448   -0.144892
  1          2.400909    1.211208   -1.658132
  1          0.765330    0.000011    2.105935
  1          2.356883    0.000019    1.495413
  1          2.400967   -1.211252   -1.658093
  6         -2.412491    0.829104   -0.630983
  6         -2.412477   -0.829136   -0.630997
  1         -2.401052    1.211522   -1.658028
  1         -3.329006    1.180197   -0.144737
  1         -3.328985   -1.180269   -0.144767
  1         -2.401013   -1.211552   -1.658043
  1          0.000013   -2.603842    1.630185
  1          0.000030   -3.136840   -0.078654
  1         -0.000006    3.136845   -0.078674
  1         -0.000013    2.603861    1.630170
------------------------------------------------------------
```

The Cartesian coordinate of optimized 6 compound at MP2/6-311++g(d,p).

```
------------------------------------------------------------
Atomic          Coordinates (Angstroms)
Number          X          Y          Z
------------------------------------------------------------
  6       -1.104668  -1.157306   0.156191
  6        0.000025  -0.729613  -0.791032
  6        1.104761  -1.157262   0.156170
  6        2.417641  -0.831771  -0.631193
  6       -1.304776  -0.000032   1.176224
  6       -1.104752   1.157276   0.156199
  6       -0.000046   0.729585  -0.791037
  6        1.104661   1.157293   0.156174
  6        2.417530   0.831897  -0.631212
  6        1.304793   0.000019   1.176215
  1        0.000031  -1.308622  -1.721052
  6        0.000082  -2.271640   0.591630
  1        3.332841  -1.177414  -0.139361
  1       -2.356402  -0.000062   1.495299
  1       -0.756014  -0.000012   2.101958
  1       -0.000062   1.308583  -1.721064
  6       -0.000054   2.271639   0.591597
  1        3.332705   1.177666  -0.139425
  1        2.401372   1.210286  -1.659725
  1        0.756035  -0.000002   2.101953
  1        2.356421   0.000060   1.495281
  1        2.401554  -1.210203  -1.659692
  6       -2.417641   0.831800  -0.631154
  6       -2.417561  -0.831894  -0.631193
  1       -2.401552   1.210224  -1.659655
  1       -3.332835   1.177448  -0.139317
  1       -3.332734  -1.177647  -0.139390
  1       -2.401419  -1.210272  -1.659709
  1        0.000101  -2.598411   1.637446
  1        0.000096  -3.139248  -0.076170
  1       -0.000073   3.139218  -0.076244
  1       -0.000049   2.598462   1.637397
------------------------------------------------------------
```

The Cartesian coordinate of optimized 6 compound at B3LYP/6-311++g(d,p).
------------------------------------------------------------
| Atomic | Coordinates (Angstroms) | | |
| Number | X | Y | Z |
------------------------------------------------------------

| Atomic Number | X | Y | Z |
|---|---|---|---|
| 6 | -1.110476 | -1.160378 | 0.152580 |
| 6 | 0.000004 | -0.729284 | -0.785753 |
| 6 | 1.110484 | -1.160373 | 0.152583 |
| 6 | 2.431293 | -0.837241 | -0.636929 |
| 6 | -1.317612 | -0.000004 | 1.177556 |
| 6 | -1.110483 | 1.160374 | 0.152583 |
| 6 | -0.000003 | 0.729285 | -0.785753 |
| 6 | 1.110476 | 1.160378 | 0.152581 |
| 6 | 2.431280 | 0.837251 | -0.636943 |
| 6 | 1.317612 | 0.000003 | 1.177557 |
| 1 | 0.000006 | -1.293069 | -1.721614 |
| 6 | 0.000004 | -2.279580 | 0.600187 |
| 1 | 3.343071 | -1.184669 | -0.147586 |
| 1 | -2.361156 | -0.000007 | 1.508067 |
| 1 | -0.761118 | -0.000003 | 2.095107 |
| 1 | -0.000005 | 1.293071 | -1.721614 |
| 6 | -0.000007 | 2.279580 | 0.600188 |
| 1 | 3.343061 | 1.184677 | -0.147605 |
| 1 | 2.423681 | 1.214941 | -1.662344 |
| 1 | 0.761117 | 0.000002 | 2.095107 |
| 1 | 2.361156 | 0.000006 | 1.508069 |
| 1 | 2.423702 | -1.214926 | -1.662331 |
| 6 | -2.431291 | 0.837241 | -0.636931 |
| 6 | -2.431281 | -0.837251 | -0.636942 |
| 1 | -2.423694 | 1.214915 | -1.662337 |
| 1 | -3.343069 | 1.184679 | -0.147595 |
| 1 | -3.343062 | -1.184667 | -0.147597 |
| 1 | -2.423688 | -1.214951 | -1.662339 |
| 1 | 0.000004 | -2.593280 | 1.646122 |
| 1 | 0.000007 | -3.157434 | -0.047507 |
| 1 | -0.000009 | 3.157435 | -0.047505 |
| 1 | -0.000007 | 2.593280 | 1.646123 |

------------------------------------------------------------

The Cartesian coordinate of optimized 6 compound at M06L/6-311++g(d,p).

```
------------------------------------------------------------
Atomic          Coordinates (Angstroms)
Number        X           Y           Z
------------------------------------------------------------
   6       1.098996    1.152214    0.153826
   6      -0.000005    0.725597   -0.784359
   6      -1.099005    1.152212    0.153831
   6      -2.410034    0.826456   -0.625043
   6       1.303579    0.000010    1.171148
   6       1.099003   -1.152208    0.153840
   6       0.000003   -0.725609   -0.784352
   6      -1.098995   -1.152218    0.153838
   6      -2.410006   -0.826470   -0.625070
   6      -1.303583    0.000001    1.171147
   1      -0.000008    1.299617   -1.715071
   6      -0.000003    2.257280    0.581900
   1      -3.324517    1.174194   -0.137209
   1       2.352602    0.000016    1.488680
   1       0.762729    0.000014    2.100084
   1       0.000005   -1.299640   -1.715057
   6       0.000010   -2.257276    0.581928
   1      -3.324525   -1.173936   -0.137112
   1      -2.407710   -1.206101   -1.651874
   1      -0.762736    0.000007    2.100085
   1      -2.352606   -0.000002    1.488675
   1      -2.407646    1.205812   -1.651948
   6       2.410015   -0.826441   -0.625061
   6       2.410018    0.826449   -0.625059
   1       2.407714   -1.206033   -1.651879
   1       3.324534   -1.173930   -0.137118
   1       3.324517    1.174088   -0.137184
   1       2.407664    1.205892   -1.651932
   1       0.000000    2.593904    1.623752
   1      -0.000005    3.128951   -0.076899
   1       0.000014   -3.128955   -0.076859
   1       0.000010   -2.593884    1.623785
------------------------------------------------------------
```

The Cartesian coordinate of optimized 7 compound at HF/6-311++g(d,p).
------------------------------------------------------------

| Atomic Number | Coordinates (Angstroms) | | |
|---|---|---|---|
| | X | Y | Z |

------------------------------------------------------------

| | | | |
|---|---|---|---|
| 6 | 1.115784 | 1.158632 | -0.025829 |
| 6 | 0.000008 | 0.725086 | -0.938166 |
| 6 | -1.115768 | 1.158638 | -0.025832 |
| 6 | -2.359868 | 0.839141 | -0.938131 |
| 6 | 1.307345 | -0.000009 | 0.987518 |
| 6 | 1.115778 | -1.158639 | -0.025836 |
| 6 | 0.000008 | -0.725077 | -0.938174 |
| 6 | -1.115774 | -1.158632 | -0.025850 |
| 6 | -2.359857 | -0.839115 | -0.938166 |
| 6 | -1.307352 | -0.000003 | 0.987507 |
| 1 | 0.000012 | 1.283854 | -1.863549 |
| 6 | 0.000010 | 2.274844 | 0.419590 |
| 17 | -3.914894 | 1.644991 | -0.625716 |
| 17 | 2.859318 | -0.000025 | 1.925120 |
| 1 | 0.692688 | -0.000011 | 1.832417 |
| 1 | 0.000012 | -1.283835 | -1.863563 |
| 6 | -0.000004 | -2.274850 | 0.419566 |
| 17 | -3.914878 | -1.645004 | -0.625830 |
| 1 | -2.145689 | -1.127016 | -1.952585 |
| 1 | -0.692717 | -0.000010 | 1.832421 |
| 17 | -2.859357 | 0.000002 | 1.925054 |
| 1 | -2.145729 | 1.127089 | -1.952543 |
| 6 | 2.359880 | -0.839132 | -0.938130 |
| 6 | 2.359866 | 0.839124 | -0.938151 |
| 1 | 2.145736 | -1.127067 | -1.952545 |
| 17 | 3.914905 | -1.644993 | -0.625739 |
| 17 | 3.914885 | 1.645025 | -0.625834 |
| 1 | 2.145688 | 1.127029 | -1.952568 |
| 1 | 0.000010 | 2.572995 | 1.458435 |
| 1 | 0.000013 | 3.146776 | -0.217863 |
| 1 | -0.000002 | -3.146774 | -0.217897 |
| 1 | -0.000011 | -2.573012 | 1.458408 |

------------------------------------------------------------

The Cartesian coordinate of optimized 7 compound at MP2/6-31G(d).

```
------------------------------------------------------------
Atomic          Coordinates (Angstroms)
Number          X           Y           Z
------------------------------------------------------------
   6        -1.117973   -1.160085   -0.012919
   6         0.000003   -0.727080   -0.937394
   6         1.117965   -1.160087   -0.012902
   6         2.350675   -0.845603   -0.928975
   6        -1.312812   -0.000041    0.994618
   6        -1.117974    1.160096   -0.012801
   6        -0.000010    0.727177   -0.937329
   6         1.117960    1.160102   -0.012816
   6         2.350598    0.845661   -0.928992
   6         1.312813   -0.000034    0.994615
   1         0.000005   -1.300446   -1.870699
   6         0.000000   -2.268177    0.424151
  17         3.912644   -1.624570   -0.615813
  17        -2.877787   -0.000076    1.884507
   1        -0.689601   -0.000084    1.856163
   1        -0.000018    1.300625   -1.870583
   6         0.000000    2.268157    0.424345
  17         3.912582    1.624645   -0.615956
   1         2.124381    1.134355   -1.957584
   1         0.689631   -0.000070    1.856183
  17         2.877809   -0.000083    1.884482
   1         2.124551   -1.134265   -1.957596
   6        -2.350645    0.845654   -0.928920
   6        -2.350628   -0.845581   -0.929049
   1        -2.124495    1.134359   -1.957523
  17        -3.912631    1.624596   -0.615772
  17        -3.912607   -1.624578   -0.616001
   1        -2.124445   -1.134243   -1.957657
   1        -0.000010   -2.579784    1.473285
   1         0.000002   -3.146415   -0.227074
   1        -0.000009    3.146447   -0.226809
   1         0.000005    2.579678    1.473505
------------------------------------------------------------
```

The Cartesian coordinate of optimized 7 compound at MP2/6-311++g(d,p).

```
------------------------------------------------------------
Atomic          Coordinates (Angstroms)
Number          X         Y         Z
------------------------------------------------------------
   6        -1.120207  -1.164345  -0.014190
   6        -0.000092  -0.729230  -0.943854
   6         1.120363  -1.163895  -0.014440
   6         2.351405  -0.845539  -0.931080
   6        -1.308196  -0.001111   0.990966
   6        -1.119701   1.163736  -0.012785
   6        -0.000183   0.730018  -0.943386
   6         1.120153   1.164055  -0.013379
   6         2.351894   0.846423  -0.929665
   6         1.308653  -0.000025   0.990655
   1        -0.000002  -1.307251  -1.873057
   6         0.000352  -2.272103   0.428401
  17         3.911842  -1.626739  -0.618620
  17        -2.869031  -0.001401   1.890423
   1        -0.679606  -0.001940   1.849153
   1        -0.000572   1.309056  -1.872000
   6         0.000054   2.271484   0.430423
  17         3.912584   1.627041  -0.616149
   1         2.121444   1.133394  -1.958117
   1         0.679973  -0.000321   1.848734
  17         2.869345  -0.000851   1.890392
   1         2.120414  -1.131106  -1.959750
   6        -2.351692   0.846411  -0.929689
   6        -2.352386  -0.845807  -0.930150
   1        -2.120247   1.132476  -1.958179
  17        -3.911883   1.628173  -0.617555
  17        -3.913042  -1.626302  -0.617160
   1        -2.121740  -1.131730  -1.958845
   1         0.000453  -2.569937   1.481071
   1         0.000489  -3.150181  -0.222324
   1        -0.000307   3.150089  -0.219607
   1         0.000350   2.568365   1.483328
------------------------------------------------------------
```

The Cartesian coordinate of optimized 7 compound at B3LYP/6-311++g(d,p).

------------------------------------------------------------

| Atomic Number | Coordinates (Angstroms) | | |
|---|---|---|---|
| | X | Y | Z |

------------------------------------------------------------

| | | | |
|---|---|---|---|
| 6  | -1.125990 | -1.167846 | -0.028343 |
| 6  |  0.000223 | -0.728950 | -0.950785 |
| 6  |  1.126193 | -1.167892 | -0.028067 |
| 6  |  2.368826 | -0.853324 | -0.946744 |
| 6  | -1.316726 | -0.000094 |  0.983738 |
| 6  | -1.125967 |  1.167837 | -0.028141 |
| 6  |  0.000240 |  0.729086 | -0.950656 |
| 6  |  1.126221 |  1.167842 | -0.027865 |
| 6  |  2.368841 |  0.853443 | -0.946632 |
| 6  |  1.316720 | -0.000119 |  0.984050 |
| 1  |  0.000327 | -1.291641 | -1.884780 |
| 6  |  0.000017 | -2.281719 |  0.423493 |
| 17 |  3.945529 | -1.648655 | -0.623616 |
| 17 | -2.886572 | -0.000169 |  1.933804 |
| 1  | -0.692911 | -0.000176 |  1.840168 |
| 1  |  0.000354 |  1.291942 | -1.884552 |
| 6  |  0.000061 |  2.281601 |  0.423908 |
| 17 |  3.945500 |  1.648856 | -0.623490 |
| 1  |  2.157711 |  1.143771 | -1.972252 |
| 1  |  0.692679 | -0.000188 |  1.840315 |
| 17 |  2.886285 | -0.000229 |  1.934581 |
| 1  |  2.157761 | -1.143514 | -1.972417 |
| 6  | -2.368380 |  0.853389 | -0.947168 |
| 6  | -2.368415 | -0.853194 | -0.947275 |
| 1  | -2.157546 |  1.144906 | -1.972511 |
| 17 | -3.945702 |  1.647100 | -0.623096 |
| 17 | -3.945745 | -1.646943 | -0.623331 |
| 1  | -2.157610 | -1.144471 | -1.972693 |
| 1  | -0.000119 | -2.571062 |  1.473563 |
| 1  |  0.000075 | -3.165547 | -0.211504 |
| 1  |  0.000134 |  3.165549 | -0.210921 |
| 1  | -0.000071 |  2.570744 |  1.474032 |

------------------------------------------------------------

The Cartesian coordinate of optimized 7 compound at M06L/6-311++g(d,p).
------------------------------------------------------------

| Atomic Number | Coordinates (Angstroms) | | |
|---|---|---|---|
| | X | Y | Z |

------------------------------------------------------------

| | | | |
|---|---|---|---|
| 6 | -1.114892 | -1.161153 | -0.014358 |
| 6 | -0.000017 | -0.725171 | -0.935288 |
| 6 | 1.114893 | -1.161136 | -0.014397 |
| 6 | 2.345648 | -0.845718 | -0.931663 |
| 6 | -1.306281 | -0.000314 | 0.990356 |
| 6 | -1.114908 | 1.161107 | -0.013726 |
| 6 | -0.000009 | 0.725675 | -0.934869 |
| 6 | 1.114887 | 1.161113 | -0.013721 |
| 6 | 2.345614 | 0.846235 | -0.931212 |
| 6 | 1.306277 | -0.000306 | 0.990340 |
| 1 | -0.000026 | -1.295950 | -1.865954 |
| 6 | 0.000018 | -2.257997 | 0.420004 |
| 17 | 3.919323 | -1.611887 | -0.621931 |
| 17 | -2.877948 | -0.000624 | 1.899991 |
| 1 | -0.690562 | -0.000556 | 1.856894 |
| 1 | -0.000012 | 1.296990 | -1.865208 |
| 6 | -0.000013 | 2.257701 | 0.421313 |
| 17 | 3.919267 | 1.612320 | -0.621170 |
| 1 | 2.122618 | 1.138745 | -1.958025 |
| 1 | 0.690561 | -0.000560 | 1.856881 |
| 17 | 2.877948 | -0.000576 | 1.899969 |
| 1 | 2.122722 | -1.137689 | -1.958645 |
| 6 | -2.345598 | 0.846234 | -0.931278 |
| 6 | -2.345725 | -0.845769 | -0.931524 |
| 1 | -2.122490 | 1.138648 | -1.958095 |
| 17 | -3.919198 | 1.612518 | -0.621456 |
| 17 | -3.919352 | -1.611995 | -0.621681 |
| 1 | -2.122871 | -1.137643 | -1.958549 |
| 1 | 0.000040 | -2.565378 | 1.468190 |
| 1 | 0.000013 | -3.139651 | -0.221072 |
| 1 | -0.000014 | 3.139736 | -0.219239 |
| 1 | -0.000017 | 2.564461 | 1.469682 |

------------------------------------------------------------

The Cartesian coordinate of optimized 8 compound at B3LYP/6-311++g(d,p).
---------------------------------------------------------

| Atomic Number | Coordinates (Angstroms) | | |
|---|---|---|---|
| | X | Y | Z |

---------------------------------------------------------

| | | | |
|---|---|---|---|
| 6 | -1.139520 | 0.482967 | 1.184762 |
| 6 | -0.000041 | -0.484432 | 0.740086 |
| 6 | 1.139393 | 0.483217 | 1.184526 |
| 6 | 2.587122 | -0.173101 | 0.957836 |
| 6 | -1.290149 | 1.490817 | 0.000238 |
| 6 | -1.139345 | 0.483067 | -1.184612 |
| 6 | -0.000046 | -0.484596 | -0.739965 |
| 6 | 1.139560 | 0.482573 | -1.184867 |
| 6 | 2.587139 | -0.173801 | -0.957740 |
| 6 | 1.290117 | 1.490690 | -0.000558 |
| 17 | -0.000239 | -2.022518 | 1.635039 |
| 6 | -0.000005 | 0.977191 | 2.273370 |
| 17 | 3.916105 | 0.801857 | 1.666796 |
| 17 | -2.552216 | 2.865230 | 0.000245 |
| 1 | -0.589155 | 2.255125 | 0.001036 |
| 17 | -0.000470 | -2.022886 | -1.634567 |
| 6 | 0.000104 | 0.976684 | -2.273575 |
| 17 | 3.916276 | 0.800495 | -1.667291 |
| 17 | 2.967109 | -1.819548 | -1.528533 |
| 1 | 0.588950 | 2.254826 | -0.001529 |
| 17 | 2.551829 | 2.865383 | -0.000849 |
| 17 | 2.967491 | -1.818335 | 1.529851 |
| 6 | -2.586952 | -0.173393 | -0.957816 |
| 6 | -2.587082 | -0.173445 | 0.957802 |
| 17 | -2.966717 | -1.818950 | -1.529309 |
| 17 | -3.916075 | 0.801037 | -1.667242 |
| 17 | -3.916251 | 0.801094 | 1.666955 |
| 17 | -2.966949 | -1.819023 | 1.529144 |
| 1 | -0.000138 | 2.039532 | 2.511159 |
| 1 | 0.000233 | 0.375208 | 3.177243 |
| 1 | -0.000197 | 0.374502 | -3.177314 |
| 1 | 0.000347 | 2.038973 | -2.511597 |

---------------------------------------------------------

The Cartesian coordinate of optimized 8 compound at M06L/6-311++g(d,p).
------------------------------------------------------------

| Atomic Number | Coordinates (Angstroms) | | |
|---|---|---|---|
| | X | Y | Z |

------------------------------------------------------------

| | | | |
|---|---|---|---|
| 6 | 1.128738 | 0.483138 | -1.184117 |
| 6 | -0.000114 | -0.477902 | -0.737038 |
| 6 | -1.128385 | 0.484434 | -1.183182 |
| 6 | -2.562557 | -0.165811 | -0.987851 |
| 6 | 1.286710 | 1.480369 | -0.001382 |
| 6 | 1.128348 | 0.484447 | 1.183186 |
| 6 | 0.000115 | -0.477924 | 0.737006 |
| 6 | -1.128773 | 0.483037 | 1.184111 |
| 6 | -2.562539 | -0.167624 | 0.987767 |
| 6 | -1.286623 | 1.480274 | 0.001407 |
| 17 | -0.000767 | -2.005367 | -1.618307 |
| 6 | 0.000025 | 0.967669 | -2.252194 |
| 17 | -3.886577 | 0.799569 | -1.664647 |
| 17 | 2.540381 | 2.828769 | -0.002160 |
| 1 | 0.583933 | 2.247867 | -0.004022 |
| 17 | 0.000811 | -2.005421 | 1.618220 |
| 6 | -0.000104 | 0.967600 | 2.252204 |
| 17 | -3.887083 | 0.795922 | 1.665998 |
| 17 | -2.939068 | -1.806857 | 1.516268 |
| 1 | -0.583611 | 2.247518 | 0.004077 |
| 17 | -2.539976 | 2.828911 | 0.002187 |
| 17 | -2.940316 | -1.803761 | -1.519532 |
| 6 | 2.562479 | -0.165872 | 0.987920 |
| 6 | 2.562506 | -0.167565 | -0.987848 |
| 17 | 2.940054 | -1.803933 | 1.519372 |
| 17 | 3.886525 | 0.799310 | 1.664950 |
| 17 | 3.887148 | 0.795874 | -1.666013 |
| 17 | 2.938921 | -1.806817 | -1.516336 |
| 1 | 0.000677 | 2.029898 | -2.505218 |
| 1 | -0.000972 | 0.359992 | -3.155617 |
| 1 | 0.000935 | 0.359913 | 3.155620 |
| 1 | -0.000817 | 2.029826 | 2.505239 |

------------------------------------------------------------

The Cartesian coordinate of optimized 9 compound at HF/6-311++g(d,p).
-------------------------------------------------------------
| Atomic | Coordinates (Angstroms) | | |
| Number | X | Y | Z |
-------------------------------------------------------------

| | | | |
|---|---|---|---|
| 6  |  1.095816 |  0.518685 | -1.118930 |
| 6  |  0.004106 | -0.506685 | -0.733868 |
| 6  | -1.150755 |  0.386010 | -1.215877 |
| 6  | -2.584343 | -0.249688 | -0.881003 |
| 6  |  1.234165 |  1.482763 |  0.151056 |
| 6  |  1.150740 |  0.386277 |  1.215751 |
| 6  | -0.004160 | -0.506504 |  0.733945 |
| 6  | -1.095823 |  0.519064 |  1.118753 |
| 6  | -2.567715 | -0.101451 |  0.893163 |
| 6  | -1.234173 |  1.482803 | -0.151499 |
| 17 |  0.066906 | -2.033264 | -1.618074 |
| 6  | -0.003142 |  0.910458 | -2.284923 |
| 17 | -3.911073 |  0.653211 | -1.655891 |
| 17 |  2.421368 |  2.864174 |  0.206722 |
| 1  |  0.495006 |  2.149444 |  0.404965 |
| 17 | -0.067147 | -2.032888 |  1.618461 |
| 6  |  0.003166 |  0.911051 |  2.284644 |
| 17 | -3.842350 |  0.962469 |  1.559403 |
| 17 | -3.006732 | -1.644576 |  1.660391 |
| 1  | -0.495016 |  2.149441 | -0.405567 |
| 17 | -2.421346 |  2.864236 | -0.207568 |
| 17 | -2.935942 | -1.921711 | -1.358201 |
| 6  |  2.584403 | -0.249294 |  0.881124 |
| 6  |  2.567678 | -0.101907 | -0.893102 |
| 17 |  2.936486 | -1.920973 |  1.359171 |
| 17 |  3.910950 |  0.654352 |  1.655452 |
| 17 |  3.842395 |  0.961532 | -1.559924 |
| 17 |  3.006495 | -1.645446 | -1.659599 |
| 1  | -0.070913 |  1.950758 | -2.566530 |
| 1  |  0.103895 |  0.284845 | -3.154707 |
| 1  | -0.103925 |  0.285648 |  3.154575 |
| 1  |  0.070996 |  1.951412 |  2.566015 |

-------------------------------------------------------------

The Cartesian coordinate of optimized 9 compound at MP2/6-31G(d).

```
-------------------------------------------------------------
Atomic          Coordinates (Angstroms)
Number          X          Y          Z
-------------------------------------------------------------
   6       -1.094798   0.531162   1.111140
   6       -0.008592  -0.517324   0.734962
   6        1.150770   0.368172   1.231112
   6        2.563006  -0.264131   0.904687
   6       -1.231154   1.476930  -0.180172
   6       -1.150779   0.368122  -1.231130
   6        0.008595  -0.517357  -0.734945
   6        1.094796   0.531131  -1.111160
   6        2.548936  -0.074985  -0.918339
   6        1.231160   1.476952   0.180126
  17       -0.087792  -2.037951   1.611136
   6       -0.003629   0.895779   2.283847
  17        3.897698   0.627166   1.660027
  17       -2.417346   2.848187  -0.268777
   1       -0.467336   2.138630  -0.462886
  17        0.087816  -2.038024  -1.611038
   6        0.003627   0.895683  -2.283881
  17        3.815781   1.015862  -1.534633
  17        2.991191  -1.608624  -1.683127
   1        0.467357   2.138683   0.462826
  17        2.417404   2.848172   0.268711
  17        2.886713  -1.951729   1.308246
   6       -2.563021  -0.264159  -0.904691
   6       -2.548941  -0.074950   0.918337
  17       -2.886790  -1.951754  -1.308201
  17       -3.897712   0.627119  -1.660052
  17       -3.815750   1.015962   1.534590
  17       -2.991205  -1.608570   1.683163
   1        0.079514   1.942332   2.592816
   1       -0.138070   0.246629   3.150994
   1        0.138071   0.246486  -3.150992
   1       -0.079521   1.942219  -2.592910
-------------------------------------------------------------
```

The Cartesian coordinate of optimized 9 compound at B3LYP/6-311++g(d,p).

---

| Atomic Number | Coordinates (Angstroms) | | |
|---|---|---|---|
| | X | Y | Z |

---

| | | | |
|---|---|---|---|
| 6 | -1.111746 | 0.518694 | 1.143834 |
| 6 | -0.002628 | -0.506933 | 0.739517 |
| 6 | 1.157320 | 0.407667 | 1.223779 |
| 6 | 2.590317 | -0.234987 | 0.942518 |
| 6 | -1.259937 | 1.482787 | -0.122285 |
| 6 | -1.157301 | 0.406754 | -1.223970 |
| 6 | 0.002931 | -0.507287 | -0.739260 |
| 6 | 1.111738 | 0.518399 | -1.144055 |
| 6 | 2.574676 | -0.109209 | -0.952070 |
| 6 | 1.259796 | 1.483147 | 0.121581 |
| 17 | -0.050030 | -2.049428 | 1.627905 |
| 6 | 0.001882 | 0.938760 | 2.286195 |
| 17 | 3.929588 | 0.697792 | 1.683294 |
| 17 | -2.480032 | 2.877269 | -0.163581 |
| 1 | -0.525535 | 2.188735 | -0.346738 |
| 17 | 0.051512 | -2.050170 | -1.626912 |
| 6 | -0.001998 | 0.937625 | -2.286630 |
| 17 | 3.872850 | 0.944333 | -1.613611 |
| 17 | 2.997056 | -1.697171 | -1.648464 |
| 1 | 0.525270 | 2.189072 | 0.345686 |
| 17 | 2.479654 | 2.877856 | 0.162255 |
| 17 | 2.949031 | -1.917979 | 1.406912 |
| 6 | -2.590260 | -0.235931 | -0.942368 |
| 6 | -2.574858 | -0.108707 | 0.952119 |
| 17 | -2.948993 | -1.919256 | -1.405540 |
| 17 | -3.929550 | 0.696180 | -1.683940 |
| 17 | -3.872748 | 0.945925 | 1.612464 |
| 17 | -2.998276 | -1.695857 | 1.649730 |
| 1 | 0.059439 | 1.992755 | 2.552992 |
| 1 | -0.085696 | 0.316725 | 3.172155 |
| 1 | 0.085730 | 0.315174 | -3.172283 |
| 1 | -0.059813 | 1.991475 | -2.553947 |

---

The Cartesian coordinate of optimized 9 compound at M06L/6-311++g(d,p).

------------------------------------------------------------

| Atomic Number | Coordinates (Angstroms) | | |
|---|---|---|---|
| | X | Y | Z |

------------------------------------------------------------

| | | | |
|---|---|---|---|
| 6 | -1.104058 | 0.532951 | 1.124149 |
| 6 | -0.013818 | -0.505549 | 0.735606 |
| 6 | 1.137968 | 0.384349 | 1.237393 |
| 6 | 2.556124 | -0.260312 | 0.955997 |
| 6 | -1.243593 | 1.471310 | -0.164455 |
| 6 | -1.137969 | 0.384036 | -1.237456 |
| 6 | 0.013909 | -0.505663 | -0.735529 |
| 6 | 1.104044 | 0.532838 | -1.124231 |
| 6 | 2.555107 | -0.070631 | -0.962697 |
| 6 | 1.243548 | 1.471429 | 0.164199 |
| 17 | -0.096818 | -2.038984 | 1.605257 |
| 6 | -0.013861 | 0.914883 | 2.270329 |
| 17 | 3.903737 | 0.623526 | 1.689984 |
| 17 | -2.446916 | 2.842379 | -0.277687 |
| 1 | -0.496812 | 2.164400 | -0.406237 |
| 17 | 0.097316 | -2.039193 | -1.604982 |
| 6 | 0.013812 | 0.914489 | -2.270483 |
| 17 | 3.834631 | 1.014912 | -1.556048 |
| 17 | 3.007943 | -1.616643 | -1.689088 |
| 1 | 0.496741 | 2.164534 | 0.405846 |
| 17 | 2.446849 | 2.842538 | 0.277189 |
| 17 | 2.876416 | -1.954822 | 1.328608 |
| 6 | -2.556098 | -0.260659 | -0.955920 |
| 6 | -2.555197 | -0.070408 | 0.962723 |
| 17 | -2.876326 | -1.955298 | -1.327984 |
| 17 | -3.903728 | 0.622865 | -1.690246 |
| 17 | -3.834622 | 1.015502 | 1.555618 |
| 17 | -3.008441 | -1.616094 | 1.689555 |
| 1 | 0.062985 | 1.963065 | 2.565932 |
| 1 | -0.138584 | 0.277909 | 3.144656 |
| 1 | 0.138581 | 0.277378 | -3.144703 |
| 1 | -0.063115 | 1.962616 | -2.566261 |

------------------------------------------------------------

The Cartesian coordinate of optimized 10 compound at HF/6-311++g(d,p).

------------------------------------------------------------

| Atomic Number | Coordinates (Angstroms) | | |
|---|---|---|---|
| | X | Y | Z |

------------------------------------------------------------

| | | | |
|---|---|---|---|
| 6 | -1.027328 | 1.382762 | 0.062405 |
| 6 | -0.781056 | 0.000000 | -0.576236 |
| 6 | -1.027328 | -1.382763 | 0.062405 |
| 6 | 0.000000 | -2.211840 | -0.767303 |
| 6 | 0.000000 | 1.538563 | 1.209402 |
| 6 | 1.027328 | 1.382763 | 0.062405 |
| 6 | 0.781056 | 0.000000 | -0.576236 |
| 6 | 1.027328 | -1.382762 | 0.062405 |
| 6 | 0.000000 | -1.538563 | 1.209402 |
| 1 | -2.054405 | 1.706092 | 0.178054 |
| 1 | -1.210450 | 0.000000 | -1.573931 |
| 1 | -2.054405 | -1.706092 | 0.178054 |
| 1 | 0.000000 | -3.265280 | -0.513464 |
| 1 | 0.000000 | 2.559205 | 1.576916 |
| 1 | 0.000000 | 0.878905 | 2.056615 |
| 1 | 2.054405 | 1.706092 | 0.178054 |
| 1 | 1.210450 | 0.000000 | -1.573931 |
| 1 | 2.054405 | -1.706092 | 0.178054 |
| 1 | 0.000000 | -0.878905 | 2.056615 |
| 1 | 0.000000 | -2.559205 | 1.576916 |
| 1 | 0.000000 | -2.089332 | -1.846274 |
| 6 | 0.000000 | 2.211840 | -0.767303 |
| 1 | 0.000000 | 3.265280 | -0.513464 |
| 1 | 0.000000 | 2.089332 | -1.846274 |

------------------------------------------------------------

The Cartesian coordinate of optimized 10 compound at MP2/6-31G(d).
```
-----------------------------------------------------------------
Atomic          Coordinates (Angstroms)
Number          X          Y          Z
-----------------------------------------------------------------
   6         1.378880    0.063615   -1.029519
   6         0.000000   -0.575727   -0.785026
   6        -1.378881    0.063613   -1.029519
   6        -2.205955   -0.768233    0.000001
   6         1.533483    1.208841    0.000001
   6         1.378881    0.063613    1.029519
   6         0.000000   -0.575727    0.785026
   6        -1.378880    0.063615    1.029519
   6        -1.533483    1.208841   -0.000001
   1         1.708122    0.181036   -2.066532
   1         0.000000   -1.587873   -1.212398
   1        -1.708124    0.181033   -2.066532
   1        -3.269937   -0.511136    0.000001
   1         2.565620    1.576073    0.000000
   1         0.866991    2.065375    0.000003
   1         1.708124    0.181033    2.066532
   1         0.000000   -1.587873    1.212398
   1        -1.708122    0.181036    2.066532
   1        -0.866992    2.065375   -0.000003
   1        -2.565620    1.576073    0.000000
   1        -2.075501   -1.857158    0.000002
   6         2.205954   -0.768233   -0.000001
   1         3.269937   -0.511136   -0.000001
   1         2.075501   -1.857158   -0.000002
-----------------------------------------------------------------
```

The Cartesian coordinate of optimized 10 compound at MP2/6-311++g(d,p).

```
------------------------------------------------------------
Atomic         Coordinates (Angstroms)
Number         X           Y           Z
------------------------------------------------------------
  6        -1.033236    1.378570    0.063661
  6        -0.786804    0.000000   -0.580854
  6        -1.033236   -1.378563    0.063673
  6        -0.000007   -2.210596   -0.767998
  6        -0.000007    1.529222    1.212942
  6         1.033236    1.378563    0.063673
  6         0.786804    0.000000   -0.580854
  6         1.033236   -1.378570    0.063661
  6         0.000007   -1.529222    1.212942
  1        -2.069214    1.706439    0.180199
  1        -1.211535   -0.000006   -1.593274
  1        -2.069214   -1.706427    0.180224
  1        -0.000008   -3.271756   -0.500621
  1        -0.000003    2.562897    1.575640
  1        -0.000017    0.858388    2.065712
  1         2.069214    1.706427    0.180224
  1         1.211535    0.000006   -1.593274
  1         2.069214   -1.706439    0.180199
  1         0.000017   -0.858388    2.065712
  1         0.000003   -2.562897    1.575640
  1        -0.000014   -2.078100   -1.856423
  6         0.000007    2.210596   -0.767998
  1         0.000008    3.271756   -0.500621
  1         0.000014    2.078100   -1.856423
------------------------------------------------------------
```

The Cartesian coordinate of optimized 10 compound at B3LYP/6-311++g(d,p).
------------------------------------------------------------
| Atomic | Coordinates (Angstroms) | | |
| Number | X | Y | Z |
------------------------------------------------------------

| | | | |
|---|---|---|---|
| 6 | -1.386987 | 0.062119 | 1.034302 |
| 6 | 0.000001 | -0.580354 | 0.789200 |
| 6 | 1.386988 | 0.062119 | 1.034302 |
| 6 | 2.224309 | -0.770030 | -0.000001 |
| 6 | -1.538630 | 1.215619 | -0.000001 |
| 6 | -1.386988 | 0.062119 | -1.034302 |
| 6 | 0.000000 | -0.580354 | -0.789200 |
| 6 | 1.386987 | 0.062120 | -1.034302 |
| 6 | 1.538629 | 1.215620 | 0.000000 |
| 1 | -1.714526 | 0.179945 | 2.067788 |
| 1 | 0.000001 | -1.587643 | 1.217534 |
| 1 | 1.714528 | 0.179944 | 2.067787 |
| 1 | 3.283601 | -0.506866 | -0.000001 |
| 1 | -2.564612 | 1.591816 | 0.000001 |
| 1 | -0.863901 | 2.063049 | -0.000002 |
| 1 | -1.714527 | 0.179944 | -2.067787 |
| 1 | 0.000001 | -1.587643 | -1.217534 |
| 1 | 1.714527 | 0.179946 | -2.067787 |
| 1 | 0.863899 | 2.063049 | 0.000002 |
| 1 | 2.564611 | 1.591818 | 0.000000 |
| 1 | 2.102440 | -1.857094 | -0.000001 |
| 6 | -2.224309 | -0.770029 | 0.000000 |
| 1 | -3.283601 | -0.506865 | 0.000001 |
| 1 | -2.102441 | -1.857094 | 0.000001 |

------------------------------------------------------------

The Cartesian coordinate of optimized 10 compound at M06L/6-311++g(d,p).
------------------------------------------------------------

| Atomic Number | Coordinates (Angstroms) | | |
|---|---|---|---|
| | X | Y | Z |

------------------------------------------------------------

| | | | |
|---|---|---|---|
| 6 | 1.374298 | 0.062046 | -1.022849 |
| 6 | 0.000000 | -0.577053 | -0.779784 |
| 6 | -1.374299 | 0.062045 | -1.022849 |
| 6 | -2.203289 | -0.760936 | 0.000001 |
| 6 | 1.527592 | 1.204620 | 0.000001 |
| 6 | 1.374299 | 0.062045 | 1.022849 |
| 6 | 0.000000 | -0.577052 | 0.779784 |
| 6 | -1.374298 | 0.062046 | 1.022849 |
| 6 | -1.527592 | 1.204620 | -0.000001 |
| 1 | 1.702476 | 0.179214 | -2.058343 |
| 1 | 0.000000 | -1.587777 | -1.205092 |
| 1 | -1.702477 | 0.179211 | -2.058343 |
| 1 | -3.263019 | -0.497391 | 0.000001 |
| 1 | 2.557246 | 1.571779 | 0.000000 |
| 1 | 0.860510 | 2.060201 | 0.000003 |
| 1 | 1.702477 | 0.179211 | 2.058343 |
| 1 | 0.000000 | -1.587777 | 1.205092 |
| 1 | -1.702475 | 0.179214 | 2.058343 |
| 1 | -0.860510 | 2.060201 | -0.000003 |
| 1 | -2.557246 | 1.571779 | 0.000000 |
| 1 | -2.079718 | -1.849571 | 0.000002 |
| 6 | 2.203289 | -0.760936 | -0.000001 |
| 1 | 3.263019 | -0.497391 | -0.000001 |
| 1 | 2.079718 | -1.849571 | -0.000002 |

------------------------------------------------------

The Cartesian coordinate of optimized 11 compound at HF/6-311++g(d,p).

---

| Atomic Number | Coordinates (Angstroms) | | |
|---|---|---|---|
| | X | Y | Z |
| 6 | -0.751815 | 1.205975 | 1.032473 |
| 6 | -0.751815 | -0.327039 | 0.790465 |
| 6 | 0.369633 | -1.372254 | 1.032472 |
| 6 | -0.177702 | -2.410300 | 0.000000 |
| 6 | 0.247392 | 1.770607 | 0.000000 |
| 6 | -0.751815 | 1.205975 | -1.032473 |
| 6 | -0.751815 | -0.327039 | -0.790465 |
| 6 | 0.369633 | -1.372254 | -1.032472 |
| 6 | 1.463942 | -1.026246 | 0.000000 |
| 1 | -0.787031 | 1.571315 | 2.047201 |
| 1 | -1.661870 | -0.722887 | 1.224749 |
| 1 | 0.612878 | -1.647113 | 2.047199 |
| 17 | 0.447095 | -4.077856 | 0.000000 |
| 17 | 0.490345 | 3.546695 | 0.000000 |
| 1 | 1.249545 | 1.419017 | 0.000000 |
| 1 | -0.787031 | 1.571315 | -2.047201 |
| 1 | -1.661870 | -0.722887 | -1.224749 |
| 1 | 0.612878 | -1.647113 | -2.047199 |
| 1 | 1.889982 | -0.053405 | 0.000000 |
| 17 | 2.928907 | -2.059389 | 0.000000 |
| 1 | -1.246709 | -2.540594 | 0.000000 |
| 6 | -1.884342 | 1.513381 | 0.000000 |
| 17 | -2.678111 | 3.107447 | 0.000000 |
| 1 | -2.708553 | 0.820252 | 0.000000 |

---

The Cartesian coordinate of optimized 11 compound at MP2/6-31G(d).

```
----------------------------------------------------------
Atomic          Coordinates (Angstroms)
Number        X          Y          Z
----------------------------------------------------------
  6       -1.399245   -0.200368    1.035204
  6       -0.000057   -0.813359    0.795802
  6        1.399152   -0.200419    1.035208
  6        2.135328   -1.119268    0.000000
  6       -1.525447    0.939453    0.000000
  6       -1.399245   -0.200368   -1.035204
  6       -0.000057   -0.813359   -0.795802
  6        1.399152   -0.200419   -1.035208
  6        1.525477    0.939381    0.000000
  1       -1.754392   -0.086124    2.061882
  1       -0.000079   -1.820548    1.231819
  1        1.754287   -0.086185    2.061893
 17        3.905201   -1.201865    0.000000
 17       -3.065769    1.836863    0.000000
  1       -0.790874    1.727973    0.000000
  1       -1.754392   -0.086124   -2.061882
  1       -0.000079   -1.820548   -1.231819
  1        1.754287   -0.086186   -2.061893
  1        0.791004    1.727992    0.000000
 17        3.065939    1.836550    0.000000
  1        1.808600   -2.162176    0.000000
  6       -2.135379   -1.119252    0.000000
 17       -3.905246   -1.202022    0.000000
  1       -1.808577   -2.162137    0.000000
----------------------------------------------------------
```

The Cartesian coordinate of optimized 11 compound at MP2/6-311++g(d,p).

```
------------------------------------------------------------
Atomic          Coordinates (Angstroms)
Number          X         Y         Z
------------------------------------------------------------
   6        -0.750274   1.197845   1.038658
   6        -0.750274  -0.331183   0.798977
   6         0.379205  -1.361808   1.038638
   6        -0.162797  -2.408509   0.000000
   6         0.245534   1.768582   0.000000
   6        -0.750274   1.197845  -1.038658
   6        -0.750274  -0.331183  -0.798977
   6         0.379205  -1.361808  -1.038638
   6         1.471996  -1.010745   0.000000
   1        -0.789886   1.573562   2.062289
   1        -1.673398  -0.738539   1.228941
   1         0.630037  -1.644344   2.062266
  17         0.478754  -4.057430   0.000000
  17         0.450652   3.537156   0.000000
   1         1.261176   1.408021   0.000000
   1        -0.789886   1.573562  -2.062289
   1        -1.673398  -0.738539  -1.228941
   1         0.630037  -1.644344  -2.062266
   1         1.890015  -0.017369   0.000000
  17         2.917021  -2.050859   0.000000
   1        -1.249086  -2.528554   0.000000
   6        -1.888737   1.503232   0.000000
  17        -2.673820   3.088840   0.000000
   1        -2.709801   0.781923   0.000000
------------------------------------------------------------
```

The Cartesian coordinate of optimized 11 compound at B3LYP/6-311++g(d,p).
------------------------------------------------------------

| Atomic Number | Coordinates (Angstroms) | | |
|---|---|---|---|
| | X | Y | Z |

------------------------------------------------------------

| | | | |
|---|---|---|---|
| 6 | -1.407628 | -0.214448 | 1.040218 |
| 6 | 0.000072 | -0.832823 | 0.800451 |
| 6 | 1.407743 | -0.214381 | 1.040212 |
| 6 | 2.152410 | -1.133522 | 0.000000 |
| 6 | -1.523769 | 0.931586 | 0.000000 |
| 6 | -1.407628 | -0.214448 | -1.040218 |
| 6 | 0.000072 | -0.832823 | -0.800451 |
| 6 | 1.407743 | -0.214382 | -1.040213 |
| 6 | 1.523733 | 0.931677 | 0.000000 |
| 1 | -1.760805 | -0.098007 | 2.061836 |
| 1 | 0.000100 | -1.834969 | 1.235324 |
| 1 | 1.760930 | -0.097924 | 2.061825 |
| 17 | 3.947501 | -1.210892 | 0.000000 |
| 17 | -3.071382 | 1.873623 | 0.000001 |
| 1 | -0.792509 | 1.718798 | 0.000001 |
| 1 | -1.760805 | -0.098006 | -2.061836 |
| 1 | 0.000099 | -1.834969 | -1.235324 |
| 1 | 1.760930 | -0.097925 | -2.061825 |
| 1 | 0.792349 | 1.718775 | -0.000001 |
| 17 | 3.071160 | 1.874022 | -0.000001 |
| 1 | 1.849243 | -2.177973 | 0.000001 |
| 6 | -2.152340 | -1.133547 | 0.000000 |
| 17 | -3.947439 | -1.210701 | 0.000000 |
| 1 | -1.849260 | -2.178022 | -0.000001 |

------------------------------------------------------------

The Cartesian coordinate of optimized 11 compound at M06L/6-311++g(d,p).

---

| Atomic Number | Coordinates (Angstroms) | | |
|---|---|---|---|
| | X | Y | Z |
---

| | | | |
|---|---|---|---|
| 6 | -1.394760 | -0.204180 | 1.029736 |
| 6 | -0.000025 | -0.816082 | 0.791135 |
| 6 | 1.394720 | -0.204202 | 1.029738 |
| 6 | 2.128787 | -1.119003 | 0.000000 |
| 6 | -1.515854 | 0.933355 | 0.000000 |
| 6 | -1.394760 | -0.204180 | -1.029736 |
| 6 | -0.000025 | -0.816082 | -0.791135 |
| 6 | 1.394720 | -0.204202 | -1.029738 |
| 6 | 1.515864 | 0.933323 | 0.000000 |
| 1 | -1.750562 | -0.089372 | 2.052920 |
| 1 | -0.000035 | -1.823616 | 1.219028 |
| 1 | 1.750517 | -0.089398 | 2.052925 |
| 17 | 3.904196 | -1.202100 | 0.000000 |
| 17 | -3.059853 | 1.843414 | 0.000000 |
| 1 | -0.783674 | 1.724622 | 0.000000 |
| 1 | -1.750562 | -0.089372 | -2.052920 |
| 1 | -0.000035 | -1.823616 | -1.219028 |
| 1 | 1.750517 | -0.089398 | -2.052925 |
| 1 | 0.783731 | 1.724632 | 0.000000 |
| 17 | 3.059928 | 1.843275 | 0.000000 |
| 1 | 1.806580 | -2.162060 | 0.000000 |
| 6 | -2.128808 | -1.118998 | 0.000000 |
| 17 | -3.904215 | -1.202169 | 0.000000 |
| 1 | -1.806568 | -2.162045 | 0.000000 |

---

The Cartesian coordinate of optimized 12 compound at HF/6-311++g(d,p).
------------------------------------------------------------
| Atomic | Coordinates (Angstroms) | | |
| Number | X | Y | Z |
------------------------------------------------------------
| 6  | -1.402530 |  0.223654 |  1.031251 |
| 6  |  0.000013 | -0.416134 |  0.800241 |
| 6  |  1.402547 |  0.223675 |  1.031254 |
| 6  |  2.174805 | -0.662371 | -0.000005 |
| 6  | -1.493218 |  1.369653 | -0.000008 |
| 6  | -1.402530 |  0.223641 | -1.031256 |
| 6  |  0.000019 | -0.416134 | -0.800241 |
| 6  |  1.402547 |  0.223688 | -1.031250 |
| 6  |  1.493181 |  1.369693 |  0.000008 |
| 1  | -1.720839 |  0.342658 |  2.053919 |
| 17 |  0.000023 | -1.988701 |  1.684273 |
| 1  |  1.720859 |  0.342674 |  2.053921 |
| 17 |  3.950168 | -0.672510 | -0.000006 |
| 17 | -2.999319 |  2.334492 | -0.000017 |
| 1  | -0.765927 |  2.137844 | -0.000009 |
| 1  | -1.720841 |  0.342633 | -2.053924 |
| 17 |  0.000056 | -1.988701 | -1.684273 |
| 1  |  1.720857 |  0.342698 | -2.053916 |
| 1  |  0.765847 |  2.137845 |  0.000009 |
| 17 |  2.999218 |  2.334633 |  0.000017 |
| 1  |  1.922384 | -1.701218 | -0.000013 |
| 6  | -2.174794 | -0.662398 |  0.000005 |
| 17 | -3.950156 | -0.672490 |  0.000006 |
| 1  | -1.922390 | -1.701249 |  0.000013 |
------------------------------------------------------------

The Cartesian coordinate of optimized 12 compound at MP2/6-31G(d).

```
--------------------------------------------------------------
Atomic         Coordinates (Angstroms)
Number          X         Y         Z
--------------------------------------------------------------
   6        1.395550   0.233790  -1.034306
   6        0.000012  -0.402637  -0.806478
   6       -1.395547   0.233741  -1.034319
   6       -2.165549  -0.659963   0.000013
   6        1.496615   1.377179   0.000024
   6        1.395550   0.233754   1.034319
   6       -0.000003  -0.402637   0.806478
   6       -1.395548   0.233776   1.034307
   6       -1.496630   1.377164  -0.000024
   1        1.722459   0.352157  -2.069445
  17        0.000060  -1.977888  -1.668387
   1       -1.722462   0.352072  -2.069460
  17       -3.933631  -0.671098   0.000017
  17        3.016962   2.305169   0.000043
   1        0.759296   2.158413   0.000031
   1        1.722464   0.352088   2.069459
  17       -0.000026  -1.977888   1.668387
   1       -1.722457   0.352140   2.069445
   1       -0.759322   2.158409  -0.000031
  17       -3.016995   2.305123  -0.000043
   1       -1.890024  -1.709958   0.000036
   6        2.165550  -0.659951  -0.000013
  17        3.933632  -0.671104  -0.000017
   1        1.890017  -1.709944  -0.000036
--------------------------------------------------------------
```

The Cartesian coordinate of optimized 12 compound at MP2/6-311++g(d,p).
----------------------------------------------------------------
Atomic        Coordinates (Angstroms)
Number          X          Y          Z
----------------------------------------------------------------
  6       -1.393258   0.237153   1.037313
  6       -0.000657  -0.402436   0.808141
  6        1.393746   0.236661   1.037909
  6        2.163363  -0.658932   0.000269
  6       -1.491250   1.381822  -0.000735
  6       -1.393897   0.236617  -1.038112
  6       -0.000268  -0.402997  -0.808626
  6        1.393816   0.236506  -1.037623
  6        1.491737   1.381463   0.000126
  1       -1.723324   0.351823   2.070474
 17       -0.000753  -1.978183   1.667112
  1        1.723398   0.350834   2.071238
 17        3.928902  -0.677195   0.000160
 17       -3.010196   2.307883  -0.000156
  1       -0.749512   2.159449  -0.001327
  1       -1.724036   0.350633  -2.071312
 17       -0.000079  -1.978849  -1.666913
  1        1.723918   0.350827  -2.070804
  1        0.750210   2.159310   0.000166
 17        3.010898   2.307238  -0.000018
  1        1.881772  -1.706808   0.000269
  6       -2.163303  -0.658551  -0.000055
 17       -3.928825  -0.676981   0.000381
  1       -1.881703  -1.706429   0.000049
----------------------------------------------------------------

The Cartesian coordinate of optimized 12 compound at B3LYP/6-311++g(d,p).
----------------------------------------------------------
| Atomic | Coordinates (Angstroms) | | |
| Number | X | Y | Z |
----------------------------------------------------------
| 6  | -1.405499 | 0.225978  | 1.039908  |
| 6  | -0.000035 | -0.412400 | 0.811629  |
| 6  | 1.405462  | 0.225909  | 1.039910  |
| 6  | 2.184561  | -0.666764 | 0.000002  |
| 6  | -1.496162 | 1.376662  | 0.000004  |
| 6  | -1.405498 | 0.225983  | -1.039906 |
| 6  | -0.000036 | -0.412400 | -0.811629 |
| 6  | 1.405464  | 0.225904  | -1.039911 |
| 6  | 1.496245  | 1.376574  | -0.000004 |
| 1  | -1.730136 | 0.347422  | 2.069199  |
| 17 | -0.000084 | -2.012378 | 1.691795  |
| 1  | 1.730095  | 0.347342  | 2.069203  |
| 17 | 3.975040  | -0.668786 | 0.000002  |
| 17 | -3.019363 | 2.350272  | 0.000005  |
| 1  | -0.759375 | 2.154033  | 0.000007  |
| 1  | -1.730134 | 0.347432  | -2.069197 |
| 17 | -0.000093 | -2.012378 | -1.691795 |
| 1  | 1.730097  | 0.347332  | -2.069205 |
| 1  | 0.759553  | 2.154034  | -0.000007 |
| 17 | 3.019589  | 2.349959  | -0.000005 |
| 1  | 1.928435  | -1.716883 | 0.000005  |
| 6  | -2.184586 | -0.666703 | -0.000002 |
| 17 | -3.975066 | -0.668828 | -0.000002 |
| 1  | -1.928422 | -1.716813 | -0.000005 |
----------------------------------------------------------

The Cartesian coordinate of optimized 12 compound at M06L/6-311++g(d,p).
------------------------------------------------------------

| Atomic Number | Coordinates (Angstroms) | | |
|---|---|---|---|
| | X | Y | Z |

------------------------------------------------------------

| | | | |
|---|---|---|---|
| 6 | 1.393585 | 0.233682 | -1.029894 |
| 6 | 0.000006 | -0.397827 | -0.801240 |
| 6 | -1.393582 | 0.233661 | -1.029898 |
| 6 | -2.162636 | -0.655679 | 0.000005 |
| 6 | 1.489678 | 1.375806 | 0.000009 |
| 6 | 1.393586 | 0.233669 | 1.029898 |
| 6 | 0.000001 | -0.397827 | 0.801240 |
| 6 | -1.393582 | 0.233674 | 1.029895 |
| 6 | -1.489687 | 1.375796 | -0.000009 |
| 1 | 1.719575 | 0.352503 | -2.061811 |
| 17 | 0.000026 | -1.996083 | -1.649886 |
| 1 | -1.719574 | 0.352468 | -2.061817 |
| 17 | -3.933677 | -0.669391 | 0.000006 |
| 17 | 3.009818 | 2.318708 | 0.000016 |
| 1 | 0.752039 | 2.157525 | 0.000011 |
| 1 | 1.719578 | 0.352479 | 2.061816 |
| 17 | -0.000005 | -1.996082 | 1.649886 |
| 1 | -1.719571 | 0.352493 | 2.061812 |
| 1 | -0.752058 | 2.157524 | -0.000011 |
| 17 | -3.009842 | 2.318673 | -0.000016 |
| 1 | -1.885502 | -1.705001 | 0.000013 |
| 6 | 2.162638 | -0.655672 | -0.000005 |
| 17 | 3.933679 | -0.669396 | -0.000007 |
| 1 | 1.885499 | -1.704993 | -0.000013 |

------------------------------------------------------------

The Cartesian coordinate of optimized 13 compound at HF/6-311++g(d,p).
---------------------------------------------------------

| Atomic Number | Coordinates (Angstroms) | | |
|---|---|---|---|
| | X | Y | Z |

---------------------------------------------------------

| | | | |
|---|---|---|---|
| 6  | 1.580317  | 0.257925  | -1.029087 |
| 6  | 0.134035  | -0.321071 | -0.804416 |
| 6  | -1.226774 | 0.414839  | -1.035326 |
| 6  | -2.171470 | -0.296250 | -0.000001 |
| 6  | 1.801136  | 1.389423  | -0.000025 |
| 6  | 1.580319  | 0.257960  | 1.029075  |
| 6  | 0.134044  | -0.321063 | 0.804422  |
| 6  | -1.226775 | 0.414826  | 1.035337  |
| 6  | -1.136463 | 1.563436  | 0.000013  |
| 1  | 1.900520  | 0.346929  | -2.053858 |
| 17 | 0.126245  | -1.839068 | -1.769763 |
| 1  | -1.532950 | 0.569078  | -2.056693 |
| 17 | -3.841753 | 0.342540  | 0.000001  |
| 17 | 3.427439  | 2.148771  | -0.000040 |
| 1  | 1.211625  | 2.270141  | -0.000038 |
| 1  | 1.900523  | 0.346998  | 2.053843  |
| 17 | 0.126301  | -1.839049 | 1.769786  |
| 1  | -1.532954 | 0.569052  | 2.056705  |
| 1  | -0.230375 | 2.086640  | 0.000015  |
| 17 | -2.273707 | 2.936801  | 0.000022  |
| 17 | -2.435438 | -2.052558 | -0.000021 |
| 6  | 2.277921  | -0.695122 | 0.000011  |
| 17 | 4.040396  | -0.901902 | 0.000015  |
| 1  | 1.924669  | -1.702341 | 0.000031  |

---------------------------------------------------------

The Cartesian coordinate of optimized 13 compound at MP2/6-31G(d).

---

| Atomic Number | Coordinates (Angstroms) | | |
|---|---|---|---|
| | X | Y | Z |
| 6 | 1.564744 | 0.279149 | -1.031867 |
| 6 | 0.129340 | -0.308110 | -0.810782 |
| 6 | -1.229202 | 0.412991 | -1.039108 |
| 6 | -2.166403 | -0.307239 | -0.000002 |
| 6 | 1.796418 | 1.406461 | -0.000025 |
| 6 | 1.564746 | 0.279182 | 1.031854 |
| 6 | 0.129348 | -0.308101 | 0.810791 |
| 6 | -1.229205 | 0.412979 | 1.039119 |
| 6 | -1.147731 | 1.561184 | 0.000013 |
| 1 | 1.892467 | 0.366959 | -2.069651 |
| 17 | 0.133329 | -1.828548 | -1.759688 |
| 1 | -1.543697 | 0.567744 | -2.073404 |
| 17 | -3.832137 | 0.330090 | -0.000005 |
| 17 | 3.432419 | 2.126733 | -0.000038 |
| 1 | 1.200994 | 2.304032 | -0.000039 |
| 1 | 1.892472 | 0.367025 | 2.069634 |
| 17 | 0.133380 | -1.828523 | 1.759720 |
| 1 | -1.543704 | 0.567720 | 2.073416 |
| 1 | -0.224315 | 2.087602 | 0.000018 |
| 17 | -2.298965 | 2.914293 | 0.000021 |
| 17 | -2.386866 | -2.063622 | -0.000021 |
| 6 | 2.262798 | -0.680852 | 0.000008 |
| 17 | 4.018206 | -0.888622 | 0.000010 |
| 1 | 1.887437 | -1.697579 | 0.000025 |

---

The Cartesian coordinate of optimized 13 compound at MP2/6-311++g(d,p).
---------------------------------------------------------------

| Atomic Number | Coordinates (Angstroms) | | |
|---|---|---|---|
| | X | Y | Z |

---------------------------------------------------------------

| | | | |
|---|---|---|---|
| 6 | 1.556878 | 0.284458 | -1.034221 |
| 6 | 0.125849 | -0.312011 | -0.811042 |
| 6 | -1.232481 | 0.407767 | -1.043565 |
| 6 | -2.169173 | -0.308826 | -0.000156 |
| 6 | 1.787156 | 1.413264 | 0.002168 |
| 6 | 1.557840 | 0.282735 | 1.035785 |
| 6 | 0.125123 | -0.311252 | 0.812700 |
| 6 | -1.232412 | 0.410087 | 1.042051 |
| 6 | -1.143841 | 1.558599 | -0.002032 |
| 1 | 1.886524 | 0.368442 | -2.070436 |
| 17 | 0.141339 | -1.832685 | -1.758957 |
| 1 | -1.550951 | 0.557334 | -2.076027 |
| 17 | -3.830291 | 0.336148 | 0.000142 |
| 17 | 3.423261 | 2.129053 | 0.001020 |
| 1 | 1.190967 | 2.310800 | 0.005271 |
| 1 | 1.888097 | 0.363881 | 2.071996 |
| 17 | 0.139734 | -1.832272 | 1.759925 |
| 1 | -1.551733 | 0.561826 | 2.073972 |
| 1 | -0.209879 | 2.068444 | -0.003818 |
| 17 | -2.276660 | 2.923232 | -0.001030 |
| 17 | -2.389394 | -2.066166 | 0.000123 |
| 6 | 2.256778 | -0.677243 | -0.000232 |
| 17 | 4.008553 | -0.894073 | -0.001737 |
| 1 | 1.875465 | -1.691230 | -0.000939 |

---------------------------------------------------------------

The Cartesian coordinate of optimized 13 compound at B3LYP/6-311++g(d,p).
----------------------------------------------------------
| Atomic | Coordinates (Angstroms) | | |
| Number | X | Y | Z |
----------------------------------------------------------
| 6  | -1.584221 |  0.261866 |  1.037533 |
| 6  | -0.135010 | -0.316008 |  0.816544 |
| 6  |  1.228710 |  0.418393 |  1.045397 |
| 6  |  2.178845 | -0.298659 | -0.000004 |
| 6  | -1.804203 |  1.396214 | -0.000058 |
| 6  | -1.584217 |  0.261789 | -1.037561 |
| 6  | -0.134990 | -0.316028 | -0.816525 |
| 6  |  1.228702 |  0.418424 | -1.045369 |
| 6  |  1.141282 |  1.572493 |  0.000033 |
| 1  | -1.910912 |  0.352545 |  2.069141 |
| 17 | -0.125497 | -1.857616 |  1.783459 |
| 1  |  1.541257 |  0.575298 |  2.073371 |
| 17 |  3.867689 |  0.348225 | -0.000005 |
| 17 | -3.448698 |  2.166691 | -0.000089 |
| 1  | -1.205694 |  2.286557 | -0.000091 |
| 1  | -1.910902 |  0.352390 | -2.069178 |
| 17 | -0.125375 | -1.857667 | -1.783391 |
| 1  |  1.541242 |  0.575360 | -2.073341 |
| 1  |  0.225854 |  2.109705 |  0.000045 |
| 17 |  2.299283 |  2.953136 |  0.000048 |
| 17 |  2.432257 | -2.074559 | -0.000052 |
| 6  | -2.287892 | -0.699450 |  0.000021 |
| 17 | -4.066314 | -0.897504 |  0.000026 |
| 1  | -1.929758 | -1.718058 |  0.000062 |
----------------------------------------------------------

The Cartesian coordinate of optimized 13 compound at M06L/6-311++g(d,p).
------------------------------------------------------------

| Atomic Number | Coordinates (Angstroms) | | |
|---|---|---|---|
| | X | Y | Z |
| 6 | 1.565791 | 0.273756 | -1.027550 |
| 6 | 0.131597 | -0.307796 | -0.805108 |
| 6 | -1.224494 | 0.411664 | -1.035715 |
| 6 | -2.163489 | -0.298622 | -0.000038 |
| 6 | 1.792373 | 1.399926 | -0.000598 |
| 6 | 1.565837 | 0.274564 | 1.027253 |
| 6 | 0.131817 | -0.307585 | 0.805291 |
| 6 | -1.224534 | 0.411357 | 1.035986 |
| 6 | -1.140351 | 1.558729 | 0.000311 |
| 1 | 1.891665 | 0.360741 | -2.062662 |
| 17 | 0.134328 | -1.848431 | -1.742334 |
| 1 | -1.537733 | 0.566156 | -2.066581 |
| 17 | -3.834370 | 0.350898 | -0.000015 |
| 17 | 3.431304 | 2.137523 | -0.000935 |
| 1 | 1.198418 | 2.297891 | -0.000914 |
| 1 | 1.891745 | 0.362362 | 2.062284 |
| 17 | 0.135642 | -1.847929 | 1.742975 |
| 1 | -1.537849 | 0.565549 | 2.066875 |
| 1 | -0.215882 | 2.086222 | 0.000378 |
| 17 | -2.284236 | 2.924908 | 0.000502 |
| 17 | -2.415029 | -2.061245 | -0.000525 |
| 6 | 2.264515 | -0.680727 | 0.000239 |
| 17 | 4.022114 | -0.888273 | 0.000301 |
| 1 | 1.889476 | -1.697169 | 0.000696 |

------------------------------------------------------------

The Cartesian coordinate of optimized 14 compound at HF/6-311++g(d,p).
-------------------------------------------------------------

| Atomic | Coordinates (Angstroms) | | |
|---|---|---|---|
| Number | X | Y | Z |

-------------------------------------------------------------

| | | | |
|---|---|---|---|
| 6 | -1.415521 | 0.457309 | 1.031076 |
| 6 | 0.000016 | -0.206861 | 0.816264 |
| 6 | 1.415681 | 0.457035 | 1.031236 |
| 6 | 2.284991 | -0.360861 | -0.000046 |
| 6 | -1.476642 | 1.608199 | -0.000295 |
| 6 | -1.415687 | 0.457041 | -1.031235 |
| 6 | -0.000025 | -0.206865 | -0.816256 |
| 6 | 1.415511 | 0.457300 | -1.031074 |
| 6 | 1.476664 | 1.608189 | 0.000294 |
| 1 | -1.729324 | 0.578848 | 2.053991 |
| 17 | -0.000096 | -1.616382 | 1.934009 |
| 1 | 1.729503 | 0.578230 | 2.054185 |
| 17 | 4.022177 | 0.060046 | -0.000178 |
| 17 | -2.812447 | 2.800721 | -0.000260 |
| 1 | -0.684851 | 2.291224 | -0.000706 |
| 1 | -1.729504 | 0.578240 | -2.054185 |
| 17 | 0.000102 | -1.616398 | -1.933987 |
| 1 | 1.729314 | 0.578831 | -2.053991 |
| 1 | 0.684895 | 2.291238 | 0.000706 |
| 17 | 2.812518 | 2.800656 | 0.000254 |
| 17 | 2.340919 | -2.137141 | -0.000077 |
| 6 | -2.285010 | -0.360846 | 0.000045 |
| 17 | -4.022194 | 0.060066 | 0.000169 |
| 17 | -2.340974 | -2.137124 | 0.000067 |

-------------------------------------------------------------

The Cartesian coordinate of optimized 14 compound at MP2/6-31G(d).

```
------------------------------------------------------------
Atomic          Coordinates (Angstroms)
Number          X           Y           Z
------------------------------------------------------------
  6         1.409304    0.467020   -1.034178
  6         0.000134   -0.190952   -0.827053
  6        -1.409159    0.466726   -1.034197
  6        -2.267060   -0.366073   -0.000016
  6         1.484098    1.614815    0.000007
  6         1.409452    0.467089    1.034228
  6         0.000185   -0.190769    0.827326
  6        -1.409167    0.466841    1.034217
  6        -1.484483    1.614474   -0.000075
  1         1.732483    0.587853   -2.069993
 17         0.000151   -1.586660   -1.955754
  1        -1.732355    0.587399   -2.070025
 17        -4.002401    0.039262   -0.000085
 17         2.838656    2.775973   -0.000161
  1         0.679102    2.309954    0.000131
  1         1.732749    0.587957    2.070002
 17         0.000251   -1.586018    1.956594
  1        -1.732472    0.587645    2.069997
  1        -0.679928    2.310112   -0.000209
 17        -2.840001    2.774502   -0.000105
 17        -2.261026   -2.136271   -0.000256
  6         2.267351   -0.365783   -0.000022
 17         4.002761    0.039229   -0.000232
 17         2.261403   -2.135973   -0.000079
------------------------------------------------------------
```

The Cartesian coordinate of optimized 14 compound at MP2/6-311++g(d,p).

```
-----------------------------------------------------------
Atomic          Coordinates (Angstroms)
Number          X          Y          Z
-----------------------------------------------------------
   6        1.409604   0.465251  -1.038649
   6        0.002479  -0.192042  -0.831653
   6       -1.404834   0.468093  -1.036362
   6       -2.259835  -0.370421  -0.001014
   6        1.479709   1.616059  -0.004075
   6        1.406171   0.468802   1.036297
   6       -0.000704  -0.192546   0.833337
   6       -1.409769   0.464563   1.039202
   6       -1.483060   1.615240   0.004673
   1        1.737933   0.577788  -2.072461
  17       -0.000903  -1.569145  -1.985690
   1       -1.734232   0.585270  -2.069325
  17       -3.994618   0.025840  -0.003562
  17        2.827219   2.780603  -0.003339
   1        0.667212   2.303452  -0.010706
   1        1.736635   0.586311   2.068926
  17        0.002756  -1.570537   1.985927
   1       -1.739339   0.575910   2.072799
   1       -0.672923   2.305498   0.012072
  17       -2.834730   2.774985   0.004088
  17       -2.232652  -2.141553  -0.000993
   6        2.261647  -0.368493   0.000540
  17        3.996153   0.028832   0.002062
  17        2.236555  -2.139688   0.000620
-----------------------------------------------------------
```

The Cartesian coordinate of optimized 14 compound at B3LYP/6-311++g(d,p).
------------------------------------------------------------

| Atomic Number | Coordinates (Angstroms) | | |
|---|---|---|---|
| | X | Y | Z |

------------------------------------------------------------

| | | | |
|---|---|---|---|
| 6  |  1.418351 |  0.463418 | -1.041035 |
| 6  |  0.000026 | -0.198204 | -0.832134 |
| 6  | -1.418311 |  0.463341 | -1.041073 |
| 6  | -2.291152 | -0.364430 | -0.000001 |
| 6  |  1.482103 |  1.617139 |  0.000048 |
| 6  |  1.418389 |  0.463394 |  1.041101 |
| 6  | -0.000003 | -0.198112 |  0.832277 |
| 6  | -1.418356 |  0.463455 |  1.041050 |
| 6  | -1.482257 |  1.617104 | -0.000076 |
| 1  |  1.739969 |  0.586789 | -2.070332 |
| 17 | -0.000023 | -1.622847 | -1.963814 |
| 1  | -1.739949 |  0.586591 | -2.070378 |
| 17 | -4.046967 |  0.059301 | -0.000090 |
| 17 |  2.838187 |  2.817076 |  0.000038 |
| 1  |  0.680764 |  2.312536 |  0.000067 |
| 1  |  1.740063 |  0.586736 |  2.070383 |
| 17 | -0.000183 | -1.622523 |  1.964245 |
| 1  | -1.740055 |  0.586814 |  2.070322 |
| 1  | -0.681057 |  2.312654 | -0.000100 |
| 17 | -2.838667 |  2.816671 | -0.000169 |
| 17 | -2.325450 | -2.158020 |  0.000000 |
| 6  |  2.291303 | -0.364338 | -0.000018 |
| 17 |  4.047086 |  0.059510 | -0.000080 |
| 17 |  2.326002 | -2.157917 | -0.000178 |

------------------------------------------------------------

The Cartesian coordinate of optimized 14 compound at M06L/6-311++g(d,p).

---

| Atomic Number | Coordinates (Angstroms) | | |
|---|---|---|---|
| | X | Y | Z |

---

| | | | |
|---|---|---|---|
| 6 | 1.407048 | 0.463618 | -1.030593 |
| 6 | 0.000098 | -0.191732 | -0.822312 |
| 6 | -1.406886 | 0.463503 | -1.030540 |
| 6 | -2.267061 | -0.361017 | -0.000031 |
| 6 | 1.479544 | 1.610551 | -0.000104 |
| 6 | 1.407149 | 0.463801 | 1.030586 |
| 6 | 0.000135 | -0.191517 | 0.822630 |
| 6 | -1.406969 | 0.463533 | 1.030632 |
| 6 | -1.479932 | 1.610292 | 0.000030 |
| 1 | 1.728031 | 0.582573 | -2.063680 |
| 17 | -0.000036 | -1.597980 | -1.953384 |
| 1 | -1.727900 | 0.582417 | -2.063623 |
| 17 | -4.008333 | 0.049583 | -0.000179 |
| 17 | 2.830530 | 2.784904 | -0.000273 |
| 1 | 0.673699 | 2.305926 | -0.000125 |
| 1 | 1.728266 | 0.582927 | 2.063613 |
| 17 | 0.000152 | -1.597224 | 1.954365 |
| 1 | -1.728111 | 0.582468 | 2.063674 |
| 1 | -0.674505 | 2.306136 | 0.000037 |
| 17 | -2.831841 | 2.783574 | -0.000034 |
| 17 | -2.279508 | -2.141271 | -0.000303 |
| 6 | 2.267380 | -0.360742 | -0.000012 |
| 17 | 4.008673 | 0.049743 | -0.000200 |
| 17 | 2.280216 | -2.140987 | -0.000087 |

---

The Cartesian coordinate of optimized 15 compound at HF/6-311++g(d,p).

---

| Atomic Number | Coordinates (Angstroms) | | |
|---|---|---|---|
| | X | Y | Z |

---

| | | | |
|---|---|---|---|
| 6 | 1.365059 | 0.501300 | -0.988237 |
| 6 | -0.010505 | -0.239249 | -0.813968 |
| 6 | -1.449097 | 0.376738 | -1.068011 |
| 6 | -2.290559 | -0.357062 | 0.017996 |
| 6 | 1.440526 | 1.591483 | 0.144799 |
| 6 | 1.449120 | 0.376762 | 1.068020 |
| 6 | 0.010511 | -0.239214 | 0.814021 |
| 6 | -1.365066 | 0.501312 | 0.988243 |
| 6 | -1.440564 | 1.591471 | -0.144808 |
| 1 | 1.674735 | 0.710333 | -1.998724 |
| 17 | 0.048819 | -1.653816 | -1.918912 |
| 1 | -1.761881 | 0.419587 | -2.097343 |
| 17 | -4.015262 | 0.105785 | 0.084627 |
| 17 | 2.738785 | 2.820999 | 0.118879 |
| 1 | 0.630161 | 2.231189 | 0.333422 |
| 1 | 1.761921 | 0.419639 | 2.097346 |
| 17 | -0.048859 | -1.653697 | 1.919070 |
| 1 | -1.674771 | 0.710349 | 1.998721 |
| 1 | -0.630227 | 2.231208 | -0.333436 |
| 17 | -2.738902 | 2.820903 | -0.118923 |
| 17 | -2.382044 | -2.131884 | 0.023907 |
| 6 | 2.290593 | -0.357036 | -0.018002 |
| 17 | 4.015285 | 0.105836 | -0.084696 |
| 17 | 2.382175 | -2.131851 | -0.023969 |

---

The Cartesian coordinate of optimized 15 compound at MP2/6-31G(d).
------------------------------------------------------------

| Atomic Number | Coordinates (Angstroms) | | |
|---|---|---|---|
| | X | Y | Z |

------------------------------------------------------------

| | | | |
|---|---|---|---|
| 6 | -1.350613 | 0.516323 | 0.982743 |
| 6 | 0.011679 | -0.229441 | 0.823913 |
| 6 | 1.446515 | 0.376078 | 1.078094 |
| 6 | 2.274514 | -0.360473 | -0.014162 |
| 6 | -1.441184 | 1.596199 | -0.161373 |
| 6 | -1.446488 | 0.376058 | -1.078083 |
| 6 | -0.011687 | -0.229479 | -0.823859 |
| 6 | 1.350613 | 0.516289 | -0.982746 |
| 6 | 1.441079 | 1.596209 | 0.161353 |
| 1 | -1.671335 | 0.736144 | 2.003858 |
| 17 | -0.060687 | -1.630404 | 1.937273 |
| 1 | 1.765360 | 0.409681 | 2.121344 |
| 17 | 3.995250 | 0.093179 | -0.095285 |
| 17 | -2.755909 | 2.798729 | -0.122180 |
| 1 | -0.617286 | 2.242474 | -0.368370 |
| 1 | -1.765331 | 0.409601 | -2.121336 |
| 17 | 0.060478 | -1.630534 | -1.937110 |
| 1 | 1.671292 | 0.736125 | -2.003871 |
| 1 | 0.617112 | 2.242408 | 0.368335 |
| 17 | 2.755611 | 2.798959 | 0.122135 |
| 17 | 2.311469 | -2.130838 | -0.013629 |
| 6 | -2.274407 | -0.360536 | 0.014160 |
| 17 | -3.995183 | 0.093010 | 0.095210 |
| 17 | -2.311026 | -2.130912 | 0.013575 |

------------------------------------------------------------

The Cartesian coordinate of optimized 15 compound at MP2/6-311++g(d,p).
---------------------------------------------------------------

| Atomic Number | Coordinates (Angstroms) | | |
|---|---|---|---|
| | X | Y | Z |

---------------------------------------------------------------

| | | | |
|---|---|---|---|
| 6 | 1.342056 | 0.516948 | -0.978573 |
| 6 | -0.012032 | -0.242681 | -0.827502 |
| 6 | -1.446723 | 0.362502 | -1.086966 |
| 6 | -2.271850 | -0.361883 | 0.013499 |
| 6 | 1.425566 | 1.591641 | 0.179783 |
| 6 | 1.446633 | 0.362330 | 1.086791 |
| 6 | 0.011692 | -0.242191 | 0.827025 |
| 6 | -1.341688 | 0.517545 | 0.978426 |
| 6 | -1.425563 | 1.591797 | -0.180191 |
| 1 | 1.669825 | 0.742734 | -1.994950 |
| 17 | 0.069966 | -1.631331 | -1.955590 |
| 1 | -1.766794 | 0.383596 | -2.128783 |
| 17 | -3.987843 | 0.099520 | 0.106319 |
| 17 | 2.721432 | 2.809392 | 0.137116 |
| 1 | 0.588890 | 2.217365 | 0.406232 |
| 1 | 1.766270 | 0.383666 | 2.128734 |
| 17 | -0.070225 | -1.630545 | 1.955763 |
| 1 | -1.669210 | 0.743500 | 1.994826 |
| 1 | -0.589264 | 2.217908 | -0.407091 |
| 17 | -2.722117 | 2.808896 | -0.137503 |
| 17 | -2.306150 | -2.133710 | 0.012150 |
| 6 | 2.272150 | -0.361937 | -0.013314 |
| 17 | 3.987985 | 0.100074 | -0.105783 |
| 17 | 2.306884 | -2.133662 | -0.012052 |

---------------------------------------------------------------

The Cartesian coordinate of optimized 15 compound at B3LYP/6-311++g(d,p).
---------------------------------------------------------

| Atomic | Coordinates (Angstroms) | | |
|---|---|---|---|
| Number | X | Y | Z |

---------------------------------------------------------

| | | | |
|---|---|---|---|
| 6 | -1.386374 | 0.496939 | 1.012710 |
| 6 | 0.005953 | -0.212861 | 0.830957 |
| 6 | 1.443116 | 0.413571 | 1.066714 |
| 6 | 2.294204 | -0.362646 | -0.008623 |
| 6 | -1.466224 | 1.610022 | -0.095459 |
| 6 | -1.443015 | 0.413473 | -1.066676 |
| 6 | -0.005928 | -0.213016 | -0.830721 |
| 6 | 1.386339 | 0.496880 | -1.012682 |
| 6 | 1.466089 | 1.610069 | 0.095411 |
| 1 | -1.708050 | 0.676780 | 2.033926 |
| 17 | -0.035292 | -1.639432 | 1.956619 |
| 1 | 1.761999 | 0.484601 | 2.101626 |
| 17 | 4.044386 | 0.080252 | -0.057091 |
| 17 | -2.805823 | 2.826291 | -0.074640 |
| 1 | -0.656839 | 2.285842 | -0.224653 |
| 1 | -1.761820 | 0.484387 | -2.101620 |
| 17 | 0.035155 | -1.639970 | -1.955910 |
| 1 | 1.707896 | 0.676696 | -2.033938 |
| 1 | 0.656590 | 2.285760 | 0.224590 |
| 17 | 2.805395 | 2.826658 | 0.074429 |
| 17 | 2.345432 | -2.156071 | -0.010363 |
| 6 | -2.294078 | -0.362736 | 0.008595 |
| 17 | -4.044294 | 0.080078 | 0.056802 |
| 17 | -2.344973 | -2.156173 | 0.010078 |

---------------------------------------------------------

The Cartesian coordinate of optimized 15 compound at M06L/6-311++g(d,p).
------------------------------------------------------------

| Atomic | Coordinates (Angstroms) | | |
|---|---|---|---|
| Number | X | Y | Z |

------------------------------------------------------------

| | | | |
|---|---|---|---|
| 6  |  1.372667 |  0.500333 | -0.998131 |
| 6  | -0.006577 | -0.207342 | -0.821355 |
| 6  | -1.433027 |  0.409831 | -1.059695 |
| 6  | -2.270257 | -0.358956 |  0.008443 |
| 6  |  1.462336 |  1.602901 |  0.102416 |
| 6  |  1.433052 |  0.409866 |  1.059699 |
| 6  |  0.006589 | -0.207309 |  0.821396 |
| 6  | -1.372665 |  0.500343 |  0.998129 |
| 6  | -1.462397 |  1.602878 | -0.102436 |
| 1  |  1.695055 |  0.681506 | -2.022218 |
| 17 |  0.040940 | -1.614482 | -1.947768 |
| 1  | -1.750222 |  0.471892 | -2.098567 |
| 17 | -4.005702 |  0.070863 |  0.060736 |
| 17 |  2.796955 |  2.794185 |  0.078841 |
| 1  |  0.648835 |  2.278035 |  0.236418 |
| 1  |  1.750258 |  0.471967 |  2.098565 |
| 17 | -0.040992 | -1.614368 |  1.947902 |
| 1  | -1.695081 |  0.681516 |  2.022208 |
| 1  | -0.648947 |  2.278065 | -0.236451 |
| 17 | -2.797140 |  2.794021 | -0.078882 |
| 17 | -2.299667 | -2.139559 |  0.009217 |
| 6  |  2.270312 | -0.358918 | -0.008446 |
| 17 |  4.005746 |  0.070931 | -0.060793 |
| 17 |  2.299855 | -2.139517 | -0.009258 |

------------------------------------------------------------

The Cartesian coordinate of optimized 16 compound at HF/6-311++g(d,p).

---

| Atomic Number | Coordinates (Angstroms) | | |
|---|---|---|---|
| | X | Y | Z |
| 6 | 1.115866 | 1.064257 | 0.011201 |
| 6 | 0.000193 | 0.715159 | -0.901288 |
| 6 | -1.115273 | 1.064860 | 0.011205 |
| 6 | 1.280593 | -0.000352 | 1.110362 |
| 6 | 1.115278 | -1.064857 | 0.011210 |
| 6 | -0.000187 | -0.715160 | -0.901285 |
| 6 | -1.115863 | -1.064254 | 0.011204 |
| 6 | -2.242421 | 0.000501 | -0.730093 |
| 6 | -1.280593 | 0.000360 | 1.110359 |
| 1 | 0.000355 | 1.304419 | -1.811308 |
| 6 | 0.000655 | 2.294902 | 0.393154 |
| 1 | 2.319841 | -0.000630 | 1.426044 |
| 1 | 0.748347 | -0.000208 | 2.029484 |
| 1 | -0.000350 | -1.304423 | -1.811303 |
| 6 | -0.000672 | -2.294896 | 0.393162 |
| 1 | -3.242245 | 0.000783 | -0.328229 |
| 1 | -2.232533 | 0.000490 | -1.812754 |
| 1 | -0.748348 | 0.000219 | 2.029483 |
| 1 | -2.319841 | 0.000639 | 1.426038 |
| 6 | 2.242425 | -0.000518 | -0.730088 |
| 1 | 2.232541 | -0.000513 | -1.812750 |
| 1 | 3.242249 | -0.000798 | -0.328223 |
| 1 | 0.000749 | 2.627586 | 1.421641 |
| 1 | 0.000881 | 3.127761 | -0.292204 |
| 1 | -0.000895 | -3.127759 | -0.292191 |
| 1 | -0.000767 | -2.627575 | 1.421651 |

---

The Cartesian coordinate of optimized 16 compound at MP2/6-31G(d).
------------------------------------------------------------
Atomic        Coordinates (Angstroms)
Number        X          Y          Z
------------------------------------------------------------
  6      1.121727   1.068389   0.016368
  6      0.000196   0.719490  -0.905653
  6     -1.121123   1.068979   0.016376
  6      1.285347  -0.000342   1.113347
  6      1.121134  -1.068976   0.016380
  6     -0.000183  -0.719490  -0.905651
  6     -1.121717  -1.068389   0.016366
  6     -2.239870   0.000497  -0.739075
  6     -1.285343   0.000346   1.113341
  1      0.000354   1.317756  -1.824657
  6      0.000620   2.292491   0.395926
  1      2.337804  -0.000624   1.427514
  1      0.749364  -0.000190   2.046714
  1     -0.000341  -1.317758  -1.824655
  6     -0.000668  -2.292490   0.395930
  1     -3.255061   0.000784  -0.337792
  1     -2.208509   0.000485  -1.834192
  1     -0.749364   0.000195   2.046711
  1     -2.337801   0.000628   1.427505
  6      2.239885  -0.000505  -0.739064
  1      2.208532  -0.000495  -1.834180
  1      3.255074  -0.000790  -0.337774
  1      0.000718   2.626824   1.437812
  1      0.000846   3.134273  -0.299188
  1     -0.000883  -3.134273  -0.299182
  1     -0.000767  -2.626819   1.437817
------------------------------------------------------------

The Cartesian coordinate of optimized 16 compound at MP2/6-311++g(d,p).
--------------------------------------------------------------

| Atomic Number | Coordinates (Angstroms) | | |
|---|---|---|---|
| | X | Y | Z |

--------------------------------------------------------------

| | | | |
|---|---|---|---|
| 6 | 1.124666 | 1.072410 | 0.016370 |
| 6 | 0.000263 | 0.721195 | -0.912266 |
| 6 | -1.123875 | 1.073208 | 0.016366 |
| 6 | 1.284466 | -0.000467 | 1.115400 |
| 6 | 1.123896 | -1.073201 | 0.016367 |
| 6 | -0.000247 | -0.721190 | -0.912265 |
| 6 | -1.124658 | -1.072410 | 0.016363 |
| 6 | -2.246099 | 0.000661 | -0.740465 |
| 6 | -1.284488 | 0.000468 | 1.115382 |
| 1 | 0.000485 | 1.325709 | -1.825529 |
| 6 | 0.000858 | 2.298396 | 0.399445 |
| 1 | 2.338471 | -0.000825 | 1.425575 |
| 1 | 0.741289 | -0.000295 | 2.044650 |
| 1 | -0.000475 | -1.325702 | -1.825529 |
| 6 | -0.000883 | -2.298389 | 0.399451 |
| 1 | -3.254991 | 0.001029 | -0.324179 |
| 1 | -2.212889 | 0.000639 | -1.835527 |
| 1 | -0.741350 | 0.000286 | 2.044656 |
| 1 | -2.338508 | 0.000835 | 1.425509 |
| 6 | 2.246116 | -0.000680 | -0.740449 |
| 1 | 2.212916 | -0.000659 | -1.835512 |
| 1 | 3.255006 | -0.001053 | -0.324160 |
| 1 | 0.000967 | 2.621747 | 1.444420 |
| 1 | 0.001163 | 3.137418 | -0.298505 |
| 1 | -0.001176 | -3.137418 | -0.298489 |
| 1 | -0.000993 | -2.621723 | 1.444432 |

--------------------------------------------------------------

The Cartesian coordinate of optimized 16 compound at B3LYP/6-311++g(d,p).
----------------------------------------------------------------

| Atomic Number | Coordinates (Angstroms) | | |
|---|---|---|---|
| | X | Y | Z |

----------------------------------------------------------------

| | | | |
|---|---|---|---|
| 6 | 1.127994 | 1.074175 | 0.012161 |
| 6 | 0.000209 | 0.721301 | -0.907319 |
| 6 | -1.127349 | 1.074802 | 0.012171 |
| 6 | 1.295782 | -0.000367 | 1.111870 |
| 6 | 1.127361 | -1.074799 | 0.012176 |
| 6 | -0.000192 | -0.721301 | -0.907317 |
| 6 | -1.127981 | -1.074175 | 0.012158 |
| 6 | -2.266164 | 0.000526 | -0.746160 |
| 6 | -1.295777 | 0.000370 | 1.111863 |
| 1 | 0.000374 | 1.314082 | -1.825532 |
| 6 | 0.000656 | 2.308403 | 0.408224 |
| 1 | 2.342580 | -0.000665 | 1.435186 |
| 1 | 0.746172 | -0.000202 | 2.034382 |
| 1 | -0.000356 | -1.314083 | -1.825530 |
| 6 | -0.000715 | -2.308402 | 0.408227 |
| 1 | -3.269124 | 0.000826 | -0.329256 |
| 1 | -2.248733 | 0.000516 | -1.837024 |
| 1 | -0.746173 | 0.000207 | 2.034377 |
| 1 | -2.342578 | 0.000667 | 1.435172 |
| 6 | 2.266183 | -0.000533 | -0.746145 |
| 1 | 2.248759 | -0.000524 | -1.837010 |
| 1 | 3.269140 | -0.000832 | -0.329235 |
| 1 | 0.000755 | 2.623240 | 1.451032 |
| 1 | 0.000898 | 3.153893 | -0.274526 |
| 1 | -0.000944 | -3.153893 | -0.274521 |
| 1 | -0.000815 | -2.623237 | 1.451036 |

----------------------------------------------------------------

The Cartesian coordinate of optimized 16 compound at M06L/6-311++g(d,p).
---------------------------------------------------------------

| Atomic Number | Coordinates (Angstroms) | | |
|---|---|---|---|
| | X | Y | Z |
---------------------------------------------------------------
| 6 | 1.119473 | 1.060511 | 0.012924 |
| 6 | 0.000153 | 0.717142 | -0.906713 |
| 6 | -1.119005 | 1.060967 | 0.012928 |
| 6 | 1.282396 | -0.000266 | 1.108903 |
| 6 | 1.119015 | -1.060963 | 0.012933 |
| 6 | -0.000139 | -0.717142 | -0.906711 |
| 6 | -1.119464 | -1.060511 | 0.012922 |
| 6 | -2.233608 | 0.000384 | -0.723361 |
| 6 | -1.282393 | 0.000271 | 1.108896 |
| 1 | 0.000278 | 1.322043 | -1.817839 |
| 6 | 0.000481 | 2.276755 | 0.387049 |
| 1 | 2.333801 | -0.000486 | 1.420587 |
| 1 | 0.747093 | -0.000149 | 2.041993 |
| 1 | -0.000265 | -1.322045 | -1.817836 |
| 6 | -0.000527 | -2.276752 | 0.387054 |
| 1 | -3.240077 | 0.000607 | -0.308842 |
| 1 | -2.218030 | 0.000376 | -1.817950 |
| 1 | -0.747095 | 0.000155 | 2.041990 |
| 1 | -2.333800 | 0.000491 | 1.420575 |
| 6 | 2.233623 | -0.000394 | -0.723350 |
| 1 | 2.218052 | -0.000389 | -1.817939 |
| 1 | 3.240089 | -0.000615 | -0.308825 |
| 1 | 0.000556 | 2.602868 | 1.429777 |
| 1 | 0.000660 | 3.122804 | -0.298160 |
| 1 | -0.000695 | -3.122804 | -0.298152 |
| 1 | -0.000601 | -2.602860 | 1.429783 |
---------------------------------------------------------------

The Cartesian coordinate of optimized 17 compound at HF/6-311++g(d,p).

---

| Atomic | Coordinates (Angstroms) | | |
|---|---|---|---|
| Number | X | Y | Z |

---

| | | | |
|---|---|---|---|
| 6 | -1.238966 | -0.661609 | 0.981934 |
| 6 | 0.000000 | -1.280292 | 0.366088 |
| 6 | 1.238966 | -0.661609 | 0.981934 |
| 6 | 2.223859 | -0.774279 | -0.193695 |
| 6 | -1.391643 | 1.381564 | -0.273756 |
| 6 | -1.378792 | 0.107840 | -1.134038 |
| 6 | 0.000000 | -0.635696 | -1.078776 |
| 6 | 1.378792 | 0.107840 | -1.134038 |
| 6 | 1.391644 | 1.381564 | -0.273755 |
| 1 | -1.551988 | -1.086734 | 1.928578 |
| 1 | 0.000000 | -2.365525 | 0.349645 |
| 1 | 3.214033 | -0.368586 | -0.012925 |
| 1 | -2.425154 | 1.676097 | -0.112917 |
| 1 | -0.898282 | 2.241988 | -0.692024 |
| 1 | -1.763338 | 0.240077 | -2.139063 |
| 1 | 0.000000 | -1.407019 | -1.842261 |
| 1 | 1.763338 | 0.240076 | -2.139063 |
| 1 | 0.898282 | 2.241988 | -0.692024 |
| 1 | 2.425154 | 1.676096 | -0.112917 |
| 6 | -2.223859 | -0.774279 | -0.193695 |
| 1 | -3.214033 | -0.368586 | -0.012925 |
| 1 | -2.331113 | -1.803413 | -0.526148 |
| 1 | 2.331113 | -1.803413 | -0.526148 |
| 6 | 0.788369 | 0.841133 | 1.041269 |
| 6 | -0.788369 | 0.841134 | 1.041269 |
| 1 | 1.551988 | -1.086734 | 1.928578 |
| 1 | 1.169975 | 1.366908 | 1.908586 |
| 1 | -1.169975 | 1.366908 | 1.908586 |

---

The Cartesian coordinate of optimized 17 compound at MP2/6-31G(d).

```
-------------------------------------------------------------
Atomic          Coordinates (Angstroms)
Number          X           Y           Z
-------------------------------------------------------------
  6         -1.241141   -0.664098    0.981074
  6          0.000000   -1.280168    0.366478
  6          1.241141   -0.664098    0.981074
  6          2.220542   -0.776041   -0.199860
  6         -1.388430    1.381417   -0.268894
  6         -1.375039    0.113214   -1.133322
  6          0.000000   -0.626758   -1.078064
  6          1.375039    0.113214   -1.133322
  6          1.388430    1.381418   -0.268894
  1         -1.557306   -1.096045    1.936426
  1          0.000000   -2.377566    0.344120
  1          3.222590   -0.369336   -0.021044
  1         -2.432228    1.678011   -0.104538
  1         -0.888531    2.252818   -0.687222
  1         -1.761884    0.250313   -2.149188
  1          0.000001   -1.408781   -1.848488
  1          1.761884    0.250314   -2.149188
  1          0.888530    2.252819   -0.687221
  1          2.432227    1.678012   -0.104539
  6         -2.220542   -0.776040   -0.199860
  1         -3.222590   -0.369335   -0.021043
  1         -2.320099   -1.813598   -0.541458
  1          2.320098   -1.813599   -0.541459
  6          0.786762    0.836722    1.042978
  6         -0.786761    0.836722    1.042978
  1          1.557307   -1.096045    1.936426
  1          1.172679    1.364499    1.922110
  1         -1.172679    1.364499    1.922110
-------------------------------------------------------------
```

The Cartesian coordinate of optimized 17 compound at MP2/6-311++g(d,p).
---------------------------------------------------------------

| Atomic Number | Coordinates (Angstroms) | | |
|---|---|---|---|
| | X | Y | Z |

---------------------------------------------------------------

| | | | |
|---|---|---|---|
| 6 | -1.243612 | -0.665753 | 0.983970 |
| 6 | -0.000001 | -1.285138 | 0.368299 |
| 6 | 1.243617 | -0.665756 | 0.983972 |
| 6 | 2.223962 | -0.774779 | -0.201039 |
| 6 | -1.386259 | 1.383914 | -0.269722 |
| 6 | -1.374721 | 0.113583 | -1.137229 |
| 6 | -0.000005 | -0.630324 | -1.080660 |
| 6 | 1.374719 | 0.113590 | -1.137230 |
| 6 | 1.386257 | 1.383918 | -0.269717 |
| 1 | -1.562546 | -1.095935 | 1.937735 |
| 1 | -0.000003 | -2.381275 | 0.344872 |
| 1 | 3.222215 | -0.359877 | -0.022188 |
| 1 | -2.431690 | 1.676306 | -0.107902 |
| 1 | -0.879525 | 2.252824 | -0.684575 |
| 1 | -1.762214 | 0.249680 | -2.151711 |
| 1 | -0.000003 | -1.412558 | -1.849865 |
| 1 | 1.762212 | 0.249691 | -2.151711 |
| 1 | 0.879523 | 2.252832 | -0.684562 |
| 1 | 2.431689 | 1.676308 | -0.107899 |
| 6 | -2.223958 | -0.774778 | -0.201032 |
| 1 | -3.222208 | -0.359871 | -0.022175 |
| 1 | -2.320569 | -1.813185 | -0.540721 |
| 1 | 2.320567 | -1.813187 | -0.540731 |
| 6 | 0.788060 | 0.837802 | 1.046456 |
| 6 | -0.788060 | 0.837802 | 1.046453 |
| 1 | 1.562552 | -1.095942 | 1.937733 |
| 1 | 1.174450 | 1.364856 | 1.924293 |
| 1 | -1.174452 | 1.364855 | 1.924290 |

---------------------------------------------------------------

The Cartesian coordinate of optimized 17 compound at B3LYP/6-311++g(d,p).
------------------------------------------------------------

| Atomic Number | Coordinates (Angstroms) | | |
|---|---|---|---|
| | X | Y | Z |

------------------------------------------------------------

| | | | |
|---|---|---|---|
| 6 | -1.245882 | -0.670594 | 0.984532 |
| 6 | 0.000000 | -1.289456 | 0.366078 |
| 6 | 1.245882 | -0.670594 | 0.984531 |
| 6 | 2.234646 | -0.775865 | -0.197343 |
| 6 | -1.392694 | 1.388043 | -0.270195 |
| 6 | -1.384504 | 0.112820 | -1.139473 |
| 6 | 0.000000 | -0.633917 | -1.089836 |
| 6 | 1.384504 | 0.112820 | -1.139473 |
| 6 | 1.392693 | 1.388043 | -0.270195 |
| 1 | -1.564005 | -1.101519 | 1.935203 |
| 1 | 0.000000 | -2.382442 | 0.345162 |
| 1 | 3.229403 | -0.361748 | -0.012326 |
| 1 | -2.432299 | 1.693045 | -0.111998 |
| 1 | -0.884068 | 2.251548 | -0.687929 |
| 1 | -1.774785 | 0.250054 | -2.149659 |
| 1 | 0.000000 | -1.408755 | -1.860840 |
| 1 | 1.774785 | 0.250055 | -2.149659 |
| 1 | 0.884067 | 2.251549 | -0.687928 |
| 1 | 2.432298 | 1.693047 | -0.111998 |
| 6 | -2.234646 | -0.775865 | -0.197342 |
| 1 | -3.229403 | -0.361748 | -0.012326 |
| 1 | -2.346839 | -1.809790 | -0.537898 |
| 1 | 2.346838 | -1.809790 | -0.537898 |
| 6 | 0.791873 | 0.841796 | 1.050504 |
| 6 | -0.791873 | 0.841796 | 1.050504 |
| 1 | 1.564006 | -1.101519 | 1.935203 |
| 1 | 1.178252 | 1.366928 | 1.925574 |
| 1 | -1.178252 | 1.366928 | 1.925574 |

------------------------------------------------------------

The Cartesian coordinate of optimized 17 compound at M06L/6-311++g(d,p).
------------------------------------------------------------

| Atomic Number | Coordinates (Angstroms) | | |
|---|---|---|---|
| | X | Y | Z |

------------------------------------------------------------

| | | | |
|---|---|---|---|
| 6 | -1.238859 | -0.662707 | 0.978086 |
| 6 | 0.000000 | -1.278836 | 0.365816 |
| 6 | 1.238859 | -0.662707 | 0.978086 |
| 6 | 2.217536 | -0.770855 | -0.197051 |
| 6 | -1.384181 | 1.377096 | -0.269211 |
| 6 | -1.371594 | 0.110718 | -1.130659 |
| 6 | 0.000000 | -0.628674 | -1.074863 |
| 6 | 1.371594 | 0.110718 | -1.130659 |
| 6 | 1.384182 | 1.377096 | -0.269211 |
| 1 | -1.556433 | -1.091912 | 1.931103 |
| 1 | 0.000000 | -2.373524 | 0.341018 |
| 1 | 3.214821 | -0.358502 | -0.017569 |
| 1 | -2.426249 | 1.675898 | -0.111068 |
| 1 | -0.882100 | 2.246568 | -0.685518 |
| 1 | -1.759920 | 0.246095 | -2.143248 |
| 1 | 0.000000 | -1.409484 | -1.842631 |
| 1 | 1.759920 | 0.246095 | -2.143249 |
| 1 | 0.882101 | 2.246568 | -0.685518 |
| 1 | 2.426250 | 1.675897 | -0.111068 |
| 6 | -2.217537 | -0.770855 | -0.197051 |
| 1 | -3.214821 | -0.358501 | -0.017569 |
| 1 | -2.325563 | -1.806640 | -0.535879 |
| 1 | 2.325562 | -1.806640 | -0.535879 |
| 6 | 0.785951 | 0.835674 | 1.039617 |
| 6 | -0.785950 | 0.835674 | 1.039617 |
| 1 | 1.556433 | -1.091913 | 1.931103 |
| 1 | 1.171974 | 1.362970 | 1.915437 |
| 1 | -1.171974 | 1.362971 | 1.915436 |

------------------------------------------------------------

The Cartesian coordinate of optimized 18 compound at HF/6-311++g(d,p).

---

| Atomic Number | Coordinates (Angstroms) | | |
|---|---|---|---|
| | X | Y | Z |
---

| 6  |  1.226596 | -0.152909 | -1.237704 |
| 6  |  0.000038 | -0.867058 | -0.710256 |
| 6  | -1.226560 | -0.152973 | -1.237698 |
| 6  | -2.281331 | -0.394208 | -0.150221 |
| 6  |  1.351891 |  1.747361 |  0.329118 |
| 6  |  1.471240 |  0.358491 |  0.945942 |
| 6  |  0.000036 | -0.392060 |  0.830902 |
| 6  | -1.471203 |  0.358423 |  0.945951 |
| 6  | -1.351980 |  1.747292 |  0.329097 |
| 1  |  1.511841 | -0.433508 | -2.238842 |
| 17 |  0.000086 | -2.589655 | -1.177936 |
| 17 | -3.861693 |  0.260832 | -0.679131 |
| 17 |  2.766001 |  2.870351 |  0.373543 |
| 1  |  0.714704 |  2.350072 |  0.910776 |
| 1  |  1.862192 |  0.313890 |  1.948090 |
| 17 |  0.000076 | -1.645945 |  2.103359 |
| 1  | -1.862147 |  0.313808 |  1.948102 |
| 1  | -0.714872 |  2.350098 |  0.910743 |
| 17 | -2.766249 |  2.870085 |  0.373454 |
| 6  |  2.281373 | -0.394134 | -0.150234 |
| 17 |  3.861747 |  0.260856 | -0.679169 |
| 17 |  2.688380 | -2.068914 |  0.310158 |
| 17 | -2.688363 | -2.068984 |  0.310168 |
| 6  | -0.794876 |  1.358379 | -1.066303 |
| 6  |  0.794837 |  1.358424 | -1.066294 |
| 1  | -1.511801 | -0.433591 | -2.238831 |
| 1  | -1.180468 |  1.991171 | -1.849721 |
| 1  |  1.180403 |  1.991246 | -1.849699 |

---

The Cartesian coordinate of optimized 18 compound at MP2/6-31G(d).
-------------------------------------------------------------

| Atomic | Coordinates (Angstroms) | | |
| --- | --- | --- | --- |
| Number | X | Y | Z |

-------------------------------------------------------------

| | | | |
| --- | --- | --- | --- |
| 6  | 0.005686  | -0.007790 | 0.007712 |
| 6  | 0.003452  | -0.006432 | 1.520241 |
| 6  | 1.437994  | -0.007656 | 1.999715 |
| 6  | 1.361703  | -0.745208 | 3.341244 |
| 6  | 0.422960  | -2.393050 | -0.451702 |
| 6  | -0.844782 | -2.074369 | 0.328893 |
| 6  | -0.422837 | -1.538516 | 1.826918 |
| 6  | 0.862990  | -2.074273 | 2.703800 |
| 6  | 2.006466  | -2.392978 | 1.750421 |
| 1  | 0.132814  | 0.978653  | -0.444897 |
| 17 | -0.904883 | 1.373479  | 2.172991 |
| 17 | 2.962549  | -0.712142 | 4.132451 |
| 17 | 0.301100  | -3.070086 | -2.114775 |
| 1  | 0.923295  | -3.229567 | -0.010324 |
| 1  | -1.593286 | -2.867896 | 0.382229 |
| 17 | -1.851218 | -1.805033 | 2.854045 |
| 1  | 0.574212  | -2.867790 | 3.396422 |
| 1  | 1.747213  | -3.229464 | 1.135610 |
| 17 | 3.544262  | -3.070182 | 2.395101 |
| 6  | -1.290357 | -0.745343 | -0.346831 |
| 17 | -1.530981 | -0.712430 | -2.116251 |
| 17 | -2.826422 | -0.130002 | 0.303948 |
| 17 | 0.256004  | -0.130087 | 4.590522 |
| 6  | 2.089014  | -1.038843 | 1.000238 |
| 6  | 1.160424  | -1.038926 | -0.291176 |
| 1  | 1.907458  | 0.978821  | 2.023215 |
| 1  | 3.118298  | -0.778856 | 0.739344 |
| 1  | 1.735407  | -0.778962 | -1.183867 |

-------------------------------------------------------------

The Cartesian coordinate of optimized 18 compound at MP2/6-311++g(d,p).
----------------------------------------------------------------

| Atomic Number | Coordinates (Angstroms) | | |
|---|---|---|---|
| | X | Y | Z |

----------------------------------------------------------------

| | | | |
|---|---|---|---|
| 6 | 1.226814 | -0.150984 | -1.242742 |
| 6 | 0.000482 | -0.868079 | -0.718145 |
| 6 | -1.228077 | -0.152720 | -1.242330 |
| 6 | -2.270675 | -0.395527 | -0.145162 |
| 6 | 1.350499 | 1.746163 | 0.331527 |
| 6 | 1.463004 | 0.358365 | 0.949823 |
| 6 | -0.000336 | -0.386769 | 0.830614 |
| 6 | -1.460858 | 0.360659 | 0.949898 |
| 6 | -1.352325 | 1.747916 | 0.327573 |
| 1 | 1.523804 | -0.437585 | -2.253216 |
| 17 | 0.002713 | -2.582994 | -1.180130 |
| 17 | -3.842069 | 0.267573 | -0.678828 |
| 17 | 2.760306 | 2.861505 | 0.381584 |
| 1 | 0.690053 | 2.348704 | 0.920553 |
| 1 | 1.857296 | 0.303670 | 1.965653 |
| 17 | 0.000099 | -1.652111 | 2.082126 |
| 1 | -1.854617 | 0.308149 | 1.966006 |
| 1 | -0.692708 | 2.355737 | 0.912014 |
| 17 | -2.765374 | 2.859098 | 0.375479 |
| 6 | 2.271527 | -0.393780 | -0.147650 |
| 17 | 3.841055 | 0.272957 | -0.680617 |
| 17 | 2.650710 | -2.068730 | 0.315931 |
| 17 | -2.646707 | -2.070596 | 0.319615 |
| 6 | -0.797743 | 1.357549 | -1.068041 |
| 6 | 0.795951 | 1.358354 | -1.065831 |
| 1 | -1.525905 | -0.441219 | -2.251987 |
| 1 | -1.190444 | 1.999681 | -1.859165 |
| 1 | 1.190495 | 2.002042 | -1.854781 |

----------------------------------------------------------------

The Cartesian coordinate of optimized 18 compound at B3LYP/6-311++g(d,p).
------------------------------------------------------------
| Atomic | Coordinates (Angstroms) | | |
| Number | X | Y | Z |
------------------------------------------------------------
| 6  |  1.234877 | -0.149861 | -1.248330 |
| 6  | -0.000085 | -0.866681 | -0.724641 |
| 6  | -1.234920 | -0.149669 | -1.248314 |
| 6  | -2.288832 | -0.393993 | -0.151209 |
| 6  |  1.355389 |  1.751450 |  0.330294 |
| 6  |  1.477929 |  0.361123 |  0.952691 |
| 6  | -0.000006 | -0.394806 |  0.838239 |
| 6  | -1.477854 |  0.361246 |  0.952717 |
| 6  | -1.355279 |  1.751564 |  0.330306 |
| 1  |  1.528603 | -0.431516 | -2.255767 |
| 17 | -0.000425 | -2.607657 | -1.189888 |
| 17 | -3.890184 |  0.273293 | -0.678683 |
| 17 |  2.787644 |  2.890278 |  0.379178 |
| 1  |  0.709995 |  2.366701 |  0.916317 |
| 1  |  1.872237 |  0.314065 |  1.962678 |
| 17 | -0.000092 | -1.667792 |  2.109535 |
| 1  | -1.872151 |  0.314201 |  1.962710 |
| 1  | -0.709901 |  2.366821 |  0.916338 |
| 17 | -2.787584 |  2.890333 |  0.379132 |
| 6  |  2.288880 | -0.394121 | -0.151262 |
| 17 |  3.890152 |  0.273236 | -0.678872 |
| 17 |  2.691648 | -2.090417 |  0.311778 |
| 17 | -2.691254 | -2.090405 |  0.311715 |
| 6  | -0.799428 |  1.368345 | -1.071716 |
| 6  |  0.799549 |  1.368234 | -1.071736 |
| 1  | -1.528709 | -0.431279 | -2.255746 |
| 1  | -1.190122 |  2.009699 | -1.857500 |
| 1  |  1.190309 |  2.009545 | -1.857522 |
------------------------------------------------------------

The Cartesian coordinate of optimized 18 compound at M06L/6-311++g(d,p).
---------------------------------------------------------------

| Atomic Number | Coordinates (Angstroms) | | |
|---|---|---|---|
| | X | Y | Z |

---------------------------------------------------------------

| | | | |
|---|---|---|---|
| 6  |  1.227156 | -0.143250 | -1.240907 |
| 6  |  0.000077 | -0.857631 | -0.723950 |
| 6  | -1.227129 | -0.143434 | -1.240920 |
| 6  | -2.271402 | -0.390453 | -0.148404 |
| 6  |  1.351149 |  1.744761 |  0.331313 |
| 6  |  1.462878 |  0.360828 |  0.947545 |
| 6  |  0.000004 | -0.385730 |  0.824175 |
| 6  | -1.462963 |  0.360701 |  0.947529 |
| 6  | -1.351235 |  1.744652 |  0.331328 |
| 1  |  1.523602 | -0.423068 | -2.250449 |
| 17 |  0.000417 | -2.585431 | -1.175570 |
| 17 | -3.856850 |  0.272410 | -0.671407 |
| 17 |  2.774453 |  2.861796 |  0.378523 |
| 1  |  0.701420 |  2.361176 |  0.916407 |
| 1  |  1.854727 |  0.305905 |  1.961108 |
| 17 |  0.000124 | -1.666349 |  2.066057 |
| 1  | -1.854821 |  0.305760 |  1.961088 |
| 1  | -0.701452 |  2.361027 |  0.916408 |
| 17 | -2.774427 |  2.861825 |  0.378658 |
| 6  |  2.271327 | -0.390341 | -0.148347 |
| 17 |  3.856865 |  0.272443 | -0.671200 |
| 17 |  2.655094 | -2.073710 |  0.314521 |
| 17 | -2.655567 | -2.073704 |  0.314551 |
| 6  | -0.794504 |  1.360537 | -1.059760 |
| 6  |  0.794374 |  1.360640 | -1.059752 |
| 1  | -1.523511 | -0.423292 | -2.250469 |
| 1  | -1.187081 |  2.003455 | -1.846718 |
| 1  |  1.186880 |  2.003592 | -1.846719 |

---------------------------------------------------------------

**Table S1.** The C-H inter-nuclear distances (Å), the X-C-H angles (degrees) and C-H stretching frequencies (cm$^{-1}$) of the CH$_2$/CClH unit, all computed at HF/6-311++G(d,p) level.

| Structure | C-H Inter-nuclear distances [a] | X-C-H bond angle[b] | Anti-symmetrical C-H stretching vibrations | Symmetrical C-H vibrations stretching |
|---|---|---|---|---|
| *1* | 1.077 (1.087) | 107.3 | 3282 | 3341 |
| *2* | 1.073 (1.086) | 107.0 | 3317 | 3387 |
| *3* | 1.072 | 106.7 | 3328 | 3403 |
| *4* | 1.076 (1.089) | 106.8 | 3287 | 3352 |
| *5* | 1.069 (1.088) | 105.4 | 3341 | 3438 |
| *6* | 1.062 (1.086) | 103.3 | 3388 | 3509 |
| *7* | 1.045 | 94.9 | 3564 | 3732 |
| *9* | 1.027 | 88.2 | 3739 | 3873 |
| *10* | 1.074 (1.085) | 108.1 | 3305 | 3377 |
| *11* | 1.062 | 101.5 | 3422 | 3516 |
| *12* | 1.058 | 100.8 | 3461 | 3576 |
| *13* | 1.053 | 99.2 | 3457 | 3693 |
| *14* | 1.046 | 97.5 | 3581 | 3751 |
| *15* | 1.050 | 98.3 | 3547 | 3661 |
| *16* | 1.062 (1.086) | 103.2 | 3371 | 3516 |
| *17* | 1.077 (1.087) | 106.1 | 3272 | 3328 |
| *18* | 1.052 | 96.0 | 3495 | 3620 |

[a] The C-H distances of "outer" CH bonds are given in the parenthesis.

[b] For structures 1-6, 10, 16, and 17 X is hydrogen whereas it is chlorine in the rest of structures.

**Table S2.** The C-H inter-nuclear distances (Å), the X-C-H angles (degrees) and C-H stretching frequencies (cm$^{-1}$) of the CH$_2$/CClH unit, all computed at B3LYP/6-311++G(d,p) level.

| Structure | C-H Inter-nuclear distances [a] | X-C-H bond angle [b] | Anti-symmetrical C-H stretching vibrations | Symmetrical C-H vibrations stretching |
|---|---|---|---|---|
| *1*  | 1.086 (1.095) | 107.6 | 3132 | 3168 |
| *2*  | 1.083 (1.094) | 107.2 | 3161 | 3205 |
| *3*  | 1.082         | 107.0 | 3170 | 3217 |
| *4*  | 1.085 (1.096) | 107.1 | 3133 | 3173 |
| *5*  | 1.079 (1.096) | 105.8 | 3179 | 3243 |
| *6*  | 1.073 (1.095) | 103.7 | 3218 | 3302 |
| *7*  | 1.060         | 94.9  | 3348 | 3452 |
| *8*  | 1.037         | 85.1  | 3546 | 3738 |
| *9*  | 1.043         | 87.1  | 3492 | 3599 |
| *10* | 1.083 (1.093) | 108.4 | 3149 | 3195 |
| *11* | 1.074         | 101.6 | 3230 | 3282 |
| *12* | 1.071         | 100.9 | 3261 | 3324 |
| *13* | 1.067         | 99.2  | 3249 | 3423 |
| *14* | 1.061         | 97.5  | 3360 | 3452 |
| *15* | 1.062         | 97.9  | 3347 | 3425 |
| *16* | 1.074 (1.096) | 103.6 | 3192 | 3294 |
| *17* | 1.086 (1.095) | 106.2 | 3120 | 3154 |
| *18* | 1.067         | 95.7  | 3283 | 3347 |

[a] The C-H distances of "outer" CH bonds are given in the parenthesis.

[b] For structures 1-6, 10, 16, and 17 X is hydrogen whereas it is chlorine in the rest of structures.

**Table S3.** The C-H inter-nuclear distances (Å), the X-C-H angles (degrees) and C-H stretching frequencies (cm$^{-1}$) of the CH$_2$/CClH unit, all computed at MP2/6-31G(d) level.

| Structure | C-H Inter-nuclear distances [a] | X-C-H bond angle[b] | Anti-symmetrical C-H stretching vibrations | Symmetrical C-H vibrations stretching |
|---|---|---|---|---|
| *1* | 1.088 (1.098) | 107.6 | 3217 | 3261 |
| *2* | 1.086 (1.097) | 107.3 | 3236 | 3288 |
| *3* | 1.085 | 107.1 | --- | --- |
| *4* | 1.088 (1.100) | 107.0 | 3208 | 3258 |
| *5* | 1.081 (1.100) | 105.5 | 3254 | 3329 |
| *6* | 1.076 (1.098) | 103.2 | 3285 | 3381 |
| *7* | 1.063 | 96.3 | --- | --- |
| *9* | 1.049 | 89.2 | --- | --- |
| *10* | 1.085 (1.096) | 108.3 | 3235 | 3290 |
| *11* | 1.078 | 102.7 | 3294 | 3365 |
| *12* | 1.074 | 101.9 | 3324 | 3408 |
| *13* | 1.070 | 100.3 | 3300 | 3525 |
| *14* | 1.064 | 98.6 | 3426 | 3543 |
| *15* | 1.067 | 99.5 | 3389 | 3470 |
| *16* | 1.076 (1.099) | 103.2 | 3266 | 3379 |
| *17* | 1.088 (1.097) | 106.1 | 3199 | 3237 |
| *18* | 1.070 | 96.8 | --- | --- |

[a] The C-H distances of "outer" CH bonds are given in the parenthesis.

[b] For structures 1-6, 10, 16, and 17 X is hydrogen whereas it is chlorine in the rest of structures.

**Table S4.** The C-H inter-nuclear distances (Å) and X-C-H angles (degrees) of the CH$_2$/CClH unit, all computed at MP2/6-311++G(d,p) level.

| Structure | C-H Inter-nuclear distances [a] | X-C-H bond angle [b] |
|---|---|---|
| *1* | 1.087 (1.098) | 108.1 |
| *2* | 1.085 (1.097) | 107.9 |
| *3* | 1.085 | 107.7 |
| *4* | 1.088 (1.100) | 107.7 |
| *5* | 1.082 (1.100) | 106.2 |
| *6* | 1.076 (1.099) | 103.8 |
| *7* | 1.064 | 96.3 |
| *10* | 1.085 (1.095) | 108.9 |
| *11* | 1.078 | 102.9 |
| *12* | 1.075 | 102.3 |
| *13* | 1.071 | 101.0 |
| *14* | 1.064 | 98.9 |
| *15* | 1.069 | 100.1 |
| *16* | 1.076 (1.099) | 103.9 |
| *17* | 1.088 (1.098) | 106.7 |
| *18* | 1.071 | 96.8 |

[a] The C-H distances of "outer" CH bonds are given in the parenthesis.

[b] For structures 1-6, 10, 16, and 17 X is hydrogen whereas it is chlorine in the rest of structures.